\documentclass{article}
\usepackage[a4paper, top=3cm, bottom=3cm, left=2.25cm, right=2.25cm]{geometry}
\usepackage{color}
\usepackage{ragged2e}
\usepackage{refcount}
\usepackage{titlesec}
\usepackage{graphicx}
\usepackage{etoolbox}
\usepackage{cite}
\usepackage[T1]{fontenc}
\usepackage{float}
\usepackage[toc,page]{appendix}
\usepackage{amssymb}
\usepackage{amsbsy}
\usepackage{dsfont}
\usepackage{lipsum}
\usepackage{mathptmx}
\usepackage{braket}
\usepackage[normalem]{ulem} 
\usepackage{slashed}
\usepackage[usenames,dvipsnames]{xcolor}
\usepackage{caption}
\usepackage{amsmath}
\usepackage{subcaption}
\usepackage[usenames,dvipsnames]{xcolor}
\usepackage[utf8]{inputenc}
\usepackage{latexsym}
\usepackage{enumitem}
\usepackage{bbding}
\usepackage{scrextend}
\deffootnote{0.5em}{0em}{\textsuperscript{\thefootnotemark}\,}

\DeclareMathSymbol{\ii}{\mathalpha}{letters}{"10}
\DeclareMathSymbol{\jj}{\mathalpha}{letters}{"11}

\usepackage{scalerel}
\usepackage{euscript}
\usepackage{cancel}
\usepackage{units}
\usepackage{lettrine}
\usepackage{braket}
\usepackage{bm}
\usepackage{booktabs}
\usepackage{cancel}
\usepackage{hyperref}
\usepackage{lscape}
\usepackage{abstract}
\usepackage{authblk}
\usepackage[usenames,dvipsnames]{xcolor}

\allowdisplaybreaks

\newcommand{\ga}{\gamma}

\newcommand{\ep}{\epsilon}

\titleformat*{\section}{\normalfont\fontsize{16}{19}\bfseries}
\titleformat*{\subsection}{\normalfont\fontsize{14}{17}\itshape}
\titleformat*{\subsubsection}{\normalfont\fontsize{14}{17}\selectfont}

\begin{document}

\title{{{Confronting the Lippmann-Schwinger equation \\ and the $N/D$ method for coupled-wave 
separable potentials}}}

\author[a]{M.~S.~S\'anchez\thanks{mario.sanchez2@um.es}}
\author[a]{J.~A.~Oller\thanks{oller@um.es}}
\author[b]{D.~R.~Entem\thanks{entem@usal.es}}
\affil[a]{\it Departamento de F\'{\i}sica, Universidad de Murcia, E-30071 Murcia, Spain}
\affil[b]{\it Grupo de F\'{\i}sica Nuclear and IUFFyM, Universidad de Salamanca, E-37008 Salamanca, Spain}

\maketitle

\begin{abstract}
We study a family of separable potentials with and without added contact interactions by solving the associated Lippmann-Schwinger equation with two coupled partial waves. The matching of the resulting amplitude matrix with the effective-range expansion is studied in detail. When a counterterm is included in the potential we also carefully discuss its renormalization. Next, we use the matrix $N/D$ method and study whether the amplitude matrices from the potentials considered admit an $N/D$ representation in matrix form. As a  novel result we show that it is typically not possible to find such matrix representation for the coupled partial-wave case. However, a separate $N/D$ representation for each coupled partial wave --- a valid option known in the literature --- is explicitly implemented and numerically solved in cases where the matrix $N/D$ method is unavailable. 
\end{abstract}

\newpage
\tableofcontents

\newpage

\section{Introduction}
\label{sec.240103.0}

\bigskip
The $N/D$  
method 
was introduced by Chew and Mandelstam~\cite{Chew:1960iv} to study 
pion-pion ($\pi\pi$)  
interactions at low energies. In the complex $s$ plane, with $s$ the usual Mandelstam variable, a partial-wave amplitude was expressed as the quotient $n(s)/d(s)$. Here, $n(s)$ has only a left-hand cut (LHC) for $s<s_L$ and $d(s)$ has only a unitarity or right-hand cut (RHC) for $s>s_R$, with $s_L$ and $s_R$ real numbers (for $\pi\pi$ scattering, $s_L=0$ and $s_R=4m_\pi^2$, where $m_\pi$ is the pion mass in the isospin limit).  Since pions are spinless, in this pioneering study there was no coupling between partial-wave amplitudes.

\bigskip
The extension of the $N/D$ method to coupled channels, quite straightforward by means of a matrix formalism,  
was originally given 
by Bjorken~\cite{Bjorken:1960zz}. Adapting to the notation used by us afterwards, one expresses the amplitude matrix as $\bm{T}=\bm{D}^{-1}\bm{N}$, where $\bm{D}$ and $\bm{N}$ are $n\times n$ matrices, $n$ being the number of coupled channels and/or partial waves --- this is what we call a {\it matrix $N/D$ representation} of the amplitude. However, time-reversal invariance requires the amplitude matrix to be symmetric, an aspect that is not transparent in the matrix $N/D$ form. Soon afterwards, 
Bjorken and Nauenberg~\cite{Bau:1960zz} addressed this issue and arrived to a partial solution, showing that --- provided the matrix representation given above --- the function $\bm{D}\,(\bm{T}-\bm{T}^\top)\,\bm{D}^\top$ is analytic everywhere in the  complex $s$ plane; thus, if one assumes its asymptotic vanishing, then $\bm{T}=\bm{T}^\top$ in the whole $s$ plane~\footnote{If there are bound states (zeroes of $\bm{D}$), then consider instead the matrix $|\bm{D}|\,\bm{T}$, where in the notation we use throughout the manuscript $|\bm{M}|$ refers to the determinant of $\bm{M}$.}. However, the limitation of this result lies precisely on the assumption that such a combination vanishes at infinity. This assumption could be natural when purely {\it regular} interactions are considered --- but not for {\it singular} interactions which commonly are the ones one has to deal with when tackling the problem within a quantum effective field theory (EFT) approach. Indeed, the EFT potential typically diverges at infinity with increasing power of the momentum as the order of the EFT calculation increases.

\bigskip
As the matter of our investigation is non-relativistic potential scattering, here we do not consider the $s$ plane but rather the so-called $A$ plane  
--- we take $A\equiv p^2$ with $p$ the magnitude of the center-of-mass momentum. Our prototype case is  nucleon-nucleon ($NN$) scattering, where there is only one channel 
mixing two different spin-triplet waves with common total angular momentum $J$ and respective orbital angular momentum $\ell = (J\mp1)$ and $\ell' = (J\pm1)$ 
--- the most prominent case in this respect, on which we focus throughout this work, is $J=1$, corresponding to the ${^3S_1}\!-{^3D_1}$ partial-wave mixing.
Since the seminal works of Weinberg~\cite{Weinberg:1990rz,Weinberg:1991um} it has become increasingly popular to address $NN$ scattering by applying the rules (`power counting') of Chiral Perturbation Theory (ChPT) to calculate the potential \footnote{Of course, in this framework the leading finite-range part of the $NN$ interaction is due to one-pion exchange (OPE).}, and then solving a Lippmann-Schwinger (or Schr{\"o}dinger) equation to get the non-perturbative $T$ matrix --- an avenue originally explored in Refs.~\cite{Ordonez:1992xp,Ordonez:1993tn,Ordonez:1995rz}.  For recent progress in this regard, the interested reader can consult Ref.~\cite{RodriguezEntem:2020jgp}, while Refs. \cite{Epelbaum:2008ga,Machleidt:2011zz,Hammer:2020ju} provide comprehensive reviews. 

\bigskip
For decades, different versions of the $N/D$ method have been applied to study the $NN$ system. The discontinuity along the LHC was calculated within some sort of perturbative, phenomenological approach. For instance, Ref.~\cite{Noyes:1959zz} studied only $S$ waves with OPE as the single source of the LHC discontinuity.
Additional heavier mesons were also considered in Refs.~\cite{Scotti:1963zz,Scotti:1965zz} as inputs for the LHC discontinuity, along the lines of the meson-exchange theory for nuclear forces that was so popular at that time. In turn, Ref.~\cite{Klar:1989jpg} modeled the LHC discontinuity by means of OPE together with one or two \textit{ad hoc} poles. More recently, in the set of works 
\cite{Oller:2003px,Albaladejo:2011bu,Albaladejo:2012sa,Guo:2013rpa,Oller:2014uxa} the discontinuity along the LHC was obtained from the perturbative calculation of the $NN$ potential within ChPT up to next-to-next-to-leading order.

\bigskip
A breakthrough in the applications of the $N/D$ method to the $NN$ sector (and to any other coupled-wave  
system) came \cite{Entem:2016ipb,Oller:2018zts} with the {\it exact} calculation of the LHC discontinuity of the $T$ matrix from the knowledge of the spectral-decomposable potential, which gave rise to the so-called {\it exact} $N/D$ method. From the discontinuity of a regular potential along the LHC one can calculate the same $T$ matrix as the one obtained by solving the associated Lippmann-Schwinger equation. Conversely, when singular potentials are considered, for the uncoupled case it was shown  \cite{Entem:2016ipb,Entem:2021kvs} that one recovers the standard results \cite{PavonValderrama:2005gu,PavonValderrama:2005wv,PavonValderrama:2005uj} obtained from a regularized Lippmann-Schwinger equation 
after the cancellation via counterterms of the loop divergences (renormalization) and the subsequent removal of the regulator (e.g. a 
momentum cutoff sent to infinity).
Remarkably,  one can also go beyond these solutions and generate new ones that  cannot be obtained by adding counterterms to the potential \cite{Entem:2007jg}, both for attractive \cite{Entem:2016ipb,Oller:2018zts} and repulsive \cite{Entem:2021kvs}  singular interactions. This is very welcome, since such new solutions account for microscopic physics that is needed for an accurate reproduction of experimental data. Concerning coupled $NN$ ${^3S_1}\!-{^3D_1}$ scattering 
governed \textit{only} by the OPE potential, it was shown in Ref.~\cite{Oller:2018zts} that one can employ a matrix $N/D$ method to achieve the renormalized solution obtained from the Lippmann-Schwinger equation after removing the cutoff \cite{PavonValderrama:2005wv}. But, contrarily to the uncoupled case, no extra solutions were obtained in Ref.~\cite{Oller:2018zts} despite the (unpublished) efforts dedicated on this respect by one of the authors.

\bigskip
This fact triggered from our side  
a further study of the coupled partial-wave case within the matrix $N/D$ method. The strategy we have adopted is to confront the matrix $N/D$ method 
with the case study of separable potentials that give rise to explicitly solvable mixed-wave scattering. The results of such efforts have lead to this work. We have encountered an interesting limitation in the application of the matrix $N/D$ method that, at the best of our knowledge, has not been previously realized in the literature. As we show below, we have found an infinite family of regular potentials coupling two partial-wave amplitudes with orbital angular momentum 0 and 2 that do \textit{not} admit a matrix $N/D$ representation. This fact hinders the applicability of this method, and it  
might be behind its failure to generate extra solutions in addition to the standard ones, as indicated above. However, the algebraic solutions to the Lippmann-Schwinger equation one finds do admit an $N/D$ representation where one applies the $N/D$ method to each partial-wave amplitude separately, as required by analyticity and unitarity. At the practical level one solves these equations iteratively by enforcing them to satisfy unitarity in coupled channels after each iteration, thus generating the input for the next round. This is repeatedly implemented until convergence is reached  \cite{Noyes:1960abc,Albaladejo:2012sa,Guo:2013rpa,Oller:2014uxa}. This method,  in addition to its applicability  for these potentials for which the matrix $N/D$ representation cannot be used, has also the benefit of leading to considerably simpler dispersive integral equations (IEs) as compared to the matrix $N/D$ method.

\bigskip
The contents of this manuscript are structured as follows. After this introduction, we solve explicitly the Lippmann-Swinger equation and its matching with the effective-range expansion (ERE) in Secs. \ref{sec.240103.1} and \ref{sec.240103.2}, respectively, for a particular potential introduced in Sec. \ref{sec.240103.1}. Next, Secs. \ref{sec.231228.1} and \ref{sec.231222.1} are devoted to work out the matrix $N/D$ representation of the amplitude due to this potential, as well as its matching with the ERE within the matrix $N/D$ method. Two types of $N/D$ dispersion relations (DRs) are presented --- the case with $\bm{N}(0)=\bm{1}$ (Sec. \ref{sec.231228.1}) and that with $\bm{D}(0)=\bm{1}$ (Sec. \ref{sec.231222.1}). These two scenarios are mutually exclusive as we discuss below. 
Then, in Sec. \ref{sec.240101.1} we generalize the potential introduced in Sec. \ref{sec.240103.1} and consider a whole family of separable potentials for which one cannot find a consistent matrix $N/D$ representation. This is also the case when the potential of Sec. \ref{sec.240103.1} is supplemented with a momentum-independent, $S$-wave contact interaction, as we elaborate in Sec. \ref{sec.240103.5}. This motivates us to introduce and apply the $N/D$ method to each coupled partial wave separately in Sec. \ref{sec.240103.6}. It is shown that the exact $T$ matrix obtained by solving the Lippmann-Schwinger equation straightforwardly admits such an $N/D$ form in coupled waves, as it must. This $N/D$ method for coupled partial waves is  worked out numerically in agreement with the exact result. We finally give our conclusions and outlook in Sec. \ref{sec.240103.7}.

\bigskip
\bigskip
\bigskip

\section{Lippmann-Schwinger equation and loop integrals} 
\setcounter{equation}{0}   
\def\theequation{\arabic{section}.\arabic{equation}}
\label{sec.240103.1}

\bigskip
In this section we will present the interaction to which we will stick in the subsequent sections (until Sec. \ref{sec.240101.1}) together with the amplitude  
due to such an interaction as found from the Lippmann-Schwinger equation. Also we will lay the ground for the following sections by giving explicit expressions for the associated loop integrals and noticing some of their basic properties.

\bigskip
Our starting point is a coupled-wave system composed by two spin-1/2 particles (our prototype example is $NN$ scattering) governed by the separable interaction 
\begin{eqnarray}
V_{\,\ell'\ell}(Q',Q) &=& \frac{\gamma_{\,\ell'\ell}}{M_N}\,g_{\,\ell'}(Q')\,g_{\,\ell}(Q)\,, \label{orpot}\end{eqnarray}
in terms of the vertex function
\begin{eqnarray}
g_\ell(Q) &=& \frac{(Q-L)^{1/4}\,Q^{\,\ell/2}}{(Q+\Lambda_0^2)^{\,1+\ell/2}}\,. \label{gi}
\end{eqnarray}
In this context, $\ell = \lbrace0,2\rbrace$ is the orbital angular momentum, both $Q$ and $\gamma_{\,\ell'\ell}=\gamma_{\,\ell\ell'}$ have dimensions of momentum squared, $L\leqslant0$ sets the onset of the LHC exhibited by the interaction \footnote{It is well-known \cite{Oller:2018zts} that $L=-m_\pi^2/4$ for  $NN$ scattering.}, and the parameter $\Lambda_0$, which may be regarded as the mass of some exchanged heavy particle and is introduced to ensure the regularity of the interaction, fulfills the condition $\Lambda_0^2>-L$. 

\bigskip
The potential \eqref{orpot} enters the coupled Lippmann-Schwinger equation
\begin{eqnarray}
T_{\,\ell'\ell}(Q',Q;A) &=& V_{\,\ell'\ell}(Q',Q) \,+\, \frac{M_N}{2\pi^2} \sum_{m=0,2}\int_0^\infty dq\,V_{\,\ell'm}(Q',q^2) \frac{q^2}{q^2-A} T_{\,m\ell}(q^2,Q;A)\,,\label{LSij}
\end{eqnarray}
where $A$ has a non-vanishing imaginary part, $\mathfrak{I}(A)\neq0$, so as to avoid a vanishing denominator in the previous integrand. 
  Plugging into Eq. \eqref{LSij} the ansatz 
\begin{eqnarray}
T_{\,\ell'\ell}(Q',Q;A) &=& \frac{\tau_{\,\ell'\ell}(A)}{M_N}\,g_{\,\ell'}(Q')\,g_{\,\ell}(Q)\,, \label{tij}
\end{eqnarray} 
which is separable in the squared-momentum variables as well, one can easily solve for $\tau_{\,\ell'\ell}(A)$, finding 
\begin{eqnarray}
\tau_{\,00}(A) &=& \left[ \frac{1-\gamma_{\,22}\,I_{\,2}(A)}{\gamma_{\,00}\,-\,|\bm{\gamma}\,|\,I_{\,2}(A)} \,-\, I_{\,0}(A)\right]^{-1}\,; \label{tau0}\\
\tau_{\,02}(A) &=& \gamma_{\,02}\left[1\,-\,\gamma_{\,00}\,I_{\,0}(A)\,-\,\Big(\,\gamma_{\,22}\,-\,|\bm{\gamma}\,|\,I_{\,0}(A)\,\Big)\,I_{\,2}(A)\right]^{-1} \quad\!\!\!=\!\!\!\quad \tau_{\,20}(A)\,; \label{tau1}\\
\tau_{\,22}(A) &=& \left[ \frac{1-\gamma_{\,00}\,I_{\,0}(A)}{\gamma_{\,22}\,-\,|\bm{\gamma}\,|\,I_{\,0}(A)} \,-\, I_{\,2}(A)\right]^{-1}\,, \label{tau2}\end{eqnarray}
with $|\bm{\gamma}\,| = \gamma_{\,00}\,\gamma_{\,22}\,-\,\gamma_{\,02}^{\,2}$ the determinant of the matrix
\begin{eqnarray}
\bm{\gamma} &=& \begin{pmatrix}
\gamma_{\,00} & \gamma_{\,02} \\ \gamma_{\,20} & \gamma_{\,22}
\end{pmatrix}\,. \label{gammamat}
\end{eqnarray}
 
\bigskip
In Eqs. \eqref{tau0}--\eqref{tau2} we introduced the loop integral
\begin{eqnarray}
I_{\,\ell}(A) &=& \frac{1}{2\pi^2}\int_0^{\,\infty} dq\,\frac{q^2\,g_\ell(q^2)^2}{q^2-A}\notag\\
&=& -\,\frac{1}{4\pi^2\,(1+\frac{A}{\Lambda_0^2})^{\,\ell+2}\,\Lambda_0^2}\,\Bigg[ \frac{1}{(1+x)^\ell}\,\sum_{m=0}^{\ell+1}\Bigg(\mathfrak{a}_{\,\ell m} \,+\, \frac{\text{arctanh}\,\sqrt{1+x}}{\sqrt{1+x}}\,\mathfrak{b}_{\,\ell m}\Bigg)\,\frac{A^{m}}{\Lambda_0^{2m}}\notag \\
&& +\, \frac{2\,A^{\ell}\sqrt{-\frac{A}{\Lambda_0^2}\,(\frac{A}{\Lambda_0^2}-x)}}{\Lambda_0^{2\ell}} \, \arctan\!\Big(\frac{\sqrt{\frac{A}{\Lambda_0^2}-x}}{\sqrt{-\frac{A}{\Lambda_0^2}}}\Big)\Bigg]\,, \label{loop}
\end{eqnarray}
where the $A$-independent coefficients $\mathfrak{a}_{\,\ell m}$ and $\mathfrak{b}_{\,\ell m}$ are $\ell$th- and $(\!\ell\!+\!1\!)$th-degree polynomials in the dimensionless parameter $x\equiv L/\Lambda_0^2$~\,($-1<x\leqslant0$), namely
\begin{eqnarray}
\mathfrak{a}_{\,00} \,\,=\,\, -1\,, &&\!\!\!\!\!\!\! \mathfrak{a}_{\,01} \,\,=\,\, -1\,;\label{a0m} \\
\mathfrak{a}_{\,20} \,\,=\,\, -\tfrac{1}{3}-\tfrac{7}{12}x-\tfrac{1}{8}x^2\,, &&\!\!\!\!\!\!\! \mathfrak{a}_{\,21}\,\, = \,\, -\tfrac{3}{2}-\tfrac{5}{2}x-\tfrac{5}{8}x^2\,, \quad \mathfrak{a}_{\,22}\,\, = \,\, -3-\tfrac{21}{4}x-\tfrac{15}{8}x^2\,, \quad \mathfrak{a}_{\,23}\,\, = \,\, -\tfrac{11}{6}-\tfrac{10}{3}x-\tfrac{11}{8}x^2\,; ~~~~~~~~ \label{a2m} \\ 
\mathfrak{b}_{\,00} \,\,=\,\, x\,, &&\!\!\!\!\!\!\! \mathfrak{b}_{\,01} \,\,=\,\, -\,x-2\,; \label{b0m}\\
\mathfrak{b}_{\,20}\,\,=\,\, \tfrac{1}{8}x^3\,, &&\!\!\!\!\!\!\! \mathfrak{b}_{\,21}\,\, = \,\, \tfrac{1}{4}x^2+\tfrac{5}{8}x^3\,, \quad \mathfrak{b}_{\,22}\,\, = \,\, x+\tfrac{5}{2}x^2+\tfrac{15}{8}x^3\,, \quad \mathfrak{b}_{\,23}\,\, = \,\, -2-5x-\tfrac{15}{4}x^2-\tfrac{5}{8}x^3\,. \label{b2m}  
\end{eqnarray}

\smallskip
\noindent If one assumes $A=\mathfrak{R}(A)\,\pm\,i\epsilon$ with $\mathfrak{R}(A)\geqslant0$ and $\epsilon\to0^+$ (scattering case), then Eq. \eqref{loop} becomes
\begin{eqnarray}
I_{\,\ell}(A) &=& \mathfrak{R}\left[I_{\,\ell}(A)\right] \,+\,i\,\frac{{\sqrt[{\rm I}]{A}}}{4\pi}\,g_\ell(A)^2\,, \label{compex}
\end{eqnarray}
where ${\sqrt[{\rm I}]{A}}$ 
is the square root of $A$ in the first Riemann sheet, defined with the cut on the positive real semi-axis [$0\leqslant{\rm arg}(A)<2\pi$], and
\begin{eqnarray}
\mathfrak{R}[I_{\,\ell}(A)] &=& -\,\frac{1}{4\pi^2\,(1+\frac{A}{\Lambda_0^2})^{\,\ell+2}\,\Lambda_0^2}\,\Bigg[ \frac{1}{(1+x)^\ell}\,\sum_{m=0}^{\ell+1}\Bigg(\mathfrak{a}_{\,\ell m} \,+\, \frac{\text{arctanh}\,\sqrt{1+x}}{\sqrt{1+x}}\,\mathfrak{b}_{\,\ell m}\Bigg)\,\frac{A^{m}}{\Lambda_0^{2m}}\notag \\
&& +\, \frac{2\,A^{\ell}\sqrt{\frac{A}{\Lambda_0^2}\,(\frac{A}{\Lambda_0^2}-x)}}{\Lambda_0^{2\ell}} \, \text{arctanh}\Big(\frac{\sqrt{\frac{A}{\Lambda_0^2}}}{\sqrt{\frac{A}{\Lambda_0^2}-x}}\Big)\Bigg]\,. \label{reloop}
\end{eqnarray}
Around threshold ($0<A\ll-L<\Lambda_0^2$),  Eq. \eqref{reloop} is amenable to the series representation
\begin{eqnarray}
\mathfrak{R}[I_{\,\ell}(A)] &=& \sum_{n=0}^\infty \mathfrak{c}_{\,\ell}^{[n]}\,\frac{A^n}{\Lambda_0^{2n}}\,, \label{Aexp}
\end{eqnarray}
the coefficient  
$\mathfrak{c}_{\,\ell}^{[n]}$ being proportional to the $n$th derivative of $I_{\,\ell}(A)$ evaluated at the origin, namely 
\begin{eqnarray}
\mathfrak{c}_{\,\ell}^{[n]} &=& \frac{\Lambda_0^{2n}}{n!}\,I_{\,\ell}^{\,(n)}(0) \quad\!\!\!=\!\!\!\quad \frac{\sqrt{-x}}{2\pi^2\,\Lambda_0^2}\left[\frac{_2F_1(-\tfrac{1}{2},\ell-n+\tfrac{1}{2};-n-\tfrac{1}{2};-\tfrac{1}{x})}{\left(3+2n\right)\begin{pmatrix}\ell+1\\\ell-n-\frac{1}{2}\end{pmatrix}} \,-\, \frac{_2F_1(\ell+2,n+1;n+\tfrac{5}{2};-\tfrac{1}{x})}{\left(3+2n\right)\left(1+2n\right)\begin{pmatrix}-\frac{1}{2}\\n\end{pmatrix}\left(-x\right)^{n+3/2}} \right]\,, \notag\\ \label{cln}
\end{eqnarray}
where \footnote{Even though the integral 

$$I_{\,\ell}^{\,(n)}(0) \quad\!\!\!=\quad\!\!\! \frac{n!}{2\pi^2}\int_0^{\,\infty} dq\,q^{\,-2n}\,g_\ell(q^2)^2
$$

\medskip
is ill-defined (infrared divergent) for $n>\ell$, Eq. \eqref{cln} holds true for each non-negative integer $n$ as its analytic continuation. Furthermore, in the latter equation the factorial of a real number $a$ should be understood as $\Gamma(a+1)$.}
\begin{eqnarray}_2F_1(s,t;u;z) &=& \sum_{n=0}^\infty\frac{(s)_n\,(t)_n}{(u)_n}\,\frac{z^n}{n!}\end{eqnarray}
is a hypergeometric function.  In this equation we have also introduced the Pochhammer symbol $(z)_m=z(z+1)\cdots(z+m-1)$, $(z)_0=1$, for any complex number $z$. Explicitly, the two first coefficients in the series \eqref{Aexp} read
\begin{eqnarray}
\mathfrak{c}_{\,\ell}^{[0]} &=& -\frac{1}{4\pi^2\Lambda_0^2\left(1+x\right)^\ell}\left[\mathfrak{a}_{\,\ell0}\,+\,\frac{\text{arctanh}\,\sqrt{1+x}}{\sqrt{1+x}}\,\mathfrak{b}_{\,\ell0}\right]\,; \label{cl0} \\
\mathfrak{c}_{\,\ell}^{[1]} &=& -\frac{1}{4\pi^2\Lambda_0^2\left(1+x\right)^\ell}\left[\Big(2\,\delta_{\,\ell0}\,-\,\left(\ell+2\right)\mathfrak{a}_{\,\ell0}\,+\,\mathfrak{a}_{\,\ell1}\Big)\,+\,\frac{\text{arctanh}\,\sqrt{1+x}}{\sqrt{1+x}}\,\Big(\!-\left(\ell+2\right)\mathfrak{b}_{\,\ell0}\,+\,\mathfrak{b}_{\,\ell1}\Big)\right]\,, \label{cl1}
\end{eqnarray}
with the $\mathfrak{a}_{\,\ell m}$ and $\mathfrak{b}_{\,\ell m}$ coefficients given in Eqs. \eqref{a0m}--\eqref{b2m}. Conversely, at high energies ($A\gg\Lambda_0^2$) Eqs. \eqref{compex} and \eqref{reloop} give up to $\mathcal{O}(1/A)$
\begin{eqnarray}
I_{\,\ell}(A) &=& -\frac{1}{4\pi^2\,A}\,\Bigg[\log\Big(\!-\frac{4}{x}\,\frac{A}{\Lambda_0^2}\Big)\,+\,\frac{1}{\left(1+x\right)^\ell}\Big(\mathfrak{a}_{\,\ell,\ell+1}\,+\,\frac{\text{arctanh}\,\sqrt{1+x}}{\sqrt{1+x}}\,\mathfrak{b}_{\,\ell,\ell+1}\Big)\,-\,i\pi\,+\,\dots\Bigg]\,. \label{asy}
\end{eqnarray}
Finally we point out that, even though it may not be obvious from Eq. \eqref{loop}, both loop integrals converge to finite values when evaluated at $A=-\Lambda_0^2$, as
\begin{eqnarray}
\lim_{A\to-\Lambda_0^2}I_{\,\ell}(A) &=& -\frac{1}{4\pi^2\Lambda_0^2\left(1+x\right)^{\ell+1}}\,\Bigg[\widetilde{\mathfrak{a}}_{\,\ell}\,+\,\Big(\frac{x~\text{arctanh}\,\sqrt{1+x}}{\sqrt{1+x}}\,-\,1\Big)\,\widetilde{\mathfrak{b}}_{\,\ell}\Bigg]\,, \label{atC}
\end{eqnarray}
where
\begin{eqnarray}
\mathfrak{\widetilde{a}}_{\,0} \,\,=\,\, -\tfrac{1}{2}\,, && \mathfrak{\widetilde{b}}_{\,0} \,\,=\,\, \tfrac{1}{4}x\,; \\
\mathfrak{\widetilde{a}}_{\,2} \,\,=\,\, -\tfrac{1}{4}-\tfrac{17}{24}x-\tfrac{59}{96}x^2\,, && \mathfrak{\widetilde{b}}_{\,2}\,\, = \,\, \tfrac{5}{64}x^3\,. 
\end{eqnarray}

\bigskip
\bigskip
\bigskip

\section{Matrix formalism and ERE}
\setcounter{equation}{0}   
\def\theequation{\arabic{section}.\arabic{equation}}
\label{sec.240103.2}

\bigskip
In this section we will re-express the solution of the Lippmann-Schwinger equation given in Sec. \ref{sec.240103.1} by means of a matrix formalism. Besides, we will make advantage of such a reformulation to give explicit relationships between the parameters of the potential \eqref{orpot} and the ERE parameters that emerge from it at low energies.

\bigskip
Employing the symmetric matrices
\begin{eqnarray}
\bm{I}(A) &=& \begin{pmatrix}
I_{\,0}(A) & 0 \\ 0 & I_{\,2}(A)
\end{pmatrix}\,; \\
\bm{\tau}(A) &=& \begin{pmatrix}
\tau_{\,00}(A) & \tau_{\,02}(A) \\ \tau_{\,20}(A) & \tau_{\,22}(A)
\end{pmatrix}\,,
\end{eqnarray}
together with Eq. \eqref{gammamat}, one can condense Eqs. \eqref{tau0}--\eqref{tau2} as the simple matrix identity
\begin{eqnarray}
\bm{\tau}(A) &=& [\bm{\gamma}^{\,-1} \,-\,\bm{I}(A)]^{\,-1}\,.
\end{eqnarray}
Consequently, 
\begin{eqnarray}
\bm{T}(A) &=& [\bm{V}(A)^{\,-1} \,-\,\bm{G}(A)]^{\,-1}\,, \label{mateq}
\end{eqnarray}
where
\begin{eqnarray}
\bm{V}(A) &=& \begin{pmatrix}
V_{\,00}(A,A) & V_{\,02}(A,A) \\ V_{\,20}(A,A) & V_{\,22}(A,A)
\end{pmatrix}\,; \\
\bm{G}(A) &=& \begin{pmatrix}
\frac{M_N\,I_{\,0}(A)}{g_{0}(A)^2} & 0 \\ 0 & \frac{M_N\,I_{\,2}(A)}{g_{2}(A)^2}
\end{pmatrix}\,; \\
\bm{T}(A) &=& \begin{pmatrix}
T_{\,00}(A,A;A) & T_{\,02}(A,A;A) \\ T_{\,20}(A,A;A) & T_{\,22}(A,A;A)
\end{pmatrix}\,.
\end{eqnarray}
Now, in virtue of Eq. \eqref{compex}, one gets from Eq. \eqref{mateq}
\begin{eqnarray}
\bm{M}(A) &\equiv& \bm{P}(A)\left[\frac{4\pi}{M_N}\,{\bm{T}}^{-1}(A)\,+\, i\,\sqrt[{\rm I}]{A}\,\bm{1}\right]\bm{P}(A) \notag\\
&=&\begin{pmatrix}
1 & 0 \\ 0 & A
\end{pmatrix}\,\begin{pmatrix}
\frac{4\pi}{g_0(A)^2}\left[\frac{\gamma_{\,22}}{|\bm{\gamma}\,|}-\mathfrak{R}\left(I_{\,0}(A)\right)\right] & -\,\frac{4\pi}{g_0(A)\,g_2(A)}\,\frac{\gamma_{\,02}}{|\bm{\gamma}\,|} \\ -\,\frac{4\pi}{g_0(A)\,g_2(A)}\,\frac{\gamma_{\,02}}{|\bm{\gamma}\,|} & \frac{4\pi}{g_2(A)^2}\left[\frac{\gamma_{\,00}}{|\bm{\gamma}\,|}-\mathfrak{R}\left(I_{\,2}(A)\right)\right]
\end{pmatrix}\,\begin{pmatrix}
1 & 0 \\ 0 & A
\end{pmatrix}\,, \label{unitarity}
\end{eqnarray}
where $\bm{1}$ is the $2\times2$ identity matrix, and  $\bm{P}(A)$ is the coupled-channel centrifugal factor given by 
\begin{eqnarray}
\bm{P}(A) &=& \text{diag}\,(A^{\ell_1/2},\ldots,A^{\ell_N/2})\,. \label{centrifugal}
\end{eqnarray} 

\bigskip
Consider the average orbital angular momentum $\overline{\ell}\equiv{(\ell'+\ell)}/{2}$, which is 0, 1, and 2 respectively for $\ell'=\ell=0$, $\ell'\neq\ell$, and $\ell'=\ell=2$. Introduce the dimensionless coefficient
\begin{eqnarray}
\mathfrak{p}_{\,\ell'\ell}^{[n]} &=& \sum_{k=0}^{\overline{\ell}+2}\frac{\begin{pmatrix}\overline{\ell}+2 \\ k\end{pmatrix}\begin{pmatrix}-\frac{1}{2} \\ n-k\end{pmatrix}}{(-x)^{\,n-k}}\,, \label{plm}
\end{eqnarray}
which enters the series representation
\begin{eqnarray}
[g_{\,\ell'}(A)\,g_{\,\ell}(A)]^{-1} &=& \frac{\Lambda_0^{3+2\overline{\ell}}}{\sqrt{-x}\,\,A^{\overline{\ell}}}\,\sum_{n=0}^\infty \mathfrak{p}_{\,\ell'\ell}^{[n]}\,\frac{A^n}{\Lambda_0^{2n}}\,. \label{vertex}
\end{eqnarray}
Using Eq. \eqref{vertex} together with Eq. \eqref{Aexp}, near threshold Eq. \eqref{unitarity} gives the generalized (matrix) ERE
\begin{eqnarray} \bm{M}(A) &=& \sum_{n=0}^\infty \bm{v}^{[n]}\,A^n\,, \label{gere} \end{eqnarray}
where the matrix element
\begin{eqnarray}
{v}_{\,\ell'\ell}^{[n]} &=& \frac{4\pi\,\Lambda_0^{3+2(\overline{\ell}-n)}}{\sqrt{-x}} \left( (\bm{\gamma}^{-1})_{\,\ell'\ell} \,\mathfrak{p}_{\,\ell'\ell}^{[n]} \,-\, \delta_{\,\ell'\ell}\sum_{k=0}^n \mathfrak{p}_{\,\ell'\ell}^{[k]}\,\mathfrak{c}_{\,\ell}^{[n-k]}\right)\,, \label{vlpln}
\end{eqnarray}
whose mass dimension is $[1+2(\overline{\ell}-n)]$, is called the $\ell'\ell$-channel $(n\!-\!1)$th shape parameter for $n\geqslant2$. Conversely, the scattering-length and effective-range matrices are given by
\begin{eqnarray}
\bm{a} &=& -\,(\,\bm{v}^{[0]}\,)^{-1}\,; \\
\bm{r} &=& 2\,\bm{v}^{[1]}\,. \label{aandr}
\end{eqnarray}
Coming back to the case of $SD$-wave coupling, we take \begin{eqnarray}
{\bm{a}} &=& \begin{pmatrix}
a_{\,00} & a_{\,02} \\ a_{\,20} & a_{\,22}\end{pmatrix}\,; \label{SLmat}\\ {\bm{r}} &=& \begin{pmatrix}
r_{\,00} & r_{\,02} \\ r_{\,20} & r_{\,22}\end{pmatrix}\,, \label{ERmat}
\end{eqnarray}
where the mass dimensions of $a_{\,\ell'\ell}$ and $r_{\,\ell'\ell}$ are $(-1\!-\!2\overline{\ell})$ and $(-1\!+\!2\overline{\ell})$, respectively. Specializing Eqs. \eqref{plm} and \eqref{vlpln} yields
\begin{eqnarray}
a_{\,00} &=& \frac{\sqrt{-x}}{4\pi\,\Lambda_0^3}\left[\mathfrak{c}_{\,0}^{[0]}\,-\,\frac{1-\mathfrak{c}_{\,2}^{[0]}\,\gamma_{\,22}}{\gamma_{\,00}\,-\,|\bm{\gamma}\,|\,\mathfrak{c}_{\,2}^{[0]}} \right]^{-1}\,, \label{a00} \\
a_{\,02} &=& \frac{\gamma_{\,02}}{\Lambda_0^2\left(\gamma_{\,00}\,-\,|\bm{\gamma}\,|\,\mathfrak{c}_{\,2}^{[0]}\right)}\,a_{\,00}\quad\!\!\!=\!\!\!\quad a_{\,20}\,, \\
a_{\,22} &=& \frac{\gamma_{\,22}\,-\,|\bm{\gamma}\,|\,\mathfrak{c}_{\,0}^{[0]}}{\Lambda_0^2\,\gamma_{\,02}}\,a_{\,02}\;; \\ 
\notag\\
r_{\,00} &=& \frac{4\pi\,\Lambda_0}{\sqrt{-x}}\left[\Big(4+\frac{1}{x}\Big)\,\Big(\frac{\gamma_{\,22}}{|\bm{\gamma}\,|}\,-\,\mathfrak{c}_{\,0}^{[0]}\Big)\,-\,2\,\mathfrak{c}_{\,0}^{[1]}\right]\,, \\
r_{\,02} &=& -\frac{4\pi\,\Lambda_0^3}{\sqrt{-x}}\,\Big(6+\frac{1}{x}\Big)\,\frac{\gamma_{\,02}}{|\bm{\gamma}\,|}\quad\!\!\!=\!\!\!\quad r_{\,20}\,, \\
r_{\,22} &=& \frac{4\pi\,\Lambda_0^5}{\sqrt{-x}}\left[\Big(8+\frac{1}{x}\Big)\,\Big(\frac{\gamma_{\,00}}{|\bm{\gamma}\,|}\,-\,\mathfrak{c}_{\,2}^{[0]}\Big)\,-\,2\,\mathfrak{c}_{\,2}^{[1]}\right]\,, \label{r22}
\end{eqnarray}
cf. Eqs. \eqref{cl0} and \eqref{cl1}.

\bigskip
\bigskip
\bigskip

\section{$\pmb{N/D}$ method and DRs: $\pmb{N(0)=1}$ normalization} 
\setcounter{equation}{0}   
\def\theequation{\arabic{section}.\arabic{equation}}
\label{sec.231228.1}

\bigskip

In the following two sections we will deduce the DRs fulfilled by the matrices $\bm{N}(A)$ and $\bm{D}(A)$ in the framework of the matrix $N/D$ method. Firstly we will show by construction that there is such a matrix representation for the on-shell amplitude $\bm{T}(A)$ stemming from the interaction \eqref{orpot} --- namely, these $\bm{N}(A)$ and $\bm{D}(A)$ matrices do exist and verify the matrix relationship $\bm{T}(A) = \bm{D}(A)^{-1}\bm{N}(A)$. Such a matrix product is not commutative in general, as it will be explicitly shown later in Sec. \ref{sec.231222.1}.

\bigskip
Since there is always an ambiguity in the definitions of $\bm{N}(A)$ and $\bm{D}(A)$ due to the fact that, given any invertible matrix $\bm{m}_I(A)$, the matrices $\overline{\bm{N}}(A)=\bm{m}_I(A)\bm{N}(A)$ and $\overline{\bm{D}}(A)=\bm{m}_I(A)\bm{D}(A)$ fulfill $\bm{T}(A) = \overline{\bm{D}}(A)^{-1}\overline{\bm{N}}(A)$ as well, one needs to impose an extra (normalization) condition either on $\bm{N}(A)$ or $\bm{D}(A)$ for the sake of uniqueness. While in the literature (see e.g. Ref. \cite{Oller:2018zts}) it is most customary to impose $\bm{D}(0)=\bm{1}$ so as to have the simple relation $\bm{T}(0)=\bm{N}(0)$, we show in what follows that for the particular case of the potential \eqref{orpot} the choice $\bm{N}(0)=\bm{1}$ --- on which we focus in the present section, deferring the $\bm{D}(0)=\bm{1}$ study to Sec. \ref{sec.231222.1} --- happens to be more natural and simpler. The last part of the section will be devoted to rewrite the derived DRs in terms of scattering observables, namely the ERE parameters, and to justify from dimensional arguments why those observables should or should not be taken as inputs in this framework.

\bigskip
Given what we learned in the previous two sections, our starting point here will be the split
\begin{eqnarray}
\bm{N}(A) &=& \frac{A^2\,\sqrt{A-L}}{M_N}\,\bm{1}\,;  \label{nmat} \\
\bm{D}(A) &=& A^2\,\bm{\rho}(A)\,[\bm{\gamma}^{\,-1} -\bm{I}(A)]\,\bm{\rho}(A)\,, \label{dmat}
\end{eqnarray}
with the auxiliary matrix
\begin{eqnarray}
\bm{\rho}(A) &=& (A+\Lambda_0^2)\begin{pmatrix}
1 & 0 \\ 0 & 1+\frac{\Lambda_0^2}{A}
\end{pmatrix}\,,
\end{eqnarray}
so that
\begin{eqnarray}
\bm{T}(A) &=& {\bm{D}(A)}^{-1}\,\bm{N}(A) \!\!\quad=\quad\!\! \bm{N}(A)\,{\bm{D}(A)}^{-1}\,, \label{TofA}
\end{eqnarray}
where the mass dimensions of $\bm{N}(A)$ and $\bm{D}(A)$ are consistently 4 and 6, respectively. One immediately notes the following: 

\begin{itemize}
\item $\bm{N}(A)$ 
  has no RHC, 
  since it does not contain loop integrals leading to the unitarity term. 
  Its discontinuity along the LHC follows from Eq.~\eqref{TofA} as
  \begin{eqnarray}
\label{231221.1}
\mathfrak{I}[\bm{N}(A)]&=&\bm{D}(A)\,\bm{\Delta}(A)\,,\quad A<L\,,
  \end{eqnarray}
with $\bm{\Delta}(A)$ the LHC discontinuity of the amplitude, namely 
\begin{eqnarray}
\bm{\Delta}(A) &=& ({2i})^{-1}\,[  \bm{T}(A+i\epsilon) - \bm{T}(A-i\epsilon) ] \quad\!\!\!=\!\!\!\quad \mathfrak{I}\left[\bm{T}(A+i\epsilon)\right] \quad\!\!\!=\!\!\!\quad \bm{D}^{-1}(A)\,\mathfrak{I}\left[\bm{N}(A+i\epsilon)\right]\,,\quad A<L\,.
\end{eqnarray}
Explicitly, our case gives \footnote{Note that, unlike what happens with  
chiral potentials,  
characterized by their pion tail, the interaction \eqref{orpot} does not have a spectral representation. Hence, the strategy pioneered in Ref. \cite{Oller:2018zts} to obtain $\bm{\Delta}(A)$ directly from the LHC discontinuity of the potential is in principle 
not available in this case. One thus needs to compute $\bm{\Delta}(A)$ from the knowledge of the amplitude itself.}
\begin{eqnarray} 
{\Delta}_{\,\ell'\ell}(A) &=& \frac{\sqrt{L-A}\,A^{\,\overline{\ell}}\,\tau_{\,\ell'\ell}(A)}{{M_N}\,(A+\Lambda_0^2)^{\,2+\overline{\ell}}}\,,\quad A<L\,,\end{eqnarray}
which is indeed real, 
and has a pole of degree $2+\overline{\ell}$ at $A=-\Lambda_0^2<L$. 

\item $\bm{D}(A)$ 
  has no LHC, 
  as its energy dependence 
  does not display 
  square roots with a cut on the negative real semi-axis. 
  Its discontinuity along the RHC can be calculated from Eq.~\eqref{TofA} 
  and reads
  \begin{eqnarray}
\label{231221.2}
\mathfrak{I}[\bm{D}(A)]&=&-\frac{M_N\sqrt{A}}{4\pi} \bm{N}(A)\,,\quad A>0\,. \label{fromunit}
  \end{eqnarray}
  Here we have made use of the unitarity requirement on the inverse of the partial-wave $T$ matrix \footnote{For detailed and pedagogical accounts on this respect, see e.g. Refs. \cite{Oller:2019rej,Oller:2019opk,Oller:2018zts}.},
  \begin{eqnarray}
\label{231221.3}
\mathfrak{I}[\bm{ T}(A)^{-1}]&=&-\frac{M_N\sqrt{A}}{4\pi}\bm{1}\,,\quad A>0\,.
  \end{eqnarray}
  \end{itemize}
Two more relevant observations are in order:
\begin{itemize}
\item Introducing $\bm{\rho}(A)$ in Eq. \eqref{dmat} allowed us to 
conveniently 
reshuffle the amplitude poles at $A=-\Lambda_0^2$ brought by the vertex functions [see Eqs. \eqref{gi} and \eqref{tij}] as zeros of $\bm{D}(A)$ at the same location. 
  However, by doing this  
  the degree of divergence of the matrix element $D_{\ell'\ell}(A)$ increases correspondingly, and more subtractions are required  (see below). 
\item The infrared quadratic divergence that $\bm{D}(A)$ would exhibit due to the $1/A$ factor in $\rho_{\,22}(A)$ is circumvented by                        including the $A^2$ factors in Eqs. \eqref{nmat} and \eqref{dmat}.
\end{itemize}  
Nevertheless, from Eq. \eqref{dmat}, 
\begin{eqnarray}
\bm{D}(0) &=& {\Lambda_0^8}\left[\frac{\gamma_{\,00}}{|\bm{\gamma}\,|}\,-\,\mathfrak{c}_{\,2}^{[0]}\right]\begin{pmatrix}
0 & 0 \\ 0 & 1
\end{pmatrix}\,,
\end{eqnarray}
which is not invertible. It is thus clear that this approach is \textit{incompatible} with imposing the usual normalization $\bm{D}(0)=\bm{1}$. In virtue of the general threshold condition
\begin{eqnarray}
T_{\,\ell'\ell}({A\to0}) &=& -\,\frac{4\pi}{M_N}\,a_{\,\ell'\ell}\,A^{\overline{\ell}}\,, \label{threscond}
\end{eqnarray}
one sees that the amplitude matrix becomes non-invertible at threshold [$|\bm{T}(0)| = {|\bm{N}(0)|}/{|\bm{D}(0)|} = 0$], which leads to two possible (mutually exclusive)  
zero-energy normalizations: 
\begin{itemize}
\item $\bm{D}(0)$ is invertible, but then $\bm{N}(0)$ is not (its determinant vanishes), and we can choose
\begin{eqnarray}
\bm{N}({A\to0}) \!\!\!\quad=\quad\!\!\! -\,\frac{4\pi}{M_N}\,\begin{pmatrix}
a_{\,00} & a_{\,02}\,A \\ a_{\,02}\,A & a_{\,22}\,A^2 
\end{pmatrix}\,;  &&  \bm{D}(0) \!\!\!\quad=\quad\!\!\! \bm{1} \,. \label{firstsce}
\end{eqnarray} 
\item $\bm{N}(0)$ is invertible and then $\bm{D}(0)$ is not (its determinant 
  diverges), and we can fix
\begin{eqnarray}
\bm{N}(0) \!\!\!\quad=\quad\!\!\! \bm{1}\,;  &&  \bm{D}({A\to0}) \!\!\!\quad=\quad\!\!\! -\,\frac{M_N}{4\pi\,|\bm{a}|}\,\begin{pmatrix}
a_{\,22} & -{a_{\,02}}/{A} \\ -{a_{\,02}}/{A} & {a_{\,00}}/{A^2} \label{sndsce} 
\end{pmatrix}\,.
\end{eqnarray} 
\end{itemize}
In order to keep $\bm{N}(A)\propto\bm{1}$ at each $A$ as in Eq. \eqref{nmat}, one is led to the latter scenario. 
We thus modify the previous normalization \eqref{nmat}--\eqref{dmat} and work instead with
\begin{eqnarray}
\bm{N}(A) &=& \frac{\sqrt{A-L}}{\sqrt{-L}}\,\bm{1}\,;  \label{nmat2} \\
\bm{D}(A) &=& \frac{M_N}{\sqrt{-L}}\,\bm{\rho}(A)\left[\bm{\gamma}^{\,-1} \,-\,\bm{I}(A)\right]\bm{\rho}(A)\,, \label{dmat2}
\end{eqnarray}
where 0 and 2 are the mass dimensions, respectively. 
This new split does verify the condition $\bm{N}(0)=\bm{1}$.

\bigskip
One important advantage of $\bm{N}(A)$ being both pole-free and proportional to the identity matrix at each $A$ is the easiness to find a proper DR for it. Since $N_{\,\ell'\ell}(A\to\infty) \propto \sqrt{A}\,\delta_{\,\ell'\ell}$, this DR has to include at least one subtraction --- this is chosen to be at $z=0$ 
since we take a zero-energy condition \eqref{sndsce}. We thus introduce the auxiliary function [$\mathfrak{I}(A) \neq 0$]
\begin{eqnarray}
f_{\,\ell'\ell}(z;A) &=& \frac{\sqrt{z-L}}{\sqrt{-L}\,z\,(z-A)}\,\delta_{\,\ell'\ell}\,, \end{eqnarray}
which has two single poles at $p=\lbrace0,\,A\rbrace$. Hence,
\begin{eqnarray}
  \oint_{\mathcal{{C}}}dz\,{f}_{\,\ell'\ell}(z;A) \!\!&=&\!\! 2\pi \,i\,
  \sum_{{z_i\in p}}
  \text{Res}\left[{f}_{\,\ell'\ell}(z;A);\,z={z_i}\right]  \quad\!\!\!\!
  =  \!\!\!\!\quad \frac{2\pi}{i\,A} \Big(1\,-\,\frac{\sqrt{A-L}}{\sqrt{-L}}\Big)\,\delta_{\,\ell'\ell} \quad\!\!\!\!
= \!\!\!\!\quad
  \frac{2\pi}{i\,A}\left[N_{\ell'\ell}(0)\,-\,N_{\ell'\ell}(A)\right]\,, ~~~~~~~~
\end{eqnarray} 
with 
$\mathcal{{C}}$ any closed positively oriented contour that encloses both poles. Now, let $\mathcal{{C}}$ be the infinite-radius circumference centered at the origin of the complex plane that engulfs the LHC. Direct evaluation yields ($\epsilon\to0^+$) 
\begin{eqnarray}\oint_{\mathcal{{C}}}dz\,{f}_{\,\ell'\ell}(z;A) \!&=&\! \int_{-\infty}^{\,L} d\omega_L\,{f}_{\,\ell'\ell}(\omega_L+i\epsilon;A)\,\,+\,\,\int_{L}^{\,-\infty} d\omega_L\,{f}_{\,\ell'\ell}(\omega_L-i\epsilon;A) \notag \\\!&=&\! \frac{2\,i\,\delta_{\,\ell'\ell}}{\sqrt{-L}}\,\int_{-\infty}^{\,L}d\omega_L\, \frac{\mathfrak{I}\,[{\sqrt{(\omega_L-L)\,+\,i\epsilon}}]}{{\omega_L\,(\omega_L-A)}}\quad\!\!\!\!=\!\!\!\!\quad \frac{2\pi}{i\,A} \Big(1\,-\,\frac{\sqrt{A-L}}{\sqrt{-L}}\Big)\,\delta_{\,\ell'\ell}\,, \end{eqnarray}
which is of course consistent. In matrix notation, our DR reads
\begin{eqnarray}
\bm{N}(A) &=& \bm{1} \,+\, \frac{A}{\pi}\int_{-\infty}^{\,L}d\omega_L\,\frac{\mathfrak{I}\left[\bm{N}(\omega_L+i\epsilon)\right]}{\omega_L\,(\omega_L-A)} \quad\!\!\!=\!\!\!\quad \bm{1} \,+\, \frac{A}{\pi}\int_{-\infty}^{\,L}d\omega_L\,\frac{\bm{D}(\omega_{\,L})\,\bm{\Delta}(\omega_{\,L})}{\omega_L\,(\omega_L-A)}\,. \label{NDR}
\end{eqnarray}

\bigskip
To build an appropriate DR for $\bm{D}(A)$, first consider that the matrix element given by Eq. \eqref{dmat2} explicitly reads
\begin{eqnarray}
D_{\,\ell'\ell}(A) &=& \frac{M_N}{\sqrt{-L}}\,\frac{(A+\Lambda_0^2)^{2+\overline{\ell}}}{A^{\overline{\ell}}}\left[(\bm{\gamma}^{\,-1})_{\ell'\ell}\,-\,I_{\,\ell}(A)\,\delta_{\,\ell'\ell}\right]\,. \label{Dlpl} 
\end{eqnarray}
Recalling the asymptotic vanishing of the loop integrals \eqref{asy}, we see that $D_{\,\ell'\ell}(A)$ diverges as $A^2$ when $A\to\infty$, i.e. $A^{3/2}$ times faster than $N_{\,\ell'\ell}(A)$. Hence, let us impose in the sought DR {two} extra subtractions at the conveniently chosen point $C\equiv-\Lambda_0^2$. Introduce the auxiliary function of complex argument [$\mathfrak{I}(A) \neq 0$]
\begin{eqnarray}
\widetilde{f}_{\,\ell'\ell}(z;A) &=& \frac{D_{\,\ell'\ell}(z)}{z\,(z-A)\,(z+\Lambda_0^2)^2}\,, \label{auxiliary}
\end{eqnarray}
which exhibits a single pole at $z=A$ and an $(\overline{\ell}+\!1)$th-degree pole at $z=0$, cf. Eq.~\eqref{Dlpl}. 
The residues are
\begin{eqnarray}
\text{Res}\left[\widetilde{f}_{\,\ell'\ell}(z;A);\,z=A\right] &=& \lim_{z\to A}\left(z-A\right)\widetilde{f}_{\,\ell'\ell}(z;A)\quad\!\!\!\!=\!\!\!\!\quad \frac{M_N}{\sqrt{-L}}\,\frac{(A+\Lambda_0^2)^{\overline{\ell}}}{A^{\overline{\ell}+1}}\left[(\bm{\gamma}^{\,-1})_{\ell'\ell}\,-\,I_{\,\ell}(A)\,\delta_{\,\ell'\ell}\right]\,; \label{resA} \\
\text{Res}\left[\widetilde{f}_{\,\ell'\ell}(z;A);\,z=0\right] &=& \frac{1}{\overline{\ell}!}\,{\displaystyle\lim_{z\to 0}}\,\frac{d^{\overline{\ell}}}{dz^{\overline{\ell}}}\,\Big(z^{\overline{\ell}+1}\,\widetilde{f}_{\,\ell'\ell}(z;A)\Big) \notag\\&=& \frac{M_N}{\sqrt{-L}}\,\sum_{i=0}^{\overline{\ell}}\,\sum_{j=0}^{\overline{\ell}-i} \,\frac{{\displaystyle\lim_{z\to 0}}\,\frac{d^i}{dz^i}(z+\Lambda_0^2)^{\overline{\ell}} \,\frac{d^j}{dz^j}(z-A)^{-1} \,\frac{d^{\overline{\ell}-i-j}}{dz^{\overline{\ell}-i-j}}\Big((\bm{\gamma}^{\,-1})_{\ell'\ell}\,-\,I_{\,\ell}(z)\,\delta_{\,\ell'\ell}\Big)}{i!\,j!\,(\overline{\ell}-i-j)!} \notag\\&=& \frac{\overline{\ell}!\,M_N}{A\,\sqrt{-L}}\,\Bigg[ \frac{\delta_{\,\ell'\ell}}{\ell!\,(\frac{A}{\Lambda_0^2})^\ell}\,\Bigg(\Big(1+\frac{A}{\Lambda_0^2}\Big)^\ell\,I_{\,\ell}(A)\,-\,J_{\,\ell}(A)\Bigg) \,-\,\frac{(1+\frac{A}{\Lambda_0^2})^{\overline{\ell}}}{\overline{\ell}!}\,\frac{(\bm{\gamma}^{\,-1})_{\ell'\ell}}{(\frac{A}{\Lambda_0^2})^{\overline{\ell}}}\Bigg]\,, \label{res0}
\end{eqnarray}
with the auxiliary integral
\begin{eqnarray}
\!\!\!\!\!\!J_{\,\ell}(A) \!\!\!&\equiv&\!\!\! \frac{A^{\,\ell+1}}{2\pi^2}\int_0^{\,\infty} dq\,\frac{g_\ell(q^2)^2}{q^{\,2\ell}\,(q^2-A)}\,\Big(1+\frac{q^2}{\Lambda_0^2}\Big)^\ell \quad\!\!\!\!\!\!=\quad\!\!\!\!\!\! \Big(1+\frac{A}{\Lambda_0^2}\Big)^\ell\,I_{\ell}(A) - \ell!\,A^\ell\,\sum_{i=0}^\ell\frac{\Lambda_0^{\,-2i}}{i!\,(\ell-i)!}\, \sum_{j=0}^{\ell-i}\frac{I_{\,\ell}^{\,(\ell-i-j)}(0)}{(\ell-i-j)!\,A^{\,j}}\,.~~~~~~~~ \label{Jl}
\end{eqnarray}
Note that the last equality follows from the fact that the $n$th derivative of the loop integral $I_{\,\ell}(A)$ ($n\leqslant\ell$) evaluated at threshold is
\begin{eqnarray}
I_{\,\ell}^{\,(n)}(0) &=& \frac{1}{2\pi^2}\int_0^{\,\infty} dq\,q^2\,g_\ell(q^2)^2\,\lim_{A\to0}\,\frac{d^{\,n}}{dA^{\,n}}(q^2-A)^{-1} \quad\!\!\!=\!\!\!\quad \frac{n!}{2\pi^2}\int_0^{\,\infty} dq\,q^{\,-2n}\,g_\ell(q^2)^2\,.
\end{eqnarray}
Integrating our auxiliary function \eqref{auxiliary} over any closed positively oriented contour $\mathcal{\widetilde{C}}$ such that both poles $\widetilde{p}=\lbrace0,\,A\rbrace$ are enclosed within it thus gives simply
\begin{eqnarray}
\oint_{\mathcal{\widetilde{C}}}dz\,\widetilde{f}_{\,\ell'\ell}(z;A) &=& 2\pi \,i\,\sum_{z_i\in \widetilde{p}}\text{Res}\left[\widetilde{f}_{\,\ell'\ell}(z;A);\,z=z_i\right] \quad\!\!\!=\!\!\!\quad -\,2\pi\,i\,\delta_{\,\ell'\ell}\,\frac{M_N}{A\,\sqrt{-L}}\,\Big(\frac{A}{\Lambda_0^2}\Big)^{\,-\ell}\,J_{\,\ell}(A)\,. \label{eqres}
\end{eqnarray} 
Now, let $\mathcal{\widetilde{C}}$ be the infinite-radius circumference centered at the origin of the complex plane that engulfs the positive real semi-axis (namely, the RHC). Independently of the residue theorem, direct evaluation of our path integral yields ($\epsilon\to0^+$)
\begin{eqnarray}
\!\!\!\!\!\!\!\!\!\!\!\!\!\!\!\!\!\!\!\!\!\!\!\oint_{\mathcal{\widetilde{C}}}dz\,\widetilde{f}_{\,\ell'\ell}(z;A) \!\!&=&\!\! \int_{\infty}^{\,0} d\omega_R\,\widetilde{f}_{\,\ell'\ell}(\omega_R-i\epsilon;A)\,\,+\,\,\int_{0}^{\,\infty} d\omega_R\,\widetilde{f}_{\,\ell'\ell}(\omega_R+i\epsilon;A) \quad\!\!\!\!\!=\!\!\!\!\!\quad 2\,i\,\int_{0}^{\,\infty} d\omega_R\,\mathfrak{I}\,[\widetilde{f}_{\,\ell'\ell}(\omega_R+i\epsilon;A)]\,. \label{direct}
\end{eqnarray}
But, given Eqs. \eqref{Dlpl} and \eqref{auxiliary},
\begin{eqnarray}
\mathfrak{I}\,[\widetilde{f}_{\,\ell'\ell}(\omega_R+i\epsilon;A)] &=& -\,\delta_{\,\ell'\ell}\,\frac{M_N}{\sqrt{-L}}\,\frac{(\omega_R+\Lambda_0^2)^\ell}{\omega_R^{\,\ell+1}\left(\omega_R-A\right)}\,\mathfrak{I}\left[I_{\,\ell}(\omega_R+i\epsilon)\right] \notag \\
&=& -\,\delta_{\,\ell'\ell}\,\frac{M_N}{4\pi\,\sqrt{-L}}\,\frac{(\omega_R+\Lambda_0^2)^\ell}{\omega_R^\ell\,\sqrt{\omega_R}\,\left(\omega_R-A\right)}\,g_\ell(\omega_R)^2\,,
\end{eqnarray}
where Eq. \eqref{compex} was recalled in the last step. Plugging the previous result into Eq. \eqref{direct} and performing the change of variable $\omega_R = q^2$ leads to
\begin{eqnarray}
\oint_{\mathcal{\widetilde{C}}}dz\,\widetilde{f}_{\,\ell'\ell}(z;A) &=& \frac{\delta_{\,\ell'\ell}}{i\,\pi}\,\frac{M_N}{\sqrt{-L}}\int_{0}^{\,\infty} \,dq\,\frac{(q^2+\Lambda_0^2)^\ell}{q^{\,2\ell}\,\left(q^2-A\right)}\,g_\ell(q^2)^2 \quad\!\!\!=\!\!\!\quad -\,2\pi\,i\,\delta_{\,\ell'\ell}\,\frac{M_N}{A\,\sqrt{-L}}\,\Big(\frac{A}{\Lambda_0^2}\Big)^{\,-\ell}\,J_{\,\ell}(A)\,, \label{eqdir}
\end{eqnarray}
in virtue of the definition of $J_{\,\ell}(A)$ \eqref{Jl}. Hence, we got a consistency check, Eqs. \eqref{eqres} and \eqref{eqdir} matching each other as they have to.

\bigskip
In matrix notation, combining Eqs. \eqref{eqres} and \eqref{direct} gives
\begin{eqnarray}
\oint_{\mathcal{\widetilde{C}}}dz\,\frac{\bm{D}(z)}{z\,(z-A)\,(z+\Lambda_0^2)^2} &=& 2\pi \,i\left[\frac{\bm{D}(A)}{A\,(A+\Lambda_0^2)^2\,} \,+\, \bm{R}_0(A)\right]\quad\!\!\!=\!\!\!\quad 2\,i\,\int_{0}^{\,\infty} d\omega_R\,\frac{\mathfrak{I}\left[\bm{D}(\omega_R+i\epsilon)\right]}{\omega_R\left(\omega_R-A\right)\,(\omega_R+\Lambda_0^2)^2}\,, \label{matnot}
\end{eqnarray}
with the residue matrix
\begin{eqnarray}
\bm{R}_0(A) &=& \lim_{z\to0}\,\begin{pmatrix}
\frac{D_{\,00}(z)}{(z-A)\,(z+\Lambda_0^2)^2} & \frac{d}{dz}\frac{z\,D_{\,02}(z)}{(z-A)\,(z+\Lambda_0^2)^2}\\ \frac{d}{dz}\frac{z\,D_{\,02}(z)}{(z-A)\,(z+\Lambda_0^2)^2} & \frac{1}{2}\frac{d^2}{dz^2}\frac{z^2\,D_{\,22}(z)}{(z-A)\,(z+\Lambda_0^2)^2}
\end{pmatrix}\,. \label{R0}
\end{eqnarray} 
This matrix can be evaluated explicitly by recalling Eq. \eqref{Dlpl} --- in the zero-energy limit $A\to0$, $D_{\,\ell'\ell}(A)/(A+\Lambda_0^2)^2$ is finite, diverges linearly, and diverges quadratically
for $\overline{\ell}=0$, $1$, and $2$ 
respectively. Namely, 
\begin{eqnarray}
\frac{D_{\,00}(A)}{(A+\Lambda_0^2)^2} \,\to\, \frac{M_N}{4\pi}\alpha_{\,00}^{[0]}\,; \qquad \!\!\!
\frac{D_{\,02}(A)}{(A+\Lambda_0^2)^2} \,\to\, \frac{M_N}{4\pi}\left(\alpha_{\,02}^{[0]}\,+\,\frac{\alpha_{\,02}^{[1]}}{A}\right)\,; \qquad  \!\!\!
\frac{D_{\,22}(A)}{(A+\Lambda_0^2)^2} \,\to\,  \frac{M_N}{4\pi}\left(\alpha_{\,22}^{[0]}\,+\,\frac{\alpha_{\,22}^{[1]}}{A}\,+\,\frac{\alpha_{\,22}^{[2]}}{A^2}\right)\,,
\end{eqnarray}
with the coefficients
\begin{eqnarray}
\alpha_{\,00}^{[0]} = \tfrac{4\pi}{\sqrt{-L}}\left(\tfrac{\gamma_{\,22}}{|\bm{\gamma}\,|}-I_0(0)\!\right); \!\!\!\!&& \!\!\!\!\\
\alpha_{\,02}^{[0]} = -\tfrac{4\pi}{\sqrt{-L}}\tfrac{\gamma_{\,02}}{|\bm{\gamma}\,|}, \!\!\!\!&& \!\!\!\!\alpha_{\,02}^{[1]} = -\tfrac{4\pi\Lambda_0^2}{\sqrt{-L}}\tfrac{\gamma_{\,02}}{|\bm{\gamma}\,|};  \\
\alpha_{\,22}^{[0]} = \tfrac{4\pi}{\sqrt{-L}}\left(\tfrac{\gamma_{\,00}}{|\bm{\gamma}\,|}-I_2(0)-2\Lambda_0^2I_2{\,\!\!'}(0)-\tfrac{\Lambda_0^4I_2''(0)}{2}\!\right), \!\!\!\!&& \!\!\!\!\alpha_{\,22}^{[1]} \!\!\!\!\!\quad\!\!\!\!=\!\!\!\!\quad\!\!\!\! \tfrac{8\pi\Lambda_0^2}{\sqrt{-L}}\left(\tfrac{\gamma_{\,00}}{|\bm{\gamma}\,|}-I_2(0)-\tfrac{\Lambda_0^2I_2{\,\!\!'}(0)}{2}\!\right) , \quad \alpha_{\,22}^{[2]} \!\!\!\!\quad\!\!\!\!=\!\!\!\!\!\!\!\!\quad \tfrac{4\pi\Lambda_0^4}{\sqrt{-L}}\left(\tfrac{\gamma_{\,00}}{|\bm{\gamma}\,|}-I_2(0)\!\right). \notag\\ 
\end{eqnarray}
Equation \eqref{R0} thus gives
\begin{eqnarray}
  \label{231228.1}
\bm{R}_0(A) &=& -\,\frac{M_N}{4\pi\,A}\,\begin{pmatrix}
\alpha_{\,00}^{[0]} & \alpha_{\,02}^{[0]}\,+\,\frac{\alpha_{\,02}^{[1]}}{A} \\ \alpha_{\,20}^{[0]}\,+\,\frac{\alpha_{\,20}^{[1]}}{A} & \alpha_{\,22}^{[0]}\,+\,\frac{\alpha_{\,22}^{[1]}}{A}\,+\,\frac{\alpha_{\,22}^{[2]}}{A^2}
\end{pmatrix}\,,
\end{eqnarray} 
where in our case it happens that $\alpha_{\,20}^{[0]}=\alpha_{\,02}^{[0]}$ and $\alpha_{\,20}^{[1]}=\alpha_{\,02}^{[1]}$. However, we will not assume this has to be the case in the derivation that follows, as $\bm{D}(A)$ will not be symmetric in general  
\footnote{Indeed, an immediate inference from $\bm{N}(A)\propto\bm{1}$ is the commutativity between $\bm{N}(A)$ and $\bm{D}(A)^{-1}$, cf. Eq. \eqref{TofA}, from where it follows in turn that $\bm{D}(A)^\top=\bm{D}(A).$ \label{footn}}, and the eight $\alpha$ coefficients will be considered unknown \textit{a priori}. Besides, 
given Eq. \eqref{231221.2}, Eq. \eqref{matnot} becomes
\begin{eqnarray}
\bm{D}(A) &=& -\,A\,(A+\Lambda_0^2)^2\,\bm{R}_0(A)  \,-\,  \frac{M_N\,A\,(A+\Lambda_0^2)^2\,}{4\pi^2}\int_{0}^{\,\infty} \frac{d\omega_R\,\sqrt{\omega_R}\,\bm{N}(\omega_R)}{\omega_R\,(\omega_R-A)\,(\omega_R+\Lambda_0^2)^2}\notag\\
&=& -\,A\,(A+\Lambda_0^2)^2\,\bm{R}_0(A)  \,-\,  \frac{M_N\,A\,(A+\Lambda_0^2)^2\,}{4\pi^2}\,\mathfrak{h}(0;A)\,\bm{1} \,+\,\frac{M_N\,A\,(A+\Lambda_0^2)^2\,}{4\pi^3}\int_{-\infty}^{\,L}d\omega_L\,\frac{\mathfrak{h}(\omega_L;A)}{\omega_L}\,\bm{D}(\omega_{\,L})\,\bm{\Delta}(\omega_{\,L})\,, \notag \\ \label{DofA}
\end{eqnarray}
having recalled Eq. \eqref{NDR}, and introduced the algebraic RHC integral \cite{Guo:2013rpa,Oller:2014uxa}
\begin{eqnarray}
\mathfrak{h}(\omega_L;A) \!\!\!&=&\!\!\! \int_{0}^{\,\infty} \frac{d\omega_R\,\sqrt{\omega_R}}{(\omega_R-\omega_L)\,(\omega_R-A)\,(\omega_R+\Lambda_0^2)^2} \quad\!\!\!\!\!\!=\!\!\!\!\!\!\quad \frac{\pi}{2\Lambda_0}\,\frac{\sqrt{\omega_L}\,+\,\sqrt{A}+2\,i\,\Lambda_0}{(\sqrt{\omega_L}\,+\,\sqrt{A})\,(\sqrt{\omega_L}+i\,\Lambda_0)^2\,(\sqrt{A}+i\,\Lambda_0)^2}\,, \label{h} 
\end{eqnarray}
where it is understood that both $\omega_L$ and $A$ have positive (arbitrarily small) imaginary parts, so that there is no ambiguity when taking the square roots \footnote{Let us remark that Eq.~\eqref{h} is readily applied in the whole complex $A$ plane keeping in mind that the square root should be defined with ${\rm arg}(A)\in[0,2\pi)$, as it is required by the integral.}. 

\bigskip
When $A\geqslant0$ (scattering case), the real and imaginary parts of Eq. \eqref{h} are
\begin{eqnarray}
\mathfrak{R}[\mathfrak{h}(\omega_L;A)] &=& \frac{\pi}{\Lambda_0\,(\Lambda_0^2+\omega_L)^2}\left[ \frac{\Lambda_0\,\sqrt{-\omega_L}}{A-\omega_L}\,+\,\frac{(A-\Lambda_0^2)\,(\omega_L-\Lambda_0^2)\,-\,4\,\Lambda_0^4}{2\,(A+\Lambda_0^2)^2} \right]\,; \label{Reh} \\
\mathfrak{I}[\mathfrak{h}(\omega_L;A)] &=& \frac{\pi\,\sqrt{A}}{(A+\Lambda_0^2)^2\,(A-\omega_L)}\,. \label{Imh}
\end{eqnarray}
Thus, taking imaginary parts in Eq. \eqref{DofA} leads us back to the unitarity condition \eqref{fromunit}; in turn, taking real parts in Eq. \eqref{DofA} provides
\begin{eqnarray}
\frac{4\pi}{M_N}\,\mathfrak{R}[\bm{D}(A)] &=& -\,(A+\Lambda_0^2)^2\left[\frac{4\pi\,A}{M_N}\,\bm{R}_0(A)\right] +\,  \frac{A^2+3\,\Lambda_0^2\,A}{2\,\Lambda_0^3} \left[\bm{1} \,-\, \frac{\Lambda_0^4}{\pi}\int_{-\infty}^{\,L}d\omega_L\,\frac{\bm{D}(\omega_{\,L})\,\bm{\Delta}(\omega_{\,L})}{(\Lambda_0^2+\omega_L)^2\,\omega_L} \right]  \notag \\ &&\,+\,  \frac{A}{\pi} \left[\frac{A-\Lambda_0^2}{2\,\Lambda_0} \int_{-\infty}^{\,L}d\omega_L\,\frac{\bm{D}(\omega_{\,L})\,\bm{\Delta}(\omega_{\,L})}{(\Lambda_0^2+\omega_L)^2} \,+\, (A\,+\,\Lambda_0^2)^2 \int_{-\infty}^{\,L}d\omega_L\,\frac{\bm{D}(\omega_{\,L})\,\bm{\Delta}(\omega_{\,L})}{(\Lambda_0^2+\omega_L)^2\,\sqrt{-\omega_L}\left(\omega_L-A\right)} \right]\,. \notag\\ \label{realparts}
\end{eqnarray}

\bigskip

\subsection{Explicit matching with the ERE}
\def\theequation{\arabic{section}.\arabic{equation}}
\label{sec.240103.3}

\bigskip

Let us match the $N/D$ representation of the amplitude discussed above  
with the matrix ERE. As we show below, this will allow us to fix the eight parameters of $\bm{R}_0(A)$ that enter Eq.~\eqref{TofA} in terms of six ERE parameters. 

\bigskip
From Eq. \eqref{unitarity}, any 
non-negative real 
$A$ fulfills
\begin{eqnarray}
\bm{M}(A) \!\!\!&=&\!\!\! \bm{P}(A)\left[\frac{4\pi}{M_N}\,\bm{N}^{-1}(A)\,\, \mathfrak{R}\left(\bm{D}(A)\right)\right]\bm{P}(A)\quad\!\!\!\!\!\!=\!\!\!\!\!\!\quad \left[\bm{P}(A)\,\bm{N}(A)\,\bm{P}^{-1}(A)\right]^{-1}\left[\frac{4\pi}{M_N}\,\bm{P}(A)\,\,\mathfrak{R}\left(\bm{D}(A)\right)\bm{P}(A)\right]\,,
\end{eqnarray}
namely
\begin{eqnarray}
\bm{P}(A)\,\bm{N}(A)\,\bm{P}^{-1}(A)\,\bm{M}(A) &=& \frac{4\pi}{M_N}\,\bm{P}(A)\,\,\mathfrak{R}\left(\bm{D}(A)\right)\bm{P}(A)\,. \label{general}
\end{eqnarray}

\bigskip
By means of the DR fulfilled by $\bm{N}(A)$ \eqref{NDR}, the left-hand side of Eq. \eqref{general} reads in our case
\begin{eqnarray}
\bm{P}(A)\,\bm{N}(A)\,\bm{P}^{-1}(A)\,\bm{M}(A) &=& \Bigg[\bm{1}\, +\, \frac{A}{\pi}\int_{-\infty}^{L}\frac{d\omega_L}{\omega_L(\omega_L-A)} \begin{pmatrix}
\left(\bm{D}(\omega_L)\bm{\Delta}(\omega_L)\right)_{00} & {A}^{-1}\left(\bm{D}(\omega_L)\bm{\Delta}(\omega_L)\right)_{02} \\ A\left(\bm{D}(\omega_L)\bm{\Delta}(\omega_L)\right)_{20} & \left(\bm{D}(\omega_L)\bm{\Delta}(\omega_L)\right)_{22}
\end{pmatrix}\Bigg]\,\bm{M}(A) \notag\\
&=& -\,\begin{pmatrix}
1\,+\,A\,\mathcal{I}_{00}^{(2)}\,+\,A^2\,\mathcal{I}_{00}^{(3)}\,\, &\,\, \mathcal{I}_{02}^{(2)}\,+\,A\,\mathcal{I}_{02}^{(3)}\,+\,A^2\,\mathcal{I}_{02}^{(4)} \\ A^2\,\mathcal{I}_{20}^{(2)} & 1\,+\,A\,\mathcal{I}_{22}^{(2)}\,+\,A^2\,\mathcal{I}_{22}^{(3)}
\end{pmatrix}\,\bm{a}^{-1} \notag\\
&&+\, \begin{pmatrix}
1\,+\,A\,\mathcal{I}_{00}^{(2)} &  \mathcal{I}_{02}^{(2)}\,+\,A\,\mathcal{I}_{02}^{(3)} \\ 0 & 1\,+\,A\,\mathcal{I}_{22}^{(2)}
\end{pmatrix}\,\frac{\bm{r}}{2}\,A \,+\, \begin{pmatrix}
1 &  \mathcal{I}_{02}^{(2)} \\ 0 & 1
\end{pmatrix}\,{\bm{v}}\,A^2\,+\,\mathcal{O}(A^3) \,, \label{lhs}
\end{eqnarray}
where Eqs. \eqref{gere} and \eqref{aandr} have been recalled ($\bm{v}\equiv\bm{v}^{[2]}$), and the auxiliary integral over the LHC
\begin{eqnarray}
\mathcal{I}_{\ell'\ell}^{(n)} &=& \frac{1}{\pi}\int_{-\infty}^L \frac{d\omega_L}{\omega_L^{\,n}}\,\left[\bm{D}(\omega_L)\bm{\Delta}(\omega_L)\right]_{\ell'\ell} \label{calI}
\end{eqnarray}
has been 
introduced. On the other hand, Eq. \eqref{realparts} yields the low-$A$ expansion of the right-hand side of Eq. \eqref{general} ---
\begin{eqnarray}
\frac{4\pi}{M_N}\,\bm{P}(A)\,\,\mathfrak{R}\left(\bm{D}(A)\right)\bm{P}(A) &=& \bm{m}_0\,+\,\bm{m}_1\,A\,+\,\bm{m}_2\,A^2 \,+\, \mathcal{O}(A^3)\,, \label{rhs}
\end{eqnarray}
having taken
\begin{eqnarray}
\bm{m}_0 &=& \Lambda_0^4\begin{pmatrix}
\!{\alpha_{\,00}^{[0]}}_{\color{white}q} & {\alpha_{\,02}^{[1]}} \\ \!{\alpha_{\,20}^{[1]}}_{\color{white}q} & {{\alpha_{\,22}^{[2]}}}
\end{pmatrix}\,; \\
\bm{m}_1 &=& \begin{pmatrix}
\frac{3}{2\,\Lambda_0}\,+\,2\,\alpha_{\,00}^{[0]}\,\Lambda_0^2 \,-\,\Lambda_0^4\mathcal{J}_{00}^{(\frac{3}{2})}\,+\,\frac{3\,\Lambda_0^3\mathcal{J}_{00}^{(1)}}{2}\,-\,\frac{\Lambda_0\mathcal{J}_{00}^{(0)}}{2} & \alpha_{\,02}^{[0]}\,\Lambda_0^4\,+\,2\,\alpha_{\,02}^{[1]}\,\Lambda_0^2 \\ \alpha_{\,20}^{[0]}\,\Lambda_0^4\,+\,2\,\alpha_{\,20}^{[1]}\,\Lambda_0^2 & \alpha_{\,22}^{[1]}\,\Lambda_0^4\,+\,2\,\alpha_{\,22}^{[2]}\,\Lambda_0^2
\end{pmatrix}\,; \\
\bm{m}_2 &=& \begin{pmatrix}
\alpha_{00}^{[0]}+\frac{1}{2\Lambda_0^3} +\Lambda_0^4\mathcal{J}_{00}^{(\frac{5}{2})}-2\Lambda_0^2\mathcal{J}_{00}^{(\frac{3}{2})}+\frac{\Lambda_0\mathcal{J}_{00}^{(1)}}{2}+\frac{\mathcal{J}_{00}^{(0)}}{2\Lambda_0} & \alpha_{02}^{[1]}+2\alpha_{02}^{[0]}\Lambda_0^2-\Lambda_0^4\mathcal{J}_{02}^{(\frac{3}{2})}+\frac{3\Lambda_0^3\mathcal{J}_{02}^{(1)}}{2}-\frac{\Lambda_0\mathcal{J}_{02}^{(0)}}{2} \\ \alpha_{20}^{[1]}+2\alpha_{20}^{[0]}\Lambda_0^2-\Lambda_0^4\mathcal{J}_{20}^{(\frac{3}{2})}+\frac{3\Lambda_0^3\mathcal{J}_{20}^{(1)}}{2}-\frac{\Lambda_0\mathcal{J}_{20}^{(0)}}{2} & \alpha_{22}^{[0]}\Lambda_0^4+2\alpha_{22}^{[1]}\Lambda_0^2+\alpha_{22}^{[2]}\end{pmatrix}\,, \notag\\
\end{eqnarray}
with
\begin{eqnarray}
\mathcal{J}_{\ell'\ell}^{(n)} &=& \frac{1}{\pi}\int_{-\infty}^L d\omega_L\,\frac{\left[\bm{D}(\omega_L)\bm{\Delta}(\omega_L)\right]_{\ell'\ell}}{(\Lambda_0^2+\omega_L)^2\,(-\omega_L)^{\,n}}\,.
\end{eqnarray}
Matching Eqs. \eqref{lhs} and \eqref{rhs} at each $\mathcal{O}(A^n)$ with $n=0,1,2$ allows one to solve for the $\alpha$ parameters that enter the IE \eqref{DofA} through the matrix $\bm{R}_0(A)$ \eqref{R0}, finding
\begin{eqnarray}
\alpha_{\,00}^{[0]} &=& \frac{a_{02}\mathcal{I}_{02}^{(2)}\,-\,a_{22}}{\Lambda_0^4\,|\bm{a}|}\,; \label{alphafirst}\\
\alpha_{\,02}^{[0]} &=& \frac{a_{00}\left(2\mathcal{I}_{02}^{(2)}-\Lambda_0^2\mathcal{I}_{02}^{(3)}\right)-a_{02}\left(2-\Lambda_0^2\mathcal{I}_{00}^{(2)}\right)}{\Lambda_0^6\,|\bm{a}|}\,+\,\frac{r_{02}+r_{22}\mathcal{I}_{02}^{(2)}}{2\Lambda_0^4}\,; \\ 
\alpha_{\,02}^{[1]} &=& \frac{a_{02}\,-\,a_{00}\mathcal{I}_{02}^{(2)}}{\Lambda_0^4\,|\bm{a}|}\,; \\
\alpha_{\,20}^{[0]} &=& -\frac{a_{02}\left(2-\Lambda_0^2\mathcal{I}_{22}^{(2)}\right)}{\Lambda_0^6\,|\bm{a}|}\,+\,\frac{r_{02}}{2\Lambda_0^4}\,; \\ 
\alpha_{\,20}^{[1]} &=& \frac{a_{02}}{\Lambda_0^4\,|\bm{a}|}\,; \\
\alpha_{\,22}^{[0]} &=& \frac{a_{02}\Lambda_0^4\mathcal{I}_{20}^{(2)}-a_{00}\left(3-2\Lambda_0^2\mathcal{I}_{22}^{(2)}+\Lambda_0^4\mathcal{I}_{22}^{(3)}\right)}{\Lambda_0^8\,|\bm{a}|}-\frac{r_{22}\left(2-\Lambda_0^2\mathcal{I}_{22}^{(2)}\right)}{2\Lambda_0^6}+\frac{v_{22}}{\Lambda_0^4}\,; \\
\alpha_{\,22}^{[1]} &=& \frac{a_{00}\left(2-\Lambda_0^2\mathcal{I}_{22}^{(2)}\right)}{\Lambda_0^6\,|\bm{a}|}\,+\,\frac{r_{22}}{2\Lambda_0^4}\,; \\
\alpha_{\,22}^{[2]} &=& -\,\frac{a_{00}}{\Lambda_0^4\,|\bm{a}|}\,. \label{alphalast}
\end{eqnarray}
Note that, since $\bm{N}(A)$ is a diagonal matrix in this normalization, from Eq. \eqref{calI} it is clear that $\mathcal{I}_{\ell'\ell}^{(n)}(A)$ is actually non-zero only for $\ell'=\ell$, so that the symmetry conditions $\alpha_{\,20}^{[0]}=\alpha_{\,02}^{[0]}$ and $\alpha_{\,20}^{[1]}=\alpha_{\,02}^{[1]}$ are indeed verified.

\bigskip
Using Eqs. \eqref{alphafirst}--\eqref{alphalast} in Eq. \eqref{matnot} gives explicitly the IEs in terms of the ERE parameters, namely
\begin{eqnarray}
D_{00}(A) &=& F_{00}(A)
\,+\, \frac{M_N}{8\pi^2\Lambda_0^4} \int_{-\infty}^Ld\omega_L\,\Big[\,\Big(f_{00}(\omega_L;A)\,\Delta_{02}(\omega_L)\,+\,f_{02}(\omega_L;A)\,\Delta_{00}(\omega_L)\Big)\,D_{00}(\omega_L)\notag\\
&&+\, \Big(f_{00}(\omega_L;A)\,\Delta_{22}(\omega_L)\,+\,f_{02}(\omega_L;A)\,\Delta_{02}(\omega_L)\Big)\,D_{02}(\omega_L)\,\Big]\,; \\
D_{02}(A) &=& F_{02}(A)
\,+\, \frac{M_N}{8\pi^2\Lambda_0^6} \int_{-\infty}^L\frac{d\omega_L}{\omega_L}\,\Big[\,\Big(f_{20}(\omega_L;A)\,\Delta_{00}(\omega_L)\,+\,f_{22}(\omega_L;A)\,\Delta_{02}(\omega_L)\Big)\,D_{00}(\omega_L)\notag\\
&&+\, \Big(f_{20}(\omega_L;A)\,\Delta_{02}(\omega_L)\,+\,f_{22}(\omega_L;A)\,\Delta_{22}(\omega_L)\Big)\,D_{02}(\omega_L)\,\Big]\,; \\
D_{20}(A) &=& F_{20}(A)
\,+\, \frac{M_N}{8\pi^2\Lambda_0^4} \int_{-\infty}^L{d\omega_L}\,\Big[\,\Big(f_{00}(\omega_L;A)\,\Delta_{02}(\omega_L)\,+\,f_{02}(\omega_L;A)\,\Delta_{00}(\omega_L)\Big)\,D_{20}(\omega_L)\notag\\
&&+\, \Big(f_{00}(\omega_L;A)\,\Delta_{22}(\omega_L)\,+\,f_{02}(\omega_L;A)\,\Delta_{02}(\omega_L)\Big)\,D_{22}(\omega_L)\,\Big]\,; \\
D_{22}(A) &=& F_{22}(A)
\,+\, \frac{M_N}{8\pi^2\Lambda_0^6} \int_{-\infty}^L\frac{d\omega_L}{\omega_L}\,\Big[\,\Big(f_{20}(\omega_L;A)\,\Delta_{00}(\omega_L)\,+\,f_{22}(\omega_L;A)\,\Delta_{02}(\omega_L)\Big)\,D_{20}(\omega_L)\notag\\
&&+\, \Big(f_{20}(\omega_L;A)\,\Delta_{02}(\omega_L)\,+\,f_{22}(\omega_L;A)\,\Delta_{22}(\omega_L)\Big)\,D_{22}(\omega_L)\,\Big]\,, 
\end{eqnarray}
with
\begin{eqnarray}
F_{00}(A)\!\!\!&=&\!\!\!\frac{M_N}{4\pi\Lambda_0^4} \,\Big[\frac{\Lambda_0A}{2}\,(A+3\Lambda_0^2)\,-\,i\sqrt{A}\Lambda_0^4\,-\,\frac{a_{22}\,(A+\Lambda_0^2)^2}{|\bm{a}|}\Big]\,; \\
F_{02}(A)\!\!\!&=&\!\!\! \frac{M_N\,(A+\Lambda_0^2)^2}{4\pi A\Lambda_0^6} \,\Big[\frac{a_{02}}{|\bm{a}|}\,(\Lambda_0^2-A)\,+\,\frac{r_{02}}{2}\,A\Lambda_0^2\Big] \quad\!\!\!\!\!\!=\!\!\!\!\!\!\quad F_{20}(A)\,; \\
F_{22}(A)\!\!\!&=&\!\!\!\frac{M_N}{4\pi A^2\Lambda_0^8} \!\Big[\frac{A^4\Lambda_0^5}{2} \!+\!\frac{3A^3\Lambda_0^7}{2}\!-\!iA^{5/2}\Lambda_0^8\!-\!\frac{a_{00}}{|\bm{a}|}(3A^4\!+\!4A^3\Lambda_0^2\!+\!\Lambda_0^8)\!+\!(\Lambda_0A\!+\!\Lambda_0^3)^2\Big(\frac{r_{22}}{2}A(\Lambda_0^2-2A)\!+\!v_{22}A^2\Lambda_0^2\Big)\Big]\,, \notag\\ 
\end{eqnarray}
and
\begin{eqnarray}
f_{00}(\omega_L;A) &=& \frac{2\,a_{02}\,(A+\Lambda_0^2)^2}{\omega_L^2\,|\bm{a}|}\,; \\
f_{02}(\omega_L;A) &=& -\frac{\Lambda_0^3A\,(\sqrt{A}-i\Lambda_0)^2\,[\sqrt{A}\,+\,i\,(\sqrt{-\omega_L}+2\Lambda_0)]}{\omega_L\,(\sqrt{A}+i\sqrt{-\omega_L})\,(\sqrt{-\omega_L}+\Lambda_0)^2}\,; \\
f_{20}(\omega_L;A) &=& \omega_L\,\Lambda_0^2\,f_{00}(\omega_L;A)\,; \\
f_{22}(\omega_L;A) &=& \omega_L\,\Lambda_0^2\,f_{02}(\omega_L;A)\,-\,\frac{2\,(A+\Lambda_0^2)^2}{\omega_L}\,\Big[\frac{a_{00}}{|\bm{a}|}\,\Big(\frac{\Lambda_0^2}{\omega_L}-2\Big)\,+\,\frac{\Lambda_0^2}{A}\,\Big(\frac{a_{00}}{|\bm{a}|}\,-\,\frac{r_{22}}{2}A\Big)\Big]\,.
\end{eqnarray}
We see that the system of four IEs actually decouples in two systems of two IEs fulfilled by $\lbrace D_{00}(A),D_{02}(A)\rbrace$ and $\lbrace D_{20}(A),D_{22}(A)\rbrace$ respectively. 

\bigskip
Since the interaction \eqref{orpot} is not singular, but regular, one might have thought of the existence of an $ND_{01}$-type solution where the proper low-energy behavior was guaranteed without the need of any free parameters (see e.g. the pedagogical discussion in Ref. \cite{Oller:2018zts}). However, there is a subtlety here due to 
the amplitude poles at $A=-\Lambda_0^2$ --- i.e. $T_{\ell'\ell}(A) \propto (A^2+\Lambda_0^2)^{-2-\overline{\ell}}$ ---
which, provided the normalization \eqref{nmat2}--\eqref{dmat2}, generates a double, triple, and quadruple zero at $A=-\Lambda_0^2$ in $D_{00}(A)$, $D_{02}(A)$, and $D_{22}(A)$, respectively, thus a single [double] zero at $z=-\Lambda_0^2$ in $\widetilde{f}_{02}(z;A)$ [$\widetilde{f}_{22}(z;A)$] as defined in Eq. \eqref{auxiliary}. This will fix several of the subtraction constants in the three-times subtracted IE fulfilled by $\bm{D}(A)$ \eqref{DofA}.

\bigskip
Out of the nine low-energy parameters corresponding to the shape-parameter approximation --- namely the neglect of all the terms with $n\geqslant3$ within the series \eqref{gere} --- only three of them ($r_{00}$, $v_{00}$, $v_{02}$) come out as predictions, 
while the other six are to be treated as inputs. This can be understood through the study of the \textit{scaling properties} of the ERE parameters in coupled channels with orbital angular momentum $\ell_1$ and $\ell_2$ \cite{Maki:2020zsv,Alarcon:2021kpx,Alarcon:2022vtn}. By means of Eq. \eqref{unitarity}, taking $A\equiv p^2>0$ yields
\begin{eqnarray}
\frac{4\pi}{M_N}\,\bm{P}(p^2)\,{\bm{T}}^{-1}(p^2)\,\bm{P}(p^2) &=& -\,\bm{a}^{-1}\,+\,\frac{\bm{r}}{2}\,p^2\,+\,\bm{v}_2\,p^4\,+\ldots\,+\,\bm{v}_n\,p^{2n}\,+\ldots\,-\, i\,p\,\bm{P}(p^2)^2\,,
\end{eqnarray}
with $\bm{P}(p^2)$ given by fixing $N=2$ in Eq. \eqref{centrifugal}. Now, replacing $p\to\lambda p$ leads to
\begin{eqnarray}
\frac{4\pi}{M_N}\,\bm{P}(\lambda^2p^2)\,{\bm{T}}^{-1}(\lambda^2p^2)\,\bm{P}(\lambda^2p^2) &=& \lambda\,\bm{P}(\lambda^2)\left[-\,{\overline{\bm{a}}}^{\,-1}\,+\,\frac{\overline{\bm{r}}}{2}\,p^2\,+\,\overline{\bm{v}}_2\,p^4\,+\ldots\,+\,\overline{\bm{v}}_n\,p^{2n}\,+\ldots\,-\, i\,p\,\bm{P}(p^2)^2 \right]\bm{P}(\lambda^2)\,, \notag\\
\end{eqnarray}
in terms of the rescaled ERE parameters
\begin{eqnarray}
{\overline{\bm{a}}}^{\,-1} &=& \left[\lambda\, \bm{P}(\lambda^2)\,\bm{a}\,\bm{P}(\lambda^2)\right]^{-1} \quad\!\!\!=\!\!\!\quad \frac{1}{|\bm{a}|}\,\begin{pmatrix}
a_{22}/\lambda^{2\ell_1+1} & -a_{02}/\lambda^{\ell_1+\ell_2+1} \\ -a_{02}/\lambda^{\ell_1+\ell_2+1} & a_{00}/\lambda^{2\ell_2+1}
\end{pmatrix}\,; \\
{\overline{\bm{r}}} &=& \lambda\, \bm{P}(\lambda^2)^{-1}\,\bm{r}\,\bm{P}(\lambda^2)^{-1} \quad\!\!\!=\!\!\!\quad \begin{pmatrix}
r_{00}/\lambda^{2\ell_1-1} & r_{02}/\lambda^{\ell_1+\ell_2-1} \\ r_{02}/\lambda^{\ell_1+\ell_2-1} & r_{22}/\lambda^{2\ell_2-1}
\end{pmatrix}\,; \\
{\overline{\bm{v}}}_n &=& \lambda^{2n-1}\, \bm{P}(\lambda^2)^{-1}\,\bm{v}_n\,\bm{P}(\lambda^2)^{-1} \quad\!\!\!=\!\!\!\quad \begin{pmatrix}
v_{n,00}/\lambda^{2(\ell_1-n)+1} & v_{n,02}/\lambda^{\ell_1+\ell_2-2n+1} \\ v_{n,02}/\lambda^{\ell_1+\ell_2-2n+1} & v_{n,22}/\lambda^{2(\ell_2-n)+1}
\end{pmatrix}\,.
\end{eqnarray}
In particular, setting $\ell_1=0$, $\ell_2=2$ as in $S\,D$-wave coupling, 
\begin{eqnarray}
{\overline{\bm{a}}}^{\,-1} \!\!&=& \frac{1}{|\bm{a}|}\begin{pmatrix}
\lambda^{-1}\,a_{22} & -\,\lambda^{-3}\,a_{02} \\ -\,\lambda^{-3}\,a_{02} & \lambda^{-5}\,a_{00}
\end{pmatrix}\,; \label{abarminus1}
\\{\overline{\bm{r}}} &=& \begin{pmatrix}
\lambda\,r_{00} & \lambda^{-1}\,r_{02} \\ \lambda^{-1}\,r_{02} & \lambda^{-3}\,r_{22}
\end{pmatrix}\,; \label{rbar}
\\{\overline{\bm{v}}}_n &=& \begin{pmatrix}
\lambda^{2n-1}\,v_{n,00} & \lambda^{2n-3}\,v_{n,02} \\ \lambda^{2n-3}\,v_{n,02} & \lambda^{2n-5}\,v_{n,22}
\end{pmatrix}\,. \label{vbar}
\end{eqnarray}
According to the sign of the exponents $q$ of the corresponding $\lambda^q$ factors, it is apparent that the three scattering lengths ($a_{00}$, $a_{02}$, $a_{22}$) and two out of the three effective ranges ($r_{02}$, $r_{22}$) are relevant; besides, one of the three first shape parameters ($v_{2,22}$) is still relevant. All the remaining ERE parameters appear to be suppressed.  
However, this na\"ive reasoning appears to not apply in all circumstances, as we will show explicitly in Sec. \ref{sec.231222.1}.

\bigskip
\bigskip
\bigskip

\section{Imposing $\pmb{D(0)=1}$ instead}
\label{sec.231222.1}
\setcounter{equation}{0}   
\def\theequation{\arabic{section}.\arabic{equation}}

\bigskip
As we anticipated in the beginning of the previous section, it comes out as a peculiar feature of the amplitude matrix generated by the interaction \eqref{orpot} that imposing the normalization $\bm{D}(0)=\bm{1}$, c.f. Eq. \eqref{firstsce}, gives rise to DRs and matching equations that are considerably more complicated than what we have just found from the condition $\bm{N}(0)=\bm{1}$. Still, our amplitude is compatible with such alternative $N/D$ split and we will show how to construct it in what follows.

\bigskip
We start from the split given by Eqs. 
\eqref{nmat2} and \eqref{dmat2}, replacing $\bm{N}(A)\to\bm{N}_0(A)$ and $\bm{D}(A)\to\bm{D}_0(A)$ so that $\bm{N}_0(0)=\bm{1}$. Then, we may also take \footnote{Naturally, the fact that $\bm{N}_1(A)$ and $\bm{D}_1(A)$ do not commute is linked to having $\bm{D}_1(A)^\top\neq \bm{D}_1(A)$. Again, this is in contrast to what happens in the $\bm{N}(0)=\bm{1}$ scheme, c.f. footnote \ref{footn}. }
\begin{eqnarray}
\bm{T}(A) &=& \bm{D}_1(A)^{-1}\,\bm{N}_1(A) \quad\!\!\!\neq\!\!\!\quad \bm{N}_1(A)\,\bm{D}_1(A)^{-1}\,,
\end{eqnarray}
with
\begin{eqnarray}
\bm{N}_1(A) &=& \frac{\sqrt{-L}}{M_N}\, [\,\bm{\rho}(A)^{-1}\,]^2 \, \bm{N}_0(A) \quad\!\!\!=\!\!\!\quad \frac{\sqrt{A-L}}{M_N\,(A+\Lambda_0^2)^2}\begin{pmatrix}
1 & 0 \\ 0 & \frac{A^2}{(A+\Lambda_0^2)^2}
\end{pmatrix}\,; \\ 
\bm{D}_1(A) &=& \bm{\rho}(A)^{-1}\left[\bm{\gamma}^{\,-1} \,-\,\bm{I}(A)\right]\bm{\rho}(A) \quad\!\!\!=\!\!\!\quad \frac{1}{|\bm{\gamma}\,|}\begin{pmatrix}
{\gamma_{\,22}}\,-\,{|\bm{\gamma}\,|}\,I_{\,0}(A) & -\,\frac{A+\Lambda_0^2}{A}\,{\gamma_{\,02}} \\ -\,\frac{A}{A+\Lambda_0^2}\,{\gamma_{\,02}} & {\gamma_{\,00}}\,-\,{|\bm{\gamma}\,|}\,I_{\,2}(A) 
\end{pmatrix}\,.
\end{eqnarray}
We further introduce
\begin{eqnarray}
\bm{d}_1(A) &=& \bm{\rho}(A)^{-1}\left[\bm{\gamma}^{\,-1} \,-\,\bm{I}(0)\right]\bm{\rho}(A) \quad\!\!\!=\!\!\!\quad \frac{1}{|\bm{\gamma}\,|}\begin{pmatrix}
{\gamma_{\,22}}\,-\,{|\bm{\gamma}\,|}\,I_{\,0}(0) & -\,\frac{A+\Lambda_0^2}{A}\,{\gamma_{\,02}} \\ -\,\frac{A}{A+\Lambda_0^2}\,{\gamma_{\,02}} & {\gamma_{\,00}}\,-\,{|\bm{\gamma}\,|}\,I_{\,2}(0) 
\end{pmatrix}\,,
\end{eqnarray}
thus the alternative representation
\begin{eqnarray}
\bm{T}(A) &=& \bm{D}_2(A)^{-1}\,\bm{N}_2(A) \quad\!\!\!\neq\!\!\!\quad \bm{N}_2(A)\,\bm{D}_2(A)^{-1}\,,
\end{eqnarray}
where
\begin{eqnarray}
\bm{N}_2(A) \!&=&\! \bm{d}_1(A)^{-1}\,\bm{N}_1(A) \quad\!\!\!=\!\!\!\quad \frac{\sqrt{A\,-\,L}}{M_N\,|\bm{d}_1(A)|\,(A+\Lambda_0^2)^2} \begin{pmatrix}
\frac{\gamma_{\,00}}{|\bm{\gamma}\,|}-I_2(0) & \frac{A}{A+\Lambda_0^2}\,\frac{\gamma_{\,02}}{|\bm{\gamma}\,|} \\ \frac{A}{A+\Lambda_0^2}\,\frac{\gamma_{\,02}}{|\bm{\gamma}\,|} & \frac{A^2}{(A+\Lambda_0^2)^2}\left[\frac{\gamma_{\,22}}{|\bm{\gamma}\,|}-I_0(0)\right]
\end{pmatrix}\,; \\
\bm{D}_2(A) \!&=&\! \bm{d}_1(A)^{-1}\,\bm{D}_1(A) \quad\!\!\!=\!\!\!\quad \frac{1}{|\bm{d}_1(A)|} \begin{pmatrix}
\left[\frac{\gamma_{00}}{|\bm{\gamma}\,|}-I_2(0)\right] \left[\frac{\gamma_{22}}{|\bm{\gamma}\,|}-I_0(A)\right]-\frac{\gamma_{02}^{2}}{|\bm{\gamma}\,|^2} & -\frac{\gamma_{02}}{|\bm{\gamma}\,|}\frac{A+\Lambda_0^2}{A}\left[I_2(A)-I_2(0)\right] \\ -\frac{\gamma_{02}}{|\bm{\gamma}\,|}\frac{A}{A+\Lambda_0^2}\left[I_0(A)-I_0(0)\right] & \left[\frac{\gamma_{00}}{|\bm{\gamma}\,|}-I_2(A)\right] \left[\frac{\gamma_{22}}{|\bm{\gamma}\,|}-I_0(0)\right]-\frac{\gamma_{02}^{2}}{|\bm{\gamma}\,|^2} \end{pmatrix}\,, \notag\\ \label{D2}
\end{eqnarray}
with the abbreviation
\begin{eqnarray}
|\bm{d}_1(A)| &=& |\bm{d}_1(0)| \quad\!\!\!=\!\!\!\quad |\bm{D}_1(0)| \quad\!\!\!=\!\!\!\quad \left[\frac{\gamma_{\,22}}{|\bm{\gamma}\,|}-I_0(0)\right] \left[\frac{\gamma_{\,00}}{|\bm{\gamma}\,|}-I_2(0)\right]-\frac{\gamma_{\,02}^{\,2}}{|\bm{\gamma}\,|^{2}}\,.
\end{eqnarray}
Besides, take the shorthand
\begin{eqnarray}
\mathfrak{D}(A) \!\!&=&\!\! |\bm{\gamma}\,|\,|\bm{D}_1(A)| \quad\!\!\!\!\!=\!\!\!\!\!\quad 1 \,+\, |\bm{\gamma}\,|\,I_0(A)\,I_2(A) \,-\, \left[ \gamma_{\,00}\,I_0(A)\,+\, \gamma_{\,22}\,I_2(A) \right] \quad\!\!\!\!\!=\!\!\!\!\!\quad 1\,+\, |\bm{\gamma}\,\bm{\Gamma}(A)|\,+\, \langle \bm{\gamma}\,\bm{\Gamma}(A)\rangle\,, \label{DA}
\end{eqnarray}
where $\bm{\Gamma}(A)\equiv-\bm{I}(A)$, and $\langle \ldots\rangle$ stands for the trace of the corresponding matrix. 
Taking $A\to 0$ 
 in Eq. \eqref{D2},
\begin{eqnarray}
\bm{D}_2(0) &=& \begin{pmatrix}
1 & -\,\frac{\gamma_{\,02}}{\mathfrak{D}(0)}\,{\Lambda_0^2}\,I_2{\,\!\!'}(0) \\ 0 & 1 \end{pmatrix}\,, 
\end{eqnarray}
so that $|\bm{D}_2(0)|=1$. Conversely, the determinant of $\bm{N}_2(A)$ vanishes at threshold, since
\begin{eqnarray}
\bm{N}_2(0) &=& \frac{\sqrt{-L}\,\left[\gamma_{\,00}\,-\,|\bm{\gamma}\,|\,I_2(0)\right]}{M_N\,\Lambda_0^4\,\mathfrak{D}(0)}\begin{pmatrix}
1 & 0 \\ 0 & 0
\end{pmatrix}\,.
\end{eqnarray}
That is to say, this is the scenario \eqref{firstsce} where the 
$\bm{D}(A)$ and $\bm{N}(A)$ 
matrices are respectively invertible and non-invertible at zero energy. We further introduce the split
\begin{eqnarray}
\bm{T}(A) &=& \bm{D}_3(A)^{-1}\,\bm{N}_3(A) \quad\!\!\!\neq\!\!\!\quad \bm{N}_3(A)\,\bm{D}_3(A)^{-1}\,,
\end{eqnarray}
by means of
\begin{eqnarray}
\bm{N}_3(A) \!&=&\! \bm{D}_2(0)^{-1}\,\bm{N}_2(A) \quad\!\!\!=\!\!\!\quad \frac{A\,\sqrt{A\,-\,L}}{M_N\,(A+\Lambda_0^2)^3\,\mathfrak{D}(0)} \begin{pmatrix}
\frac{{\gamma_{\,00}}-{|\bm{\gamma}\,|}I_2(0)}{A/(A+\Lambda_0^2)}+\frac{\gamma_{02}^2\,I_2{\,\!\!'}(0)\Lambda_0^2}{\mathfrak{D}(0)} & {\gamma_{\,02}}\left(1+\frac{{\gamma_{\,22}}-{|\bm{\gamma}\,|}I_0(0)}{(A+\Lambda_0^2)/A}\,\frac{I_2{\,\!\!'}(0)\Lambda_0^2}{\mathfrak{D}(0)}\right) \\ {\gamma_{\,02}} & \frac{{\gamma_{\,22}}-{|\bm{\gamma}\,|}I_0(0)}{(A+\Lambda_0^2)/A}
\end{pmatrix}\,; \notag \\ \\
\bm{D}_3(A) \!&=&\! \bm{D}_2(0)^{-1}\,\bm{D}_2(A) \quad\!\!\!=\!\!\!\quad \bm{1}\,+\,\bm{\delta}_{\text{I}}(A)\left[I_0(A)-I_0(0)\right]\,+\,\bm{\delta}_{\text{II}}(A)\left[I_2(A)-I_2(0)\right]\,+\,\bm{\delta}_{\text{III}}(A)\left[\tfrac{I_2(A)-I_2(0)}{A}\,-\,I_2{\,\!\!'}(0)\right]\,, \notag\\ \label{D3} \end{eqnarray}
where the matrices
\begin{eqnarray}
\bm{\delta}_{\text{I}}(A) &=& \frac{1}{\mathfrak{D}(0)}\begin{pmatrix}
{|\bm{\gamma}\,|}I_2(0)-{\gamma_{\,00}}\,-\,\frac{A\,\gamma_{02}^2\,I_2{\,\!\!'}(0)\Lambda_0^2}{\mathfrak{D}(0)\,(A+\Lambda_0^2)} & 0 \\ -\,\frac{A}{A+\Lambda_0^2}\,\gamma_{\,02} & 0
\end{pmatrix}\,; \\
\bm{\delta}_{\text{II}}(A) &=& \frac{1}{\mathfrak{D}(0)}\begin{pmatrix}
0 & -\,{\gamma_{\,02}}\left[1+\left({\gamma_{\,22}}-{|\bm{\gamma}\,|}I_0(0)\right)\frac{I_2{\,\!\!'}(0)\Lambda_0^2}{\mathfrak{D}(0)}\right] \\ 0 & {|\bm{\gamma}\,|}I_0(0)-{\gamma_{\,22}}
\end{pmatrix}\,; \\
\bm{\delta}_{\text{III}}(A) &=& -\,\frac{\gamma_{\,02}\,\Lambda_0^2}{\mathfrak{D}(0)}\begin{pmatrix}
0 & 1 \\ 0 & 0
\end{pmatrix}
\end{eqnarray}
are finite in the zero-energy limit, so from Eq. \eqref{D3} it is manifest that $\bm{D}_3(0)=\bm{1}$ as sought. 

\bigskip
The poles located at $A=-\Lambda_0^2$ exhibited by $\bm{N}_3(A)$ are removed through the auxiliary matrix
\begin{eqnarray}
\bm{\theta}(A) &=& \begin{pmatrix}
\frac{(A+\Lambda_0^2)^4}{\Lambda_0^8} & 0 \\ 0 & \frac{(A+\Lambda_0^2)^5}{\Lambda_0^{10}}
\end{pmatrix}\,,
\end{eqnarray}
which verifies $\bm{\theta}(0)=\bm{1}$, and is used to take the definitive split
\begin{eqnarray}
\bm{T}(A) &=& \bm{D}_4(A)^{-1}\,\bm{N}_4(A) \quad\!\!\!\neq\!\!\!\quad \bm{N}_4(A)\,\bm{D}_4(A)^{-1} \label{definitive}
\end{eqnarray}
through
\begin{eqnarray}
\bm{N}_4(A) \!\!\!&=&\!\!\! \bm{\theta}(A)\bm{N}_3(A) \quad\!\!\!\!\!=\!\!\!\!\!\quad \frac{\sqrt{A-L}}{M_N\mathfrak{D}(0)} \begin{pmatrix}
\frac{(A+\Lambda_0^2)^2\left[{\gamma_{00}}-{|\bm{\gamma}\,|}I_2(0)\right]}{\Lambda_0^8}+\frac{A(A+\Lambda_0^2)\gamma_{02}^2I_2{\!\!'}(0)}{\Lambda_0^6\mathfrak{D}(0)} & {\gamma_{02}}\left(\frac{A(A+\Lambda_0^2)}{\Lambda_0^8}+\frac{A^2I_2{\!\!'}(0)\left[{\gamma_{22}}-{|\bm{\gamma}\,|}I_0(0)\right]}{\Lambda_0^6\mathfrak{D}(0)}\right) \\ \frac{A\,(A+\Lambda_0^2)^2\,\gamma_{02}}{\Lambda_0^{10}} & \frac{A^2\,(A+\Lambda_0^2)\left[{\gamma_{22}}-{|\bm{\gamma}\,|}I_0(0)\right]}{\Lambda_0^{10}} \label{N4}
\end{pmatrix}; \notag \\ \\
\bm{D}_4(A) \!\!\!&=&\!\!\! \bm{\theta}(A)\bm{D}_3(A) \quad\!\!\!\!\!=\!\!\!\!\!\quad  \bm{\theta}(A)\,+\,\bm{\widetilde{\delta}}_{\text{I}}(A)\left[I_0(A)-I_0(0)\right]\,+\,\bm{\widetilde{\delta}}_{\text{II}}(A)\left[I_2(A)-I_2(0)\right]\,+\,\bm{\widetilde{\delta}}_{\text{III}}(A)\left[\tfrac{I_2(A)-I_2(0)}{A}\,-\,I_2{\,\!\!'}(0)\right]\,, \notag\\ \label{D4} \end{eqnarray}
with
\begin{eqnarray}
\bm{\widetilde{\delta}}_{\text{I}}(A) \quad\!\!\! =  &\bm{\theta}(A)\,\bm{\delta}_{\text{I}}(A)& = \quad\!\!\!\frac{1}{\Lambda_0^8\,\mathfrak{D}(0)}\begin{pmatrix}
(A+\Lambda_0^2)^4\left({|\bm{\gamma}\,|}I_2(0)-{\gamma_{\,00}}\right)-\,\frac{A\,(A+\Lambda_0^2)^3\,\gamma_{02}^2\,I_2{\,\!\!'}(0)\Lambda_0^2}{\mathfrak{D}(0)} & 0 \\ -\,\frac{{A}{(A+\Lambda_0^2)^4}\,\gamma_{\,02}}{\Lambda_0^2} & 0
\end{pmatrix}\,; \\
\bm{\widetilde{\delta}}_{\text{II}}(A) \quad\!\!\! = & \bm{\theta}(A)\,\bm{\delta}_{\text{II}}(A)& = \quad\!\!\!\frac{1}{\Lambda_0^8\,\mathfrak{D}(0)}\begin{pmatrix}
0 & -\,(A+\Lambda_0^2)^4\,{\gamma_{\,02}}\left[1+\left({\gamma_{\,22}}-{|\bm{\gamma}\,|}I_0(0)\right)\frac{I_2{\,\!\!'}(0)\Lambda_0^2}{\mathfrak{D}(0)}\right] \\ 0 & \frac{(A+\Lambda_0^2)^5\left({|\bm{\gamma}\,|}I_0(0)-{\gamma_{\,22}}\right)}{\Lambda_0^2}
\end{pmatrix}\,; \\
\bm{\widetilde{\delta}}_{\text{III}}(A) \quad\!\!\!= & \bm{\theta}(A)\,\bm{\delta}_{\text{III}}(A) &= \quad\!\!\!-\,\frac{(A+\Lambda_0^2)^4\,\gamma_{\,02}}{\Lambda_0^6\,\mathfrak{D}(0)}\begin{pmatrix}
0 & 1 \\ 0 & 0
\end{pmatrix}\,.
\end{eqnarray}
In a nutshell, coming back to the notation without the subscripts we have found that this normalization fixes the $N$ and $D$ matrices as
\begin{eqnarray}
\bm{N}(A) &=& \frac{\sqrt{A-L}}{M_N}\,\bm{\theta}(A)\left[\lim_{A\to0}\,\bm{\mu}(A)^{-1}\,\bm{\widetilde{\mu}}(A)\right]\bm{\widetilde{\mu}}(A)^{-1}\,; \\
\bm{D}(A) &=& \bm{\theta}(A)\left[\lim_{A\to0}\,\bm{\mu}(A)^{-1}\,\bm{\widetilde{\mu}}(A)\right]\bm{\widetilde{\mu}}(A)^{-1}\,\bm{\mu}(A)\,,
\end{eqnarray}
where
\begin{eqnarray}
\bm{\mu}(A) &=& \bm{\rho}(A)\left[\bm{\gamma}^{\,-1} \,-\,\bm{I}(A)\right]\bm{\rho}(A)\,; \\
\bm{\widetilde{\mu}}(A) &=& \bm{\rho}(A)\left[\bm{\gamma}^{\,-1} \,-\,\bm{I}(0)\right]\bm{\rho}(A)\,.
\end{eqnarray}
With such redefinitions, the following asymptotic behaviors arise
\begin{eqnarray}
N_{\,0\ell}(A\to\infty) &\propto& A^{\frac{5}{2}}\,, \label{N0l}\\
N_{\,2\ell}(A\to\infty) &\propto& A^{\frac{7}{2}}\,; \label{N2l}\\
D_{\,0\ell}(A\to\infty) &\propto& A^4\,, \label{D0l}\\
D_{\,2\ell}(A\to\infty) &\propto& A^5\,. \label{D2l}
\end{eqnarray}

\bigskip
According to Eq. \eqref{N0l}, we impose three subtractions in the $N_{00}(A)$ and $N_{02}(A)$ DRs, namely
\begin{eqnarray}
\oint dz\,\frac{N_{00}(z)}{z\,(z-A)\,(z+\Lambda_0^2)^{2}} &=& 2i\int_{-\infty}^L d\omega_L\,\frac{\mathfrak{I}[N_{00}(\omega_L)]}{\omega_L\,(\omega_L-A)\,(\omega_L+\Lambda_0^2)^{2}} \notag\\
&=& \frac{2\pi\,i}{\Lambda_0^4}\left[\frac{1}{(1+\frac{A}{\Lambda_0^2})^2} \left(\frac{N_{00}(A)}{A}\,+\,(2+\tfrac{A}{\Lambda_0^2})\frac{N_{00}(-\Lambda_0^2)}{\Lambda_0^2}\right)\,+\, \frac{N\,'_{\!\!00}(-\Lambda_0^2)}{1+\frac{A}{\Lambda_0^2}}\,-\,\frac{N_{00}(0)}{A}\right]\,; \notag \\ \\
\oint dz\,\frac{N_{02}(z)}{z^3\,(z-A)} &=& 2i\int_{-\infty}^L d\omega_L\,\frac{\mathfrak{I}[N_{02}(\omega_L)]}{\omega_L^3\,(\omega_L-A)} \quad\!\!\!=\quad\!\!\! \frac{2\pi\,i}{A^3}\left[N_{02}(A)-N_{02}(0)-N\,'_{\!\!02}(0)A-\frac{1}{2}N\,''_{\!\!02}(0)A^2\right]\,. \notag\\\end{eqnarray} 
Fixing the conditions
\begin{eqnarray}
N_{00}(-\Lambda_0^2) &=& 0\,; \\
N_{02}(0) &=& 0
\end{eqnarray}
[see Eq. \eqref{N4}], and naming the free parameters
\begin{eqnarray}
\nu_{00} \,\,\,\,\,=\,\,\,\,\, N_{00}(0)\,, && \tau_{00} \,\,\,\,\,=\,\,\,\,\, \frac{1}{\Lambda_0^2}-\frac{N\,'_{\!\!00}(-\Lambda_0^2)}{N_{00}(0)}\,; \label{nutau00}\\
\nu_{02} \,\,\,\,\,=\,\,\,\,\, N\,'_{\!\!02}(0)\,, && \tau_{02} \,\,\,\,\,=\,\,\,\,\, \frac{1}{2}\,N\,''_{\!\!02}(0)\,, \label{nutau02}
\end{eqnarray}
we get the DRs
\begin{eqnarray}
N_{00}(A) &=& \left(1+\frac{A}{\Lambda_0^2}\right)\nu_{00}\left(1+\tau_{00}A\right)+\,\frac{A\,(A+\Lambda_0^2)^2}{\pi}\int_{-\infty}^L d\omega_L\,\frac{\left[\bm{D}(\omega_{\,L})\,\bm{\Delta}(\omega_{\,L})\right]_{00}}{\omega_L\,(\omega_L-A)\,(\omega_L+\Lambda_0^2)^{2}}\,; \label{N00A} \\
N_{02}(A) &=& \nu_{02}A\,+\,\tau_{02}A^2\,+\,\frac{A^3}{\pi}\int_{-\infty}^L d\omega_L\,\frac{\left[\bm{D}(\omega_{\,L})\,\bm{\Delta}(\omega_{\,L})\right]_{02}}{\omega_L^3\,(\omega_L-A)}\,. \label{N02A}
\end{eqnarray}

\bigskip
Conversely, Eq. \eqref{N2l} indicates the need for four subtractions in the $N_{20}(A)$ and $N_{22}(A)$ DRs, i.e.
\begin{eqnarray}
\oint \frac{dz\,\,N_{20}(z)}{z\,(z-A)\,(z+\Lambda_0^2)^{3}} \!\!\!\!&=&\!\!\!\! 2i\int_{-\infty}^L \frac{d\omega_L\,\mathfrak{I}[N_{20}(\omega_L)]}{\omega_L\,(\omega_L-A)\,(\omega_L+\Lambda_0^2)^{3}} \notag\\
&=& \!\!\!\!\frac{2\pi\,i}{\Lambda_0^6}\left[\!-\frac{N_{20}(0)}{A} \!+\! \frac{\frac{N_{20}(A)}{A}+(3+\frac{3A}{\Lambda_0^2}+\frac{A^2}{\Lambda_0^4})\frac{N_{20}(-\Lambda_0^2)}{\Lambda_0^2}}{(1+\frac{A}{\Lambda_0^2})^3} \!+\! \frac{(2+\frac{A}{\Lambda_0^2})\,N\,'_{\!\!20}(-\Lambda_0^2)}{(1+\frac{A}{\Lambda_0^2})^2} \!+\! \frac{\frac{\Lambda_0^2}{2}N\,''_{\!\!20}(-\Lambda_0^2)}{1+\frac{A}{\Lambda_0^2}} \!\right]\,; \notag 
\\ \\ 
\oint \frac{dz\,\,N_{22}(z)}{z^2\,(z-A)\,(z+\Lambda_0^2)^{2}} \!\!\!\!&=&\!\!\!\! 2i\int_{-\infty}^L \frac{d\omega_L\,\mathfrak{I}[N_{22}(\omega_L)]}{\omega_L^2\,(\omega_L-A)\,(\omega_L+\Lambda_0^2)^{2}} \notag\\
&=& \!\!\!\!\frac{2\pi\,i}{\Lambda_0^4}\left[-\frac{(1-\frac{2A}{\Lambda_0^2})\frac{N_{22}(0)}{A}+N\,'_{\!\!22}(0)}{A} + \frac{\frac{N_{22}(A)}{A^2}-(3+\frac{2A}{\Lambda_0^2})\,\frac{N_{22}(-\Lambda_0^2)}{\Lambda_0^4}}{(1+\frac{A}{\Lambda_0^2})^2} + \frac{\frac{N\,'_{\!\!22}(-\Lambda_0^2)}{\Lambda_0^2}}{1+\frac{A}{\Lambda_0^2}} \right]\,.  
\end{eqnarray} 
Through the constraints
\begin{eqnarray}
N_{20}(0) \!\!\!\quad=\!\!\!\quad N_{20}(-\Lambda_0^2) \!\!\!\quad= &N\,'_{\!\!20}(-\Lambda_0^2)& = \!\!\!\quad 0\,; \\
N_{22}(0) \!\!\!\quad=\!\!\!\quad N_{22}(-\Lambda_0^2) \!\!\!\quad= &N\,'_{\!\!22}(0)& =\!\!\!\quad  0
\end{eqnarray}
[see Eq. \eqref{N4}], together with the free parameters
\begin{eqnarray}
\nu_{20} &=& -\frac{N\,''_{\!\!20}(-\Lambda_0^2)}{2\Lambda_0^2}\,; \label{nu20}\\
\nu_{22} &=& \frac{N\,'_{\!\!22}(-\Lambda_0^2)}{\Lambda_0^4}\,, \label{nu22}
\end{eqnarray}
we arrive at the DRs
\begin{eqnarray}
N_{20}(A) &=& \nu_{20}\,A\,(A+\Lambda_0^2)^2\,+\,\frac{A\,(A+\Lambda_0^2)^3}{\pi}\int_{-\infty}^L d\omega_L\,\frac{\left[\bm{D}(\omega_{\,L})\,\bm{\Delta}(\omega_{\,L})\right]_{20}}{\omega_L\,(\omega_L-A)\,(\omega_L+\Lambda_0^2)^{3}}\,; \label{N20A} \\
N_{22}(A) &=& \nu_{22}\,A^2\,(A+\Lambda_0^2)\,+\,\frac{A^2\,(A+\Lambda_0^2)^2}{\pi}\int_{-\infty}^L d\omega_L\,\frac{\left[\bm{D}(\omega_{\,L})\,\bm{\Delta}(\omega_{\,L})\right]_{22}}{\omega_L^2\,(\omega_L-A)\,(\omega_L+\Lambda_0^2)^{2}}\,. \label{N22A}
\end{eqnarray}

\bigskip
Following Eq. \eqref{D0l}, five subtractions are taken in the $D_{00}(A)$ and $D_{02}(A)$ DRs, so that
\begin{eqnarray}
\oint \frac{dz\,\,D_{0\ell}(z)}{z\,(z-A)\,(z+\Lambda_0^2)^4} &=& 2i\int_0^{\,\infty}\frac{d\omega_R\,\mathfrak{I}[D_{0\ell}(\omega_R)]}{\omega_R\,(\omega_R-A)\,(\omega_R+\Lambda_0^2)^4} \notag\\
&=& \frac{2\pi\,i}{\Lambda_0^8}\,\Bigg[ -\frac{D_{0\ell}(0)}{A}\,+\,\frac{\frac{D_{0\ell}(A)}{A}+(4+\frac{6A}{\Lambda_0^2}+\frac{4A^2}{\Lambda_0^4}+\frac{A^3}{\Lambda_0^6})\frac{D_{0\ell}(-\Lambda_0^2)}{\Lambda_0^2}}{(1+\frac{A}{\Lambda_0^2})^4} 
\,+\,\frac{(3+\frac{3A}{\Lambda_0^2}+\frac{A^2}{\Lambda_0^4})D\,'_{\!\!0\ell}(-\Lambda_0^2)}{(1+\frac{A}{\Lambda_0^2})^3} \notag\\
&&\,+\,\frac{(1+\frac{A}{2\Lambda_0^2})\Lambda_0^2D\,''_{\!\!0\ell}(-\Lambda_0^2)}{(1+\frac{A}{\Lambda_0^2})^2}
\,+\,\frac{\frac{\Lambda_0^4}{6}D\,'''_{\!\!0\ell}(-\Lambda_0^2)}{1+\frac{A}{\Lambda_0^2}} \Bigg]\,.
\end{eqnarray}
Imposing the conditions
\begin{eqnarray}
&&D_{00}(0) \!\!\!\quad=\!\!\!\quad 1\,, \qquad D_{00}(-\Lambda_0^2) \!\!\!\quad=\!\!\!\quad D\,'_{\!\!00}(-\Lambda_0^2) \!\!\!\quad= \!\!\!\quad D\,''_{\!\!00}(-\Lambda_0^2) \!\!\!\quad= \!\!\!\quad 0\,; \\
&&D_{02}(0) \!\!\!\quad=\!\!\!\quad 0\,, \qquad D_{02}(-\Lambda_0^2) \!\!\!\quad=\!\!\!\quad D\,'_{\!\!02}(-\Lambda_0^2) \!\!\!\quad= \!\!\!\quad D\,''_{\!\!02}(-\Lambda_0^2) \!\!\!\quad= \!\!\!\quad D\,'''_{\!\!02}(-\Lambda_0^2) \!\!\!\quad= \!\!\!\quad 0\,,\end{eqnarray}
which are consistent with Eq. \eqref{D4}, and introducing the unfixed parameter
\begin{eqnarray}
\alpha_{00} &=& \frac{1}{\Lambda_0^2}-\frac{\Lambda_0^4}{6}D\,'''_{\!\!00}(-\Lambda_0^2)\,,
\label{alpha00}\end{eqnarray}
we find the DRs
\begin{eqnarray}
D_{00}(A) &=& \left(1+\frac{A}{\Lambda_0^2}\right)^3\left(1+\alpha_{00}A\right) \,-\, \frac{M_N}{4\pi^2}\,A\,(A+\Lambda_0^2)^4\int_0^{\,\infty}\frac{d\omega_R\,\sqrt{\omega_R}\,N_{00}(\omega_R)}{\omega_R\,(\omega_R-A)\,(\omega_R+\Lambda_0^2)^4}\,; \label{D00A} \\
D_{02}(A) &=& -\, \frac{M_N}{4\pi^2}\,A\,(A+\Lambda_0^2)^4\int_0^{\,\infty}\frac{d\omega_R\,\sqrt{\omega_R}\,N_{02}(\omega_R)}{\omega_R\,(\omega_R-A)\,(\omega_R+\Lambda_0^2)^4}\,. \label{D02A}
\end{eqnarray}

\bigskip
Given Eq. \eqref{D2l}, six subtractions are required in the $D_{20}(A)$ and $D_{22}(A)$ DRs, entailing
\begin{eqnarray}
\!\!\!\!\!\!\!\!\!\!\!\!\!\!\!\!\!\!\!\!\!\!\!\!\oint \frac{dz\,\,D_{2\ell}(z)}{z\,(z-A)\,(z+\Lambda_0^2)^5} \!\!\!\!\!&=&\!\!\!\!\! 2i\int_0^{\,\infty}\frac{d\omega_R\,\mathfrak{I}[D_{2\ell}(\omega_R)]}{\omega_R\,(\omega_R-A)\,(\omega_R+\Lambda_0^2)^5} \notag\\
\!\!\!\!\!&=&\!\!\!\!\! \frac{2\pi i}{\Lambda_0^{10}}\Bigg[ -\frac{D_{2\ell}(0)}{A}+\frac{\frac{D_{2\ell}(A)}{A}+(5+\frac{10A}{\Lambda_0^2}+\frac{10A^2}{\Lambda_0^4}+\frac{5A^3}{\Lambda_0^6}+\frac{A^4}{\Lambda_0^8})\frac{D_{2\ell}(-\Lambda_0^2)}{\Lambda_0^2}}{(1+\frac{A}{\Lambda_0^2})^5}+\frac{(4+\frac{6A}{\Lambda_0^2}+\frac{4A^2}{\Lambda_0^4}+\frac{A^3}{\Lambda_0^6})D'_{\!\!2\ell}(-\Lambda_0^2)}{(1+\frac{A}{\Lambda_0^2})^4} \notag\\
&&\!\!\!\!\!\,+\,\frac{(\frac{3}{2}+\frac{3A}{2\Lambda_0^2}+\frac{A^2}{2\Lambda_0^4})\Lambda_0^2D\,''_{\!\!2\ell}(-\Lambda_0^2)}{(1+\frac{A}{\Lambda_0^2})^3} \,+\,\frac{(\frac{1}{3}+\frac{A}{6\Lambda_0^2})\Lambda_0^4D\,'''_{\!\!2\ell}(-\Lambda_0^2)}{(1+\frac{A}{\Lambda_0^2})^2}
\,+\,\frac{\frac{\Lambda_0^6}{24}D\,''''_{\!\!2\ell}(-\Lambda_0^2)}{1+\frac{A}{\Lambda_0^2}} \Bigg]\,.
\end{eqnarray}
Guided by Eq. \eqref{D4}, we take the constraints
\begin{eqnarray}
&&\!\!\!\!\!\!\!\!\!\!\!\!\!\!\!\!\!\!\!\!\!\!\!\!\!\!\!\!\!\!D_{20}(0) \!\!\!\quad=\!\!\!\quad 0\,, \qquad D_{20}(-\Lambda_0^2) \!\!\!\quad=\!\!\!\quad D\,'_{\!\!20}(-\Lambda_0^2) \!\!\!\quad= \!\!\!\quad D\,''_{\!\!20}(-\Lambda_0^2) \!\!\!\quad= \!\!\!\quad D\,'''_{\!\!20}(-\Lambda_0^2)\!\!\!\quad= \!\!\!\quad 0\,; \\
&&\!\!\!\!\!\!\!\!\!\!\!\!\!\!\!\!\!\!\!\!\!\!\!\!\!\!\!\!\!\!D_{22}(0) \!\!\!\quad=\!\!\!\quad 1\,, \qquad D_{22}(-\Lambda_0^2) \!\!\!\quad=\!\!\!\quad D\,'_{\!\!22}(-\Lambda_0^2) \!\!\!\quad= \!\!\!\quad D\,''_{\!\!22}(-\Lambda_0^2) \!\!\!\quad= \!\!\!\quad D\,'''_{\!\!22}(-\Lambda_0^2) \!\!\!\quad= \!\!\!\quad D\,''''_{\!\!22}(-\Lambda_0^2) \!\!\!\quad= \!\!\!\quad 0\,; \end{eqnarray}
besides, we introduce the free parameter
\begin{eqnarray}
\alpha_{20} &=& -\frac{D\,''''_{\!\!20}(-\Lambda_0^2)}{24\Lambda_0^2}\,. \label{alpha20}
\end{eqnarray}
Eventually,
\begin{eqnarray}
D_{20}(A) &=& \alpha_{20}\,A\,(A+\Lambda_0^2)^4 \,-\, \frac{M_N}{4\pi^2}\,A\,(A+\Lambda_0^2)^5\int_0^{\,\infty}\frac{d\omega_R\,\sqrt{\omega_R}\,N_{20}(\omega_R)}{\omega_R\,(\omega_R-A)\,(\omega_R+\Lambda_0^2)^5}\,; \label{D20A} \\
D_{22}(A) &=& \left(1+\frac{A}{\Lambda_0^2}\right)^5 \,-\, \frac{M_N}{4\pi^2}\,A\,(A+\Lambda_0^2)^5\int_0^{\,\infty}\frac{d\omega_R\,\sqrt{\omega_R}\,N_{22}(\omega_R)}{\omega_R\,(\omega_R-A)\,(\omega_R+\Lambda_0^2)^5}\,. \label{D22A}
\end{eqnarray}

\bigskip
In summary, collecting Eqs. \eqref{nutau00}, \eqref{nutau02}, \eqref{nu20}, \eqref{nu22}, \eqref{alpha00}, and \eqref{alpha20}, the  DRs just derived depend in principle upon {eight} free parameters  
--- way too much to consider this approach acceptable in terms of predictive power! Still, for the sake of completeness we will illustrate below how to build the ERE matching in this approach.

\bigskip
Bringing Eqs. \eqref{N00A}, \eqref{N02A}, \eqref{N20A}, and \eqref{N22A} respectively into Eqs. \eqref{D00A}, \eqref{D02A}, \eqref{D20A}, and \eqref{D22A}, it turns out
\begin{eqnarray}
D_{00}(A) &=& \left(1+\frac{A}{\Lambda_0^2}\right)^3\left(1+\alpha_{00}A\right) \,-\, \frac{M_N\,\nu_{00}}{4\pi}\,\frac{A\,(A+\Lambda_0^2)^4}{\Lambda_0^2}\,\Big(\mathfrak{H}^{[01]}(0;A)+\tau_{00}\mathfrak{H}^{[11]}(0;A)\Big)\notag \\
&&+\, \frac{M_N}{4\pi^2}\,A\,(A+\Lambda_0^2)^4\int_{-\infty}^Ld\omega_L\, \frac{\mathfrak{H}^{[00]}(\omega_L;A)\left[\bm{D}(\omega_{\,L})\,\bm{\Delta}(\omega_{\,L})\right]_{00}}{\omega_L\,(\omega_L+\Lambda_0^2)^{2}}\,; \label{IED00}
\\
D_{02}(A) &=& -\, \frac{M_N}{4\pi}\,A\,(A+\Lambda_0^2)^4\,\Big(\nu_{02}\mathfrak{H}^{[12]}(0;A)+\tau_{02}\mathfrak{H}^{[22]}(0;A)\Big) \notag\\
&&+\, \frac{M_N}{4\pi^2}\,A\,(A+\Lambda_0^2)^4\int_{-\infty}^L d\omega_L\,\frac{\mathfrak{H}^{[22]}(\omega_L;A)\left[\bm{D}(\omega_{\,L})\,\bm{\Delta}(\omega_{\,L})\right]_{02}}{{\omega_L^{3}}\,}\,;
\label{IED02} \\
D_{20}(A) &=& \alpha_{20}\,A\,(A+\Lambda_0^2)^4 \,-\, \frac{M_N\,\nu_{20}}{4\pi}\,A\,(A+\Lambda_0^2)^5\mathfrak{H}^{[11]}(0;A)\notag\\
&&+\, \frac{M_N}{4\pi^2}\,A\,(A+\Lambda_0^2)^5\int_{-\infty}^L d\omega_L\,\frac{\mathfrak{H}^{[00]}(\omega_L;A)\left[\bm{D}(\omega_{\,L})\,\bm{\Delta}(\omega_{\,L})\right]_{20}}{{\omega_L}\,(\omega_L+\Lambda_0^2)^3}\,; \label{IED20}
\\
D_{22}(A) &=& \left(1+\frac{A}{\Lambda_0^2}\right)^5 \,-\, \frac{M_N\,\nu_{22}}{4\pi}\,A\,(A+\Lambda_0^2)^5\mathfrak{H}^{[22]}(0;A)\notag\\
&&+\, \frac{M_N}{4\pi^2}\,A\,(A+\Lambda_0^2)^5\int_{-\infty}^L d\omega_L\,\frac{\mathfrak{H}^{[11]}(\omega_L;A)\left[\bm{D}(\omega_{\,L})\,\bm{\Delta}(\omega_{\,L})\right]_{22}}{\omega_L^2\,(\omega_L+\Lambda_0^2)^2}\,, \label{IED22}
\end{eqnarray}
with the algebraic RHC integral
\begin{eqnarray}
\mathfrak{H}^{[jk]}(\omega_L;A) &=& \frac{1}{\pi}\int_0^{\,\infty}\frac{d\omega_R\,\sqrt{\omega_R}\,\omega_R^j}{(\omega_R-\omega_L)\,(\omega_R-A)\,(\omega_R+\Lambda_0^2)^{2+k}} \notag\\
&=& \frac{1}{A-\omega_L} \,\Bigg[\,\frac{i\,\sqrt{A}\,A^j}{(A+\Lambda_0^2)^{2+k}}\,+\, \frac{\sqrt{-\omega_L}\,\omega_L^j}{(\omega_L+\Lambda_0^2)^{2+k}} \notag\\
&&+\frac{(-1)^{j-k}}{\Lambda_0^{1-2(j-k)}}\begin{pmatrix}
\frac{1}{2}\!+\!j \\ 1\!+\!k
\end{pmatrix}\,\Big(\frac{_2F_1(1,\tfrac{3}{2}\!+\!j;\tfrac{1}{2}\!+\!j\!-\!k;-\tfrac{\Lambda_0^2}{\omega_L})}{\omega_L}-\frac{_2F_1(1,\tfrac{3}{2}\!+\!j;\tfrac{1}{2}\!+\!j\!-\!k;-\tfrac{\Lambda_0^2}{A})}{A}\Big) \,\Bigg]\,. \label{auxH}
\end{eqnarray}
Thus, assuming $A\geqslant0$, taking imaginary parts in Eqs. \eqref{IED00}--\eqref{IED22}, and recalling Eqs. \eqref{N00A}, \eqref{N02A}, \eqref{N20A}, and \eqref{N22A}, it is clear that the unitarity condition \eqref{fromunit} keeps being preserved. Conversely, taking real parts in Eqs. \eqref{IED00}--\eqref{IED22} will allow us to remove the eight free coefficients from such IEs in favor of seven physical observables, as we elaborate in what follows.

\bigskip

\subsection{ERE matching}
\def\theequation{\arabic{section}.\arabic{equation}}
\label{matchingII}

\bigskip

To learn how the free coefficients $\left\lbrace\nu_{00},\nu_{02},\nu_{20},\nu_{22},\tau_{00},\tau_{02},\alpha_{00},\alpha_{20}\right\rbrace$ depend upon several ERE parameters, first come back to general relation \eqref{general}. In this context, its left-hand side 
can be expanded in powers of small $A$ by means of the DRs \eqref{N00A}, \eqref{N02A}, \eqref{N20A}, and \eqref{N22A}, yielding
\begin{eqnarray}
\frac{M_N}{4\pi}\bm{P}(A)\bm{N}(A)\bm{P}(A)^{-1}\bm{M}(A) \!\!\!&=&\!\!\! \frac{M_N}{4\pi}\bm{m}_0^{(a)}+\frac{M_N}{4\pi}\left[\bm{m}_1^{(a)}+\bm{m}_1^{(r)}\right]A+\frac{M_N}{4\pi}\left[\bm{m}_2^{(a)}+\bm{m}_2^{(r)}+\bm{m}_2^{(v)}\right]A^2+\mathcal{O}(A^3), \label{lefthandside}
\end{eqnarray}
with the matrices
\begin{eqnarray}
\bm{m}_0^{(a)} &=& \frac{1}{|\bm{a}|} \begin{pmatrix}
\nu_{02}a_{02}-\nu_{00}a_{22} & \nu_{00}a_{02}-\nu_{02}a_{00} \\ 0 & 0
\end{pmatrix}\,; 
\\ 
\bm{m}_1^{(a)} &=& \frac{1}{|\bm{a}|} \begin{pmatrix}
\tau_{02}a_{02}-\Big[\nu_{00}(\tau_{00}+\tfrac{1}{\Lambda_0^2})+\Lambda_0^4\mathfrak{G}_{00}^{[22]}\Big]a_{22} & -\tau_{02}a_{00}+\Big[\nu_{00}(\tau_{00}+\tfrac{1}{\Lambda_0^2})+\Lambda_0^4\mathfrak{G}_{00}^{[22]}\Big]a_{02} \\ 0 & 0
\end{pmatrix}\,, 
\\ 
\bm{m}_1^{(r)} &=& \frac{1}{2} \begin{pmatrix}
\nu_{00}r_{00}+\nu_{02}r_{02} & \nu_{00}r_{02}+\nu_{02}r_{22} \\ 0 & 0
\end{pmatrix}\,; 
\\ 
\bm{m}_2^{(a)} &=& \frac{1}{|\bm{a}|} \begin{pmatrix}
\mathfrak{G}_{02}^{[40]}a_{02}-(2\Lambda_0^2\mathfrak{G}_{00}^{[22]}+\Lambda_0^4\mathfrak{G}_{00}^{[32]}+\frac{\nu_{00}\tau_{00}}{\Lambda_0^2})a_{22} & -\mathfrak{G}_{02}^{[40]}a_{00}+(2\Lambda_0^2\mathfrak{G}_{00}^{[22]}+\Lambda_0^4\mathfrak{G}_{00}^{[32]}+\frac{\nu_{00}\tau_{00}}{\Lambda_0^2})a_{02} \\ (\Lambda_0^4\mathfrak{G}_{22}^{[32]}+\Lambda_0^2\nu_{22})a_{02}-(\Lambda_0^6\mathfrak{G}_{20}^{[23]}+\Lambda_0^4\nu_{20})a_{22} & -(\Lambda_0^4\mathfrak{G}_{22}^{[32]}+\Lambda_0^2\nu_{22})a_{00}+(\Lambda_0^6\mathfrak{G}_{20}^{[23]}+\Lambda_0^4\nu_{20})a_{02}
\end{pmatrix}\,, \notag\\
\\ 
\bm{m}_2^{(r)} &=& \frac{1}{2} \begin{pmatrix}
\Big[\nu_{00}(\tau_{00}+\tfrac{1}{\Lambda_0^2})+\Lambda_0^4\mathfrak{G}_{00}^{[22]}\Big]r_{00}+\tau_{02}r_{02} & \Big[\nu_{00}(\tau_{00}+\tfrac{1}{\Lambda_0^2})+\Lambda_0^4\mathfrak{G}_{00}^{[22]}\Big]r_{02}+\tau_{02}r_{22} \\ 0 & 0
\end{pmatrix}\,, 
\\ 
\bm{m}_2^{(v)}  &=& \begin{pmatrix}
\nu_{00}v_{00}+\nu_{02}v_{02} & \nu_{00}v_{02}+\nu_{02}v_{22} \\ 0 & 0
\end{pmatrix}\,, 
\end{eqnarray}
where
\begin{eqnarray}
\mathfrak{G}_{\ell'\ell}^{[nm]} &=& \frac{1}{\pi}\int_{-\infty}^L d\omega_L\,\frac{\left[\bm{D}(\omega_{\,L})\,\bm{\Delta}(\omega_{\,L})\right]_{\ell'\ell}}{\omega_L^n\,(\omega_L+\Lambda_0^2)^m}\,.
\end{eqnarray}
Regarding the right-hand side of Eq. \eqref{general}, from the IEs \eqref{IED00}--\eqref{IED22} fulfilled by $\bm{D}$ we find
\begin{eqnarray}
\bm{P}(A)\,\mathfrak{R}[\bm{D}(A)]\,\bm{P}(A) &=& \widetilde{\bm{m}}_0\,+\,\widetilde{\bm{m}}_1A\,+\,\widetilde{\bm{m}}_2A^2\,+\,\mathcal{O}(A^3), \label{righthandside}
\end{eqnarray}
with
\begin{eqnarray}
\widetilde{\bm{m}}_0 &=& \begin{pmatrix}
1 & 0 \\ 0 & 0
\end{pmatrix}\,; \\
\widetilde{\bm{m}}_1 &=& \left[\frac{3}{\Lambda_0^2}+\alpha_{00}-\frac{M_N\Lambda_0^6}{4\pi}\nu_{00}\left(\mathfrak{R}[\mathfrak{H}^{[01]}(0;0)]+\tau_{00}\mathfrak{R}[\mathfrak{H}^{[11]}(0;0)]\right)+\frac{M_N\Lambda_0^8}{4\pi}\mathfrak{F}_{00}^{[00;12]}\right] \begin{pmatrix}
1 & 0 \\ 0 & 0
\end{pmatrix}\,; \\
\widetilde{\bm{m}}_2 &=& \begin{pmatrix}
\frac{3}{\Lambda_0^4}+\frac{3\alpha_{00}}{\Lambda_0^2} & 0 \\ \Lambda_0^8\alpha_{20} & 1
\end{pmatrix} \notag\\
&&\!\!\!\!\!\!\!\!\!\!\!\!+\,\frac{M_N\Lambda_0^8}{4\pi} \begin{pmatrix}
-\frac{\nu_{00}\left[\left(\frac{4}{\Lambda_0^2}+\tau_{00}\right)\mathfrak{R}[\mathfrak{H}^{[01]}(0;0)]+\frac{4\tau_{00}}{\Lambda_0^2}\mathfrak{R}[\mathfrak{H}^{[11]}(0;0)]+\mathfrak{R}[\mathfrak{H}^{[-11]}(0;0)]\right]}{\Lambda_0^2} & -\nu_{02}\mathfrak{R}[\mathfrak{H}^{[12]}(0;0)]-\tau_{02}\mathfrak{R}[\mathfrak{H}^{[22]}(0;0)] \\ -\Lambda_0^2\nu_{20}\mathfrak{R}[\mathfrak{H}^{[11]}(0;0)] & 0
\end{pmatrix} \notag\\
&&\!\!\!\!\!\!\!\!\!\!\!\!+\,\frac{M_N\Lambda_0^8}{4\pi} \begin{pmatrix}
\frac{4}{\Lambda_0^2}\mathfrak{F}_{00}^{[00;12]}+\mathfrak{F}_{00}^{[-10;12]} & \mathfrak{F}_{02}^{[22;30]} \\ \Lambda_0^2\mathfrak{F}_{20}^{[00;13]} & 0
\end{pmatrix}\,,
\end{eqnarray}
having taken
\begin{eqnarray}
\mathfrak{F}_{\ell'\ell}^{[jk;nm]} &=& \frac{1}{\pi}\int_{-\infty}^L d\omega_L\,\frac{\left[\bm{D}(\omega_{\,L})\,\bm{\Delta}(\omega_{\,L})\right]_{\ell'\ell}\,\mathfrak{R}[\mathfrak{H}^{[jk]}(\omega_L;0)]}{\omega_L^n\,(\omega_L+\Lambda_0^2)^m}\,.
\end{eqnarray}
Then, matching Eqs. \eqref{lefthandside} and \eqref{righthandside} at each $A$ allows one to obtain closed expressions for the four $\nu$'s, the two $\tau$'s, and the two $\alpha$'s in terms of ERE parameters. Nonetheless, most of those expressions are cumbersome and ugly, and it would be an unnecessary bore to give them here. Suffice to say that our eight coefficients can be evaluated from the knowledge of ERE parameters according to
\begin{eqnarray}
\nu_{00} &=& \nu_{00}\left(a_{00}\right)\,; \\\nu_{02} &=& \nu_{02}\left(a_{02}\right)\,; \\\nu_{20} &=& \nu_{20}\left(a_{00},a_{02},a_{22},r_{02},r_{22}\right)\,; \\
\nu_{22} &=& \nu_{22}\left(a_{00},a_{02},a_{22},r_{02},r_{22}\right)\,; \\
\tau_{00} &=& \tau_{00}\left(a_{00},a_{02},a_{22},r_{02},r_{22},\beta\right)\,; \\
\tau_{02} &=& \tau_{02}\left(a_{00},a_{02},a_{22},r_{02},r_{22},\beta\right)\,; \\
\alpha_{00} &=& \alpha_{00}\left(a_{00},a_{02},a_{22},r_{00},r_{02},r_{22},\beta\right)\,; \\
\alpha_{20} &=& \alpha_{20}\left(a_{00},a_{02},a_{22},r_{02},r_{22}\right)\,,\end{eqnarray}
with the mixing parameter
\begin{eqnarray}
\beta &\equiv& a_{00}v_{02}+a_{02}v_{22}\,.
\end{eqnarray}
That is to say, we can trade the information carried by our eight coefficients by the one brought by \textit{seven} phenomenological parameters. This is in contrast with what we saw for the normalization $\bm{N}(0)=\bm{1}$, where `only' six phenomenological parameters are needed and the relevance or irrelevance of each ERE parameter is justifiable from plain  
dimensional analysis \eqref{abarminus1}--\eqref{vbar}. To make an exhaustive study of this fact is out of the scope of this work, where we contempt ourselves in having found a possible $N/D$ representation for each type of normalization, either $\bm{N}(0)=\bm{1}$ or $\bm{D}(0)=\bm{1}$.

\bigskip
\bigskip
\bigskip

\section{A matrix $\pmb{N/D}$ representation is not always possible}
\label{sec.240101.1}
\setcounter{equation}{0}   
\def\theequation{\arabic{section}.\arabic{equation}}

\bigskip

In the previous two sections we have presented a comprehensive demonstration of the existence of a consistent matrix $N/D$ representation for the on-shell amplitude that results from our initial potential \eqref{orpot}. Conversely, in this section we will proof explicitly that an apparently slight variant of the original interaction does not admit anymore a matrix $N/D$ representation.

\bigskip
Suppose that we stick to a separable form of the potential such as Eq. \eqref{orpot}, but consider instead the alternative vertex functions
\begin{eqnarray}
g_0(A) &=& \frac{(A-L)^{1/3}}{A+\Lambda_0^2}\,; \label{g0alt} \\
g_2(A) &=& \frac{A\,(A-L)^{1/6}}{(A+\Lambda_0^2)^2}\,. \label{g2alt} 
\end{eqnarray} 
The general solution for the on-shell amplitude matrix \eqref{mateq} may be recast as
\begin{eqnarray}
\bm{T}(A) &=& \bm{n}(A)\,\bm{m}(A)^{-1}\,\bm{n}(A) \quad\!\!\!=\!\!\!\quad \widetilde{\bm{D}}(A)^{-1}\,\bm{N}(A)\,, \label{nmsplit}
\end{eqnarray}
with the matrices
\begin{eqnarray}
\bm{n}(A) &=& 
\begin{pmatrix}
g_0(A) & 0 \\ 0 & g_2(A)
\end{pmatrix}\,; \label{genn} \\
\bm{m}(A) &=& \frac{M_N
}{|\bm{\gamma}\,|}\,\begin{pmatrix}
{\gamma_{\,22}}\,-\,{|\bm{\gamma}\,|}\,I_{\,0}(A) & -\,{\gamma_{\,02}} \\ -\,{\gamma_{\,02}} & {\gamma_{\,00}}\,-\,{|\bm{\gamma}\,|}\,I_{\,2}(A) 
\end{pmatrix}\,; \label{genm} \\
\bm{\widetilde{D}}(A) &=& \bm{n}(A)\,\bm{m}(A)\,\bm{n}(A)^{-1} \quad\!\!\!=\!\!\!\quad \frac{M_N
}{|\bm{\gamma}\,|}\,\begin{pmatrix}
{\gamma_{\,22}}\,-\,{|\bm{\gamma}\,|}\,I_{\,0}(A) & -\,\frac{g_0(A)}{g_2(A)}\,{\gamma_{\,02}} \\ -\,\frac{g_2(A)}{g_0(A)}\,{\gamma_{\,02}} & {\gamma_{\,00}}\,-\,{|\bm{\gamma}\,|}\,I_{\,2}(A) 
\end{pmatrix}\,; \label{genDbar} \\
\bm{N}(A) &=& \bm{n}(A)^2 \quad\!\!\!=\!\!\!\quad \begin{pmatrix}
g_0(A)^2 & 0 \\ 0 & g_2(A)^2
\end{pmatrix}\,, \label{genN}
\end{eqnarray}
where  
$I_{\,\ell}(A)$ keeps being given by the first line of Eq. \eqref{loop}.
While $\bm{N}(A)$ is RHC-free as there are no loop integrals in Eq. \eqref{genN}, $\bm{\widetilde{D}}(A)$ gives rise not only to the RHC, but also the LHC (provided that $\gamma_{\,02}\neq0$) due to the ${g_0(A)}/{g_2(A)}$ and ${g_2(A)}/{g_0(A)}$ factors present in the off-diagonal matrix elements in Eq. \eqref{genDbar}. 

\bigskip
Now, suppose that one can find some RHC-free matrix $\bm{O}(A)$ whose LHC allows us to remove the LHC from $\bm{\widetilde{D}}(A)$, say\,\footnote{We could also assume to have $\bm{O}(A)\,\bm{\widetilde{D}}(A)\,=\,\bm{R}(A)\,\bm{\overline{\bm{D}}}(A)$, with $\bm{R}(A)$ and $\bm{\overline{\bm{D}}}(A)$ RHC- and LHC-free, respectively. But in fact this would amount to a redefinition of $\bm{O}(A)$ in Eq. \eqref{Omat}, $\bm{O}(A)\,\to\,\bm{R}(A)^{-1}\,\bm{O}(A)$. In other words, $\bm{R}(A)$ can be safely ignored without loss of generality.} 
\begin{eqnarray}
\bm{D}(A)&=&\bm{O}(A)\,\bm{\widetilde{D}}(A)\,, \label{Omat}
\end{eqnarray} 
thus to separate explicitly the LHC from the RHC in Eq. \eqref{nmsplit}, i.e.
\begin{eqnarray}
\bm{T}(A) &=& \underbrace{\bm{D}(A)^{-1}}_{\text{RHC}}\,\, \underbrace{\bm{O}(A)\,\bm{N}(A)}_{\text{LHC}}\,. \label{newT}
\end{eqnarray}
Given Eqs. \eqref{g0alt} and \eqref{g2alt}, $\bm{O}(A)$ must have the form
\begin{eqnarray}
\bm{O}(A) &=& \begin{pmatrix}
\mathfrak{f}_{00}(A)\,(A-L)^{\phi_{\,00}} & \mathfrak{f}_{02}(A)\,(A-L)^{\phi_{\,02}} \\ \mathfrak{f}_{20}(A)\,(A-L)^{\phi_{\,20}} & \mathfrak{f}_{22}(A)\,(A-L)^{\phi_{\,22}} 
\end{pmatrix}\,, \label{Omat2}
\end{eqnarray}
where $\mathfrak{f}_{\ell'\ell}(A)$ has no cuts, and ${\phi}_{\,\ell'\ell}\,\in\,(-1,1)\,\backslash\,\lbrace0\rbrace$ --- since any contribution from integer values of $\phi_{\,\ell'\ell}$ is free of cuts, allowing us to reabsorb its effect into $\mathfrak{f}_{\ell'\ell}(A)$. Substituting Eqs. \eqref{g0alt}, \eqref{g2alt}, \eqref{genDbar}, and \eqref{Omat2} into Eq. \eqref{Omat},
\begin{eqnarray}
  \label{231225.1}
\bm{D} \!\!&=&\!\! -\frac{M_N\gamma_{02}(1+\frac{\Lambda_0^2}{A})}{|\bm{\gamma}\,|}\!\!\begin{pmatrix}
\frac{\mathfrak{f}_{02}}{(1+\frac{\Lambda_0^2}{A})^2}(A-L)^{\phi_{02}-\frac{1}{6}}\!+\!\frac{\left({|\bm{\gamma}\,|}I_{0}-{\gamma_{22}}\right)\mathfrak{f}_{00}}{\gamma_{02}(1+\frac{\Lambda_0^2}{A})}(A-L)^{\phi_{00}} \!&\!\!\! \mathfrak{f}_{00}(A-L)^{\phi_{00}+\frac{1}{6}}\!+\!\frac{\left({|\bm{\gamma}\,|}I_{2}-{\gamma_{00}}\right)\mathfrak{f}_{02}}{\gamma_{02}(1+\frac{\Lambda_0^2}{A})}(A-L)^{\phi_{02}}  \\ \\ \frac{\mathfrak{f}_{22}}{(1+\frac{\Lambda_0^2}{A})^2}(A-L)^{\phi_{22}-\frac{1}{6}}\!+\!\frac{\left({|\bm{\gamma}\,|}I_{0}-{\gamma_{22}}\right)\mathfrak{f}_{20}}{\gamma_{02}(1+\frac{\Lambda_0^2}{A})}(A-L)^{\phi_{20}} \!&\!\!\! \mathfrak{f}_{20}(A-L)^{\phi_{20}+\frac{1}{6}}\!+\!\frac{\left({|\bm{\gamma}\,|}I_{2}-{\gamma_{00}}\right)\mathfrak{f}_{22}}{\gamma_{02}(1+\frac{\Lambda_0^2}{A})}(A-L)^{\phi_{22}}   
\end{pmatrix}\,, \notag\\
\end{eqnarray}
where the $A$ dependence was omitted for brevity. Now we consider four mutually exclusive scenarios:

\begin{itemize}
\item 
$\bm{\mathfrak{\bm{f_{\bm{00}}(A)}}=0}$ \textbf{and} $\bm{\phi_{\bm{02}}=\bm{\frac{\bm{1}}{\bm{6}}}}$\textbf{.} Thus,
\end{itemize}
\begin{eqnarray}
\bm{D} \!\!&=&\!\! -\frac{M_N\gamma_{02}(1+\frac{\Lambda_0^2}{A})}{|\bm{\gamma}\,|}\!\!\begin{pmatrix}
\frac{\mathfrak{f}_{02}}{(1+\frac{\Lambda_0^2}{A})^2} \!&\!\!\! \frac{\left({|\bm{\gamma}\,|}I_{2}-{\gamma_{00}}\right)\mathfrak{f}_{02}}{\gamma_{02}(1+\frac{\Lambda_0^2}{A})}(A-L)^{\frac{1}{6}}  \\ \\ \frac{\mathfrak{f}_{22}}{(1+\frac{\Lambda_0^2}{A})^2}(A-L)^{\phi_{22}-\frac{1}{6}}\!+\!\frac{\left({|\bm{\gamma}\,|}I_{0}-{\gamma_{22}}\right)\mathfrak{f}_{20}}{\gamma_{02}(1+\frac{\Lambda_0^2}{A})}(A-L)^{\phi_{20}} \!&\!\!\! \mathfrak{f}_{20}(A-L)^{\phi_{20}+\frac{1}{6}}\!+\!\frac{\left({|\bm{\gamma}\,|}I_{2}-{\gamma_{00}}\right)\mathfrak{f}_{22}}{\gamma_{02}(1+\frac{\Lambda_0^2}{A})}(A-L)^{\phi_{22}}   
\end{pmatrix}\,, \notag\\
\end{eqnarray}
\begin{itemize}
\item[] and to remove the LHC from $D_{02}(A)$ one needs to impose $\mathfrak{f}_{02}(A)=0$. Hence, $D_{00}(A)=D_{02}(A)=0$, implying $|\bm{D}(A)|=0$, i.e. $\bm{D}(A)$ is not invertible, in contradiction with Eq. \eqref{newT}.

\item 
$\bm{\mathfrak{\bm{f_{\bm{00}}(A)}}=0}$ \textbf{and} $\bm{\phi_{\bm{02}}\neq\bm{\frac{\bm{1}}{\bm{6}}}}$\textbf{.} Hence,
\end{itemize}
\begin{eqnarray}
\bm{D} \!\!&=&\!\! -\frac{M_N\gamma_{02}(1+\frac{\Lambda_0^2}{A})}{|\bm{\gamma}\,|}\!\!\begin{pmatrix}
\frac{\mathfrak{f}_{02}}{(1+\frac{\Lambda_0^2}{A})^2}(A-L)^{\phi_{02}-\frac{1}{6}} \!&\!\!\! \frac{\left({|\bm{\gamma}\,|}I_{2}-{\gamma_{00}}\right)\mathfrak{f}_{02}}{\gamma_{02}(1+\frac{\Lambda_0^2}{A})}(A-L)^{\phi_{02}}  \\ \\ \frac{\mathfrak{f}_{22}}{(1+\frac{\Lambda_0^2}{A})^2}(A-L)^{\phi_{22}-\frac{1}{6}}\!+\!\frac{\left({|\bm{\gamma}\,|}I_{0}-{\gamma_{22}}\right)\mathfrak{f}_{20}}{\gamma_{02}(1+\frac{\Lambda_0^2}{A})}(A-L)^{\phi_{20}} \!&\!\!\! \mathfrak{f}_{20}(A-L)^{\phi_{20}+\frac{1}{6}}\!+\!\frac{\left({|\bm{\gamma}\,|}I_{2}-{\gamma_{00}}\right)\mathfrak{f}_{22}}{\gamma_{02}(1+\frac{\Lambda_0^2}{A})}(A-L)^{\phi_{22}}   
\end{pmatrix}\,, \notag\\
\end{eqnarray}
\begin{itemize}
\item[] and removing the LHC from $D_{00}(A)$ demands to take $\mathfrak{f}_{02}(A)=0$, thus $D_{00}(A)=D_{02}(A)=0$, which is again incompatible with Eq. \eqref{newT}.

\item 
$\bm{\mathfrak{\bm{f_{\bm{00}}(A)}}\neq0}$ \textbf{and} $\bm{\phi_{\bm{02}}=\bm{\frac{\bm{1}}{\bm{6}}}}$\textbf{.} Therefore,
\end{itemize}
\begin{eqnarray}
\bm{D} \!\!&=&\!\! -\frac{M_N\gamma_{02}(1+\frac{\Lambda_0^2}{A})}{|\bm{\gamma}\,|}\!\!\begin{pmatrix}
\frac{\mathfrak{f}_{02}}{(1+\frac{\Lambda_0^2}{A})^2}\!+\!\frac{\left({|\bm{\gamma}\,|}I_{0}-{\gamma_{22}}\right)\mathfrak{f}_{00}}{\gamma_{02}(1+\frac{\Lambda_0^2}{A})}(A-L)^{\phi_{00}} \!&\!\!\! (A-L)^{\frac{1}{6}}\,\Big[\mathfrak{f}_{00}(A-L)^{\phi_{00}}\!+\!\frac{\left({|\bm{\gamma}\,|}I_{2}-{\gamma_{00}}\right)\mathfrak{f}_{02}}{\gamma_{02}(1+\frac{\Lambda_0^2}{A})}\Big]  \\ \\ \frac{\mathfrak{f}_{22}}{(1+\frac{\Lambda_0^2}{A})^2}(A-L)^{\phi_{22}-\frac{1}{6}}\!+\!\frac{\left({|\bm{\gamma}\,|}I_{0}-{\gamma_{22}}\right)\mathfrak{f}_{20}}{\gamma_{02}(1+\frac{\Lambda_0^2}{A})}(A-L)^{\phi_{20}} \!&\!\!\! \mathfrak{f}_{20}(A-L)^{\phi_{20}+\frac{1}{6}}\!+\!\frac{\left({|\bm{\gamma}\,|}I_{2}-{\gamma_{00}}\right)\mathfrak{f}_{22}}{\gamma_{02}(1+\frac{\Lambda_0^2}{A})}(A-L)^{\phi_{22}}   
\end{pmatrix}\,, \notag\\
\end{eqnarray}
\begin{itemize}
\item[] and 
  $D_{02}(A)$ 
  being LHC-free implies the square bracket above has to vanish, namely
\begin{eqnarray}
\mathfrak{f}_{00}(A) &=& \frac{{\gamma_{00}}-{|\bm{\gamma}\,|}I_{2}(A)}{\gamma_{02}(1+\frac{\Lambda_0^2}{A})}\,\frac{\mathfrak{f}_{02}(A)}{(A-L)^{\phi_{00}}}\,.
\end{eqnarray}
Even for $\phi_{\,00}=0$, from the above it is apparent that $\mathfrak{f}_{00}(A)$ and $\mathfrak{f}_{02}(A)$ cannot be simultaneously free of cuts since $I_2(A)$ has RHC: contradiction.

\item 
$\bm{\mathfrak{\bm{f_{\bm{00}}(A)}}\neq0}$ \textbf{and} $\bm{\phi_{\bm{02}}\neq\bm{\frac{\bm{1}}{\bm{6}}}}$\textbf{.}
  In this case, to get rid 
  of the LHC of $D_{00}(A)$ in the general Eq.~\eqref{231225.1} we need to have 
\end{itemize}
\begin{eqnarray}
\mathfrak{f}_{02}(A) &=& \left(1+\frac{\Lambda_0^2}{A}\right)\frac{{\gamma_{\,22}}-{|\bm{\gamma}\,|}I_{0}(A)}{\gamma_{\,02}\,{(A-L)^{\phi_{\,02}-\phi_{\,00}-\frac{1}{6}}}}\,\mathfrak{f}_{00}(A)\,.
\end{eqnarray}
\begin{itemize}
\item[] Even for $\phi_{\,00}=\phi_{\,02}-\frac{1}{6}$, we see that $\mathfrak{f}_{00}(A)$ and $\mathfrak{f}_{02}(A)$ cannot be free of cuts at the same time: contradiction.
\end{itemize}

\bigskip

\subsection{Generalization of the counterexample}
\label{sec.231021.3}

\bigskip
Here we consider a more general form for the vertex functions, namely
\begin{eqnarray}
g_0(A) &=& \frac{(A-L)^{\theta_0}}{A+\Lambda_0^2}\,; \label{newg0alt} \\
g_2(A) &=& \frac{A\,(A-L)^{\theta_2}}{(A+\Lambda_0^2)^2}\,. \label{newg2alt} 
\end{eqnarray} 
in terms of the exponents $0\leqslant \{\theta_0,\,\theta_2\} < \frac{1}{2}$, taking for granted that $\theta_0\neq \theta_2$ unlike Eq. \eqref{gi}.
Equation \eqref{genDbar} for $\bm{\widetilde{D}}(A)$ is still valid, and $\bm{D}(A)$ of Eq.~\eqref{231225.1} becomes now
\begin{eqnarray}
  \label{231225.2}
\bm{D} \!\!\!\!\!\!&=&\!\!\!\!\!\!\!\! -\frac{M_N\gamma_{02}(1+\frac{\Lambda_0^2}{A})}{|\bm{\gamma}\,|}\!\!\begin{pmatrix}
\frac{\mathfrak{f}_{02}}{(1+\frac{\Lambda_0^2}{A})^2}(A-L)^{\phi_{02}+\Delta\theta}\!+\!\frac{\left({|\bm{\gamma}\,|}I_{0}-{\gamma_{22}}\right)\mathfrak{f}_{00}}{\gamma_{02}(1+\frac{\Lambda_0^2}{A})}(A-L)^{\phi_{00}} \!&\!\!\! \mathfrak{f}_{00}(A-L)^{\phi_{00}-\Delta\theta}\!+\!\frac{\left({|\bm{\gamma}\,|}I_{2}-{\gamma_{00}}\right)\mathfrak{f}_{02}}{\gamma_{02}(1+\frac{\Lambda_0^2}{A})}(A-L)^{\phi_{02}}  \\ \\ \frac{\mathfrak{f}_{22}}{(1+\frac{\Lambda_0^2}{A})^2}(A-L)^{\phi_{22}+\Delta\theta}\!+\!\frac{\left({|\bm{\gamma}\,|}I_{0}-{\gamma_{22}}\right)\mathfrak{f}_{20}}{\gamma_{02}(1+\frac{\Lambda_0^2}{A})}(A-L)^{\phi_{20}} \!&\!\!\! \mathfrak{f}_{20}(A-L)^{\phi_{20}-\Delta\theta}\!+\!\frac{\left({|\bm{\gamma}\,|}I_{2}-{\gamma_{00}}\right)\mathfrak{f}_{22}}{\gamma_{02}(1+\frac{\Lambda_0^2}{A})}(A-L)^{\phi_{22}}   
\end{pmatrix}\,, \notag\\
\end{eqnarray}
where $\Delta\theta\equiv\theta_2-\theta_0$. The procedure to conclude that there is no suitable $\bm{O}(A)$ is analogous to the one followed above for the particular case $\theta_0=\frac{1}{3}$ and $\theta_2=\frac{1}{6}$  
with the replacement $\mp \frac{1}{6}\to \pm\Delta\theta ~\slashed{\in}~ \mathbb{Z}$.

\bigskip
\bigskip
\bigskip

\section{Adding short-range terms to the separable interaction}
\setcounter{equation}{0}   
\def\theequation{\arabic{section}.\arabic{equation}}
\label{sec.240103.5}

\bigskip

In this section we return to our original (regular) potential \eqref{orpot} to study the case of including contact contributions on top of it.  
Having solved the Lippmann-Schwinger equation associated to this potential and matching the on-shell amplitude matrix with the ERE, we conclude that this $T$ matrix is not compatible with a matrix $N/D$ representation, like it happens with the family of interactions we proposed in Sec. \ref{sec.240101.1}. 

\bigskip
Given the singular nature of contact interactions, an immediate consequence of such an inclusion is that the Lippmann-Schwinger equation \eqref{LSij} now requires to be regularized by hand \cite{VanKolck:1999zys}. Using a (local) sharp-cutoff prescription, we replace Eq. \eqref{LSij} by [$\mathfrak{I}(A)\neq0$]
\begin{eqnarray}
T_{\,\ell'\ell}(Q',Q;A) &=& V_{\,\ell'\ell}(Q',Q) \,+\, \frac{M_N}{2\pi^2} \sum_{m=0,2}\int_0^\Lambda dq\,V_{\,\ell'm}(Q',q^2) \frac{q^2}{q^2-A} T_{\,m\ell}(q^2,Q;A)\,,\label{LSijLambda}
\end{eqnarray}
where the ultraviolet cutoff $\Lambda$ is typically chosen to be  
$\gg {\rm max}(|\bm{a}|^{-1},|\bm{r}|^{-1},\Lambda_0)$. 
As we will see, introducing $\Lambda$ in Eq. \eqref{LSijLambda} induces an implicit cutoff running in the potential --- in particular, in its short-range part --- such that the cutoff running exhibited by the resulting amplitude is mild and controlled. Here we explore the simplest possibility in this regard --- that of one single (non-derivative) contact term supplementing only the pure $S$-wave component of the original interaction \eqref{orpot}, namely
\begin{eqnarray}
V_{\,\ell'\ell}(Q',Q) &=& \frac{1}{M_N}\left[C_0(\Lambda)\,\delta_{\,\ell'0}\,\delta_{\,\ell0}\,+\,\gamma_{\,\ell'\ell}\,g_{\,\ell'}(Q')\,g_{\,\ell}(Q)\right]\,,\label{withczero}\end{eqnarray}
where the cutoff dependence of the contact term has now been made explicit. This potential is a sum of separable potentials and the amplitude happens to have the same structure. Indeed, plugging into Eq. \eqref{LSijLambda} the ansatz 
\begin{eqnarray}
T_{\,\ell'\ell}(Q',Q;A) &=& \sum_{j'=0}^{1}\sum_{j=0}^{1}c_{\,\ell'\ell}^{\,[j'\!j]}(A)\,g_{\,\ell'}(Q')^{j'}\,g_{\,\ell}(Q)^{j}\,,
\end{eqnarray}
one can solve for $c_{\,\ell'\ell}^{\,[j'\!j]}(A)$ so that the solution reads
\begin{eqnarray}
\!\!\!\!\!\!\!\!\!\!\!\!\!\!\!\!\!\!\!\!\!\!\!\!\!\!T_{\ell'\ell}(Q',Q;A) &=& \frac{\gamma_{\ell'\ell}}{M_N}\frac{t_{\ell'\ell}(Q',Q;A)}{\left[1-\gamma_{22}\mathcal{I}_2(A)\right]\left[\tfrac{1}{C_0(\Lambda)}-\mathcal{I}_{\slashed{\pi}}(A)\right]-\left[\gamma_{00}-|\bm{\gamma}|\mathcal{I}_2(A)\right]\left[\left(\tfrac{1}{C_0(\Lambda)}-\mathcal{I}_{\slashed{\pi}}(A)\right)\mathcal{I}_0(A)+\mathcal{J}\!(A)^2\right]}\,,  \label{Tell'ell}\end{eqnarray}
with an LHC-free denominator. Yet the numerator $t_{\,\ell'\ell}(Q',Q;A)$ encodes both the LHC and the RHC. Explicitly,
\begin{eqnarray}
\!\!\!\!\!\!\!\!\!\!\!\!\!\!\!\!\!\!\!\!\!\!\!\!\!\!t_{00}(Q',Q;A)  \!\!\!&=&\!\!\!  \tfrac{1-\gamma_{\,22}\mathcal{I}_2(A)}{\gamma_{\,00}} \!-\! \Big(1\!-\!\tfrac{|\bm{\gamma}\,|}{\gamma_{\,00}}\mathcal{I}_2(A)\Big)\!\left[ \!\mathcal{I}_0(A) \!-\! \mathcal{J}\!(A)\left(g_0(Q')\!+\!g_0(Q)\right)\!-\!\Big(\tfrac{1}{C_0(\Lambda)}\!-\!\mathcal{I}_{\slashed{\pi}}(A)\Big)g_0(Q')g_0(Q)\!\right]\,; \label{t00}\\ 
\!\!\!\!\!\!\!\!\!\!\!\!\!\!\!\!\!\!\!\!\!\!\!\!\!\!t_{\,02}(Q',Q;A) \!\!\!&=&\!\!\! {\mathcal{J}\!(A)\,\,g_2(Q)\,+\,\Big(\tfrac{1}{C_0(\Lambda)}-\mathcal{I}_{\slashed{\pi}}(A)\Big)\,g_0(Q')\,g_2(Q)} \quad\!\!=\!\!\quad t_{\,20}(Q,Q';A)\,; \label{t02}
\\ 
\!\!\!\!\!\!\!\!\!\!\!\!\!\!\!\!\!\!\!\!\!\!\!\!\!\!t_{\,22}(Q',Q;A) \!\!\!&=&\!\!\! \left[\Big(1-\tfrac{|\bm{\gamma}\,|}{\gamma_{\,22}}\mathcal{I}_0(A)\Big)\,\Big(\tfrac{1}{C_0(\Lambda)}-\mathcal{I}_{\slashed{\pi}}(A)\Big)\,-\,\tfrac{|\bm{\gamma}\,|}{\gamma_{\,22}}\mathcal{J}\!(A)^2\right]g_2(Q')\,g_2(Q)\,. \label{t22} 
\end{eqnarray}

\bigskip
We have introduced the regularized loop integrals [$\mathfrak{I}(A)\neq0$]
\begin{eqnarray}
\mathcal{I}_{\slashed{\pi}}(A) &=& \frac{1}{2\pi^2}\int_0^{\,\Lambda} dq\,\frac{q^2}{q^2-A} \quad\!\!\!=\!\!\!\quad \frac{\Lambda_0}{2\pi^2}\left(\frac{1}{\sqrt{y}}\,-\,\sqrt{-\frac{A}{\Lambda_0^2}}\,\,\text{arccot}\,\sqrt{-\frac{A}{\Lambda_0^2}\,y}\,\right)\,; \label{Ipiless} \\
\mathcal{I}_{\,\ell}(A) &=& \frac{1}{2\pi^2}\int_0^{\,\Lambda} dq\,\frac{q^2\,g_\ell(q^2)^2}{q^2-A}\notag\\
&=& -\,\frac{1}{4\pi^2\,(1+\frac{A}{\Lambda_0^2})^{\,\ell+2}\,\Lambda_0^2}\,\Bigg[ \frac{1}{(1+x)^\ell}\,\sum_{m=0}^{\ell+1}\Bigg(\frac{\sqrt{1\,-\,x\,y}}{(1+y)^{\,\ell+1}}\overline{\mathfrak{a}}_{\ell m} \,+\, \frac{\text{arctanh}\,\sqrt{\frac{1\,+\,x}{1\,-\,x\,y}}}{\sqrt{1+x}}\,\mathfrak{b}_{\,\ell m}\Bigg)\,\frac{A^{m}}{\Lambda_0^{2m}}\notag \\
&& +\, \frac{2\,A^{\ell}\sqrt{-\frac{A}{\Lambda_0^2}\,(\frac{A}{\Lambda_0^2}-x)}}{\Lambda_0^{2\ell}} \, \arctan\!\Big(\,\frac{\sqrt{\frac{A}{\Lambda_0^2}-x}}{\sqrt{1\,-\,x\,y}\sqrt{-\frac{A}{\Lambda_0^2}}}\,\Big)\Bigg]\,; \label{Il} \\
\mathcal{J}\!(A) &=& \frac{1}{2\pi^2}\int_0^{\,\Lambda} dq\,\frac{q^2\,g_0(q^2)}{q^2-A}\notag\\
&=& -\,\frac{\frac{A}{\Lambda_0^2}}{2\pi^2\,(1+\frac{A}{\Lambda_0^2}) (-x)^{1/4}\,\sqrt{\Lambda_0}} \Bigg[ \frac{_2F_1(\frac{1}{2},\frac{3}{4};\frac{3}{2};\frac{1}{x\,y})}{\sqrt{-\,x\,y}} \,+\, i\,\,\Pi\Big(\sqrt{\tfrac{x}{1+x}};\tfrac{\pi}{2}\Big|-1\Big) \,+\, i\,\,\Pi\Big(\!-\!\sqrt{\tfrac{x}{1+x}};\tfrac{\pi}{2}\Big|-1\Big) \notag\\
&&\,-\,\, i\,\,\Pi\Big(\sqrt{\tfrac{x}{1+x}};\arcsin\,(1-\tfrac{1}{x\,y})^{1/4}\Big|-1\Big) \,-\, i\,\,\Pi\Big(\!-\!\sqrt{\tfrac{x}{1+x}};\arcsin\,(1-\tfrac{1}{x\,y})^{1/4}\Big|-1\Big)  \notag\\
&&\,+\,\, \frac{\sqrt{-\,\frac{x}{y}}}{\frac{A}{\Lambda_0^2}}\,\Big(\,F_1(\tfrac{1}{2};-\tfrac{1}{4},1;\tfrac{3}{2};\tfrac{1}{x\,y},\tfrac{1}{\frac{A}{\Lambda_0^2}y})\,-\, (1+\tfrac{A}{\Lambda_0^2})\,F_1(\tfrac{1}{2};-\tfrac{1}{4},1;\tfrac{3}{2};\tfrac{1}{x\,y},-\tfrac{1}{y})  \, \Big)\Bigg]\,, \label{J}
\end{eqnarray}
with $y\equiv \Lambda_0^2/\Lambda^2$. We see that Eq. \eqref{Ipiless} diverges in the cutoff-removal limit $y\to0$, which indeed motivates the implementation of some regularization technique (we remark though that a sharp momentum cutoff is not the only possible choice). Besides, in Eq. \eqref{Il} we have recycled the notation of Eqs. \eqref{a0m}--\eqref{b2m} by introducing the shorthand $\overline{\mathfrak{a}}_{\ell m}={\mathfrak{a}}_{\ell m}+\Delta{\mathfrak{a}}_{\ell m}$, where the dimensionless coefficients
\begin{eqnarray}
\Delta\mathfrak{a}_{\,00} \,\,=\,\, 0\,, &&\!\!\!\!\!\!\! \Delta\mathfrak{a}_{\,01} \,\,=\,\, 0\,; \\
\Delta\mathfrak{a}_{\,20} \,\,=\,\, \tfrac{x+4x^2}{12}\,y\,+\,\tfrac{x^2}{8}\,y^2\,, &&\!\!\!\!\!\!\! \Delta\mathfrak{a}_{\,21}\,\, = \,\, -\,\tfrac{2+x-4x^2}{4}\,y\,+\,\tfrac{2x+5x^2}{8}\,y^2\,, \\
\Delta\mathfrak{a}_{\,22}\,\, = \,\, -\,\tfrac{12+19x+4x^2}{4}\,y\,-\,\tfrac{8+12x+x^2}{8}\,y^2\,, &&\!\!\!\!\!\!\! \Delta\mathfrak{a}_{\,23}\,\, = \,\, -\,\tfrac{30+53x+20x^2}{12}\,y\,-\,\tfrac{8+14x+5x^2}{8}\,y^2 \end{eqnarray}
are all zero for $y=0$.
It is thus manifest that Eq. \eqref{Il} reduces to Eq. \eqref{loop} when the cutoff is removed, as it has to. Finally, in Eq. \eqref{J} 
\begin{eqnarray}
\!\!\!\!\!\!\!\!\!\!\!\!\!\!\!\!\!\!\!\!\!\!\!\!\!\!\!\!\!\Pi(n;\phi|m) \,\,\,=\,\,\, \int_0^{\,\phi} \frac{dt}{\sqrt{1-m\sin^2t}\,(1-n\sin^2t)}\,; \!&&\! F_1(a;b_1,b_2;c;z_1,z_2) \,\,\,=\,\,\, \sum_{m=0}^\infty\sum_{n=0}^\infty \frac{(a)_{m+n}\,(b_1)_m\,(b_2)_n}{(c)_{m+n}}\,\frac{z_1^m}{m!}\,\frac{z_2^n}{n!}
\end{eqnarray} 
stand respectively for an elliptic integral of the third kind and an Apell hypergeometric function.

\bigskip

It is worth remarking that, indeed, 
 the only loop integral of the four entering Eqs. \eqref{Tell'ell}--\eqref{t22} that diverges in the cutoff-removal limit is the `pionless' one \eqref{Ipiless} 
where no vertex function (or `form factor') appears in the integrand. This is in contrast with what happens e.g. in the Weinberg power counting (WPC) inspired on ChPT (cf. Sec. \ref{sec.240103.0}), where a Yukawa-plus-delta interaction constitutes the leading-order potential governing the uncoupled $^1S_0$ channel. The Yukawa term diverges $\propto 1/r$ as $r\to 0$ --- or, in momentum space, decays $\propto 1/A$ as $A\to\infty$ (when on shell) --- so it is not singular by itself. However, as first noticed in Ref. \cite{Kaplan:1996xu}, its interference with the supplementary contact term gives rise to a logarithmic loop divergence that is proportional to $m_\pi^2$, hence not anticipated by WPC. Conversely, the finite-range potential $V_{\,00}(A,A)$ \eqref{orpot} becomes $\propto 1/A^{3/2}$ as $A\to\infty$. By dimensional analysis, this would translate into the $S$-wave coordinate counterpart of our separable interaction approaching a constant value at the origin instead of diverging. For a discussion on the (purely non-perturbative) renormalization of a singular interaction that comes out from the truncated large-$\Lambda_0$ expansion of $V_{\,00}(A,A)$, as well as on certain related issues, see App. \ref{sec.240103.4}.

\bigskip 
Taking $A=\mathfrak{R}(A)\,\pm\,i\epsilon$ with $0\leqslant \mathfrak{R}(A) < \Lambda^2$ and $\epsilon\to0^+$, Eqs. \eqref{Ipiless}--\eqref{J} give
\begin{eqnarray}
\!\!\!\!\!\!\!\!\!\!\!\!\!\!\!\!\!\!\!\!\!\!\!\!\!\!\!\mathcal{I}_{\slashed{\pi}}(A) \,=\, \mathfrak{R}\left(\mathcal{I}_{\slashed{\pi}}(A)\right) \,+\,i\,\frac{{\sqrt{A}}'}{4\pi}\,; &\mathcal{I}_{\,\ell}(A) \,=\, \mathfrak{R}\left(\mathcal{I}_{\,\ell}(A)\right) \,+\,i\,\dfrac{{\sqrt{A}}'}{4\pi}\,g_\ell(A)^2\,;& \mathcal{J}(A) \,=\, \mathfrak{R}\left(\mathcal{J}\!(A)\right) \,+\,i\,\frac{{\sqrt{A}}'}{4\pi}\,g_0(A)\,, \label{complexexpand}
\end{eqnarray}
with
\begin{eqnarray}
\mathfrak{R}\left(\mathcal{I}_{\slashed{\pi}}(A)\right) &=& \frac{\Lambda_0}{2\pi^2}\left(\frac{1}{\sqrt{y}}\,-\,\sqrt{\frac{A}{\Lambda_0^2}}\,\,\text{arctanh}\,\sqrt{\frac{A}{\Lambda_0^2}\,y}\,\right)\,; \label{ReIpiless} \\
\mathfrak{R}\left(\mathcal{I}_{\,\ell}(A)\right) &=& -\,\frac{1}{4\pi^2\,(1+\frac{A}{\Lambda_0^2})^{\,\ell+2}\,\Lambda_0^2}\,\Bigg[ \frac{1}{(1+x)^\ell}\,\sum_{m=0}^{\ell+1}\Bigg(\frac{\sqrt{1\,-\,x\,y}}{(1+y)^{\,\ell+1}}\overline{\mathfrak{a}}_{\,\ell m} \,+\, \frac{\text{arctanh}\,\sqrt{\frac{1\,+\,x}{1\,-\,x\,y}}}{\sqrt{1+x}}\,\mathfrak{b}_{\,\ell m}\Bigg)\,\frac{A^{m}}{\Lambda_0^{2m}}\notag \\
&& +\, \frac{2\,A^{\ell}\sqrt{\frac{A}{\Lambda_0^2}\,(\frac{A}{\Lambda_0^2}-x)}}{\Lambda_0^{2\ell}} \, \text{arctanh}\Big(\,\frac{\sqrt{1\,-\,x\,y}\sqrt{\frac{A}{\Lambda_0^2}}}{\sqrt{\frac{A}{\Lambda_0^2}-x}}\,\Big)\Bigg]\,; \label{ReIl} \\
\mathfrak{R}\left(\mathcal{J}\!(A)\right) &=& -\,\frac{\frac{A}{\Lambda_0^2}}{2\pi^2\,(1+\frac{A}{\Lambda_0^2}) (-x)^{1/4}\,\sqrt{\Lambda_0}} \Bigg[ \frac{_2F_1(\frac{1}{2},\frac{3}{4};\frac{3}{2};\frac{1}{x\,y})}{\sqrt{-\,x\,y}} \,+\, i\,\,\Pi\Big(\sqrt{\tfrac{x}{1+x}};\tfrac{\pi}{2}\Big|-1\Big) \,+\, i\,\,\Pi\Big(-\sqrt{\tfrac{x}{1+x}};\tfrac{\pi}{2}\Big|-1\Big) \notag\\
&&\,-\,\, i\,\,\Pi\Big(\sqrt{\tfrac{x}{1+x}};\arcsin\,(1-\tfrac{1}{x\,y})^{1/4}\Big|-1\Big) \,-\, i\,\,\Pi\Big(-\sqrt{\tfrac{x}{1+x}};\arcsin\,(1-\tfrac{1}{x\,y})^{1/4}\Big|-1\Big)  \notag\\
&&\,+\,\, \frac{\sqrt{-\,\frac{x}{y}}}{\frac{A}{\Lambda_0^2}}\,\Big(\,\mathfrak{R}\,(\,F_1(\tfrac{1}{2};-\tfrac{1}{4},1;\tfrac{3}{2};\tfrac{1}{x\,y},\tfrac{1}{\frac{A}{\Lambda_0^2}y})\,)\,-\, (1+\tfrac{A}{\Lambda_0^2})\,F_1(\tfrac{1}{2};-\tfrac{1}{4},1;\tfrac{3}{2};\tfrac{1}{x\,y},-\tfrac{1}{y})  \, \Big)\Bigg]\,. \label{ReJ}
\end{eqnarray}
At low energies,
\begin{eqnarray}
\!\!\!\!\!\!\!\!\!\!\!\!\!\!\!\!\!\!\!\!\!\!\!\!\!\!\!\!\!\!\!\!\!\!\mathfrak{R}\left(\mathcal{I}_{\slashed{\pi}}(A)\right) \,=\, \mathtt{c}_{\slashed{\pi}}^{[0]}\,+\,\mathtt{c}_{\slashed{\pi}}^{[1]}\,\frac{A}{\Lambda_0^{2}}\,+\,\ldots\,; &\mathfrak{R}\left(\mathcal{I}_{\ell}(A)\right) \,=\, \mathtt{c}_{\ell}^{[0]}\,+\,\mathtt{c}_{\ell}^{[1]}\,\dfrac{A}{\Lambda_0^{2}}\,+\,\ldots\,;& \mathfrak{R}\left(\mathcal{J}\!(A)\right) \,=\, \mathtt{d}^{[0]}\,+\,\mathtt{d}^{[1]}\,\frac{A}{\Lambda_0^{2}}\,+\,\ldots\,,
\end{eqnarray}
where 
\begin{eqnarray}
\!\!\!\!\!\!\!\!\!\!\!\!\!\!\!\!\!\!\!\!\!\!\!\!\!\!\!\!\!\!\!\!\!\!\!\!\!\!\!\!\mathtt{c}_{\slashed{\pi}}^{[0]} \,\,\,\,=\,\,\,\, \tfrac{\Lambda_0}{2\pi^2\sqrt{y}}\,, &&\!\!\!\!\! \mathtt{c}_{\slashed{\pi}}^{[1]} \,\,\,\,=\,\,\,\, -\tfrac{\sqrt{y}\Lambda_0}{2\pi^2}; \label{cpiless01} \\ 
\!\!\!\!\!\!\!\!\!\!\!\!\!\!\!\!\!\!\!\!\!\!\!\!\!\!\!\!\!\!\!\!\!\!\!\!\!\!\!\!\mathtt{c}_{\ell}^{[0]} \,\,\,\,=\,\,\,\, -\tfrac{\frac{\sqrt{1-xy}}{(1+y)^{\ell+1}}\overline{\mathfrak{a}}_{\ell0}\,\,+\,\,\tfrac{\text{arctanh}\sqrt{\frac{1+x}{1-xy}}}{\sqrt{1+x}}\mathfrak{b}_{\ell0}}{4\pi^2\Lambda_0^2\left(1+x\right)^\ell}\,, &&\!\!\!\!\! \mathtt{c}_{\ell}^{[1]} \,\,\,\,=\,\,\,\, \tfrac{\frac{\sqrt{1-xy}}{(1+y)^{\ell+1}}\left[\left(\ell+2\right)\overline{\mathfrak{a}}_{\ell0}-\overline{\mathfrak{a}}_{\ell1}-2\delta_{\ell0}\left(1+y\right)\right]\,\,+\,\,\tfrac{\text{arctanh}\sqrt{\frac{1+x}{1-xy}}}{\sqrt{1+x}}\left[\left(\ell+2\right)\mathfrak{b}_{\ell0}-\mathfrak{b}_{\ell1}\right]}{4\pi^2\Lambda_0^2\left(1+x\right)^\ell}; \label{cl01} \\
\!\!\!\!\!\!\!\!\!\!\!\!\!\!\!\!\!\!\!\!\!\!\!\!\!\!\!\!\!\!\mathtt{d}^{[0]} \,\,\,\,=\,\,\,\, \tfrac{(-x)^{1/4}}{2\pi^2\sqrt{y}\sqrt{\Lambda_0}}\,F_1(\tfrac{1}{2};-\tfrac{1}{4},1;\tfrac{3}{2};\tfrac{1}{x\,y},-\tfrac{1}{y})\,, &&\!\!\!\!\! \mathtt{d}^{[1]} \,\,\,\,=\,\,\,\, -\tfrac{(-x)^{1/4}\sqrt{y}}{2\pi^2\sqrt{\Lambda_0}}\,F_1(-\tfrac{1}{2};-\tfrac{1}{4},1;\tfrac{1}{2};\tfrac{1}{x\,y},-\tfrac{1}{y})\,. \label{d01}
\end{eqnarray}
Again, Eq. \eqref{cl01} simplifies to Eqs. \eqref{cl0} and \eqref{cl1} in the cutoff-removal limit $y\to0$. In virtue of these results, the effective-range approximation
\begin{eqnarray}
\begin{pmatrix}
1 & 0 \\ 0 & A
\end{pmatrix}\,\left(\frac{4\pi}{M_N}\,{\bm{T}}^{-1}(A)\,+\,i\,\sqrt[{\rm I}]{A}\,\bm{1}\right)\,\begin{pmatrix}
1 & 0 \\ 0 & A
\end{pmatrix} &=& -\,\bm{a}^{-1}\,+\,\frac{\bm{r}}{2}\,A\,+\,\mathcal{O}(A^2)\,, 
\end{eqnarray}
c.f. Eqs. \eqref{SLmat} and \eqref{ERmat}, eventually yields
\begin{eqnarray}
a_{\,00} &=& \frac{\sqrt{-x}}{4\pi\,\Lambda_0^3}\left[\mathtt{c}_{\,0}^{[0]}\,-\,\frac{1-\mathtt{c}_{\,2}^{[0]}\,\gamma_{\,22}}{\gamma_{\,00}\,-\,|\bm{\gamma}\,|\,\mathtt{c}_{\,2}^{[0]}} \right]^{-1}\, \Bigg[\, 1 \,-\, \frac{C_0(\Lambda)\Big(\mathtt{d}^{[0]}\,-\, \frac{\Lambda^{3/2}}{(-x)^{1/4}}\Big(\mathtt{c}_0^{[0]}\,+\,\frac{1-\mathtt{c}_2^{[0]}\gamma_{\,22}}{|\bm{\gamma}\,|\,\mathtt{c}_2^{[0]}-\gamma_{\,00}}\Big)\Big)^2}{\Big(1\,-\,C_0(\Lambda)\,\mathtt{c}_{\slashed{\pi}}^{[0]}\Big)\,\Big(\mathtt{c}_0^{[0]}\,+\,\frac{1-\mathtt{c}_2^{[0]}\gamma_{\,22}}{|\bm{\gamma}\,|\,\mathtt{c}_2^{[0]}-\gamma_{\,00}}\Big)\,+\,C_0(\Lambda)\,{\mathtt{d}^{[0]}}^2}\,\Bigg]\,, 
\label{a00M}
\\
a_{\,02} &=& \frac{\gamma_{\,02}\,a_{\,00}}{\Lambda_0^2\left(\gamma_{\,00}\,-\,|\bm{\gamma}\,|\,\mathtt{c}_{\,2}^{[0]}\right)}\Bigg[\, 1 \,+\, \frac{\frac{\Lambda_0\,C_0(\Lambda)^{3/2}}{(-x)^{1/4}}\,\Big(\mathtt{d}^{[0]}\,-\, \frac{\Lambda^{3/2}}{(-x)^{1/4}}\Big(\mathtt{c}_0^{[0]}\,+\,\frac{1-\mathtt{c}_2^{[0]}\gamma_{\,22}}{|\bm{\gamma}\,|\,\mathtt{c}_2^{[0]}-\gamma_{\,00}}\Big)\Big)}{1\,+\,C_0(\Lambda)\,\Big(\frac{\Lambda_0^{3/2}\,\mathtt{d}^{[0]}}{(-x)^{1/4}}-\mathtt{c}_{\slashed{\pi}}^{[0]}\Big)}\,\Bigg]^{-1}\quad\!\!\!=\!\!\!\quad a_{\,20}\,, \\
a_{\,22} &=& \frac{\Big(\gamma_{\,22}\,-\,|\bm{\gamma}\,|\,\mathtt{c}_{\,0}^{[0]}\Big)\,a_{\,02}}{\Lambda_0^2\,\gamma_{\,02}}\Bigg[\, 1 \,-\, \frac{C_0(\Lambda)\,\mathtt{d}^{[0]}\,\Big(\frac{\Lambda_0^{3/2}}{(-x)^{1/4}}+\frac{\mathtt{d}^{[0]}\,|\bm{\gamma}\,|}{\gamma_{\,22}-\mathtt{c}_0^{[0]}|\bm{\gamma}\,|}\Big)}{1\,+\,C_0(\Lambda)\,\Big(\frac{\Lambda_0^{3/2}\,\mathtt{d}^{[0]}}{(-x)^{1/4}}-\mathtt{c}_{\slashed{\pi}}^{[0]}\Big)}\,\Bigg]\;; \\ 
\notag\\
r_{\,00} &=& \frac{4\pi\,\Lambda_0}{\sqrt{-x}}\frac{\left[\Big(4+\frac{1}{x}\Big)\,\Big(\frac{\gamma_{\,22}}{|\bm{\gamma}\,|}\,-\,\mathtt{c}_{\,0}^{[0]}\Big)\,-\,2\,\mathtt{c}_{\,0}^{[1]}\right]\,\left[\Big(1-\mathtt{c}_{\slashed{\pi}}^{[0]}\,C_0(\Lambda)\Big)^2 \,+\, \frac{\frac{C_0(\Lambda)}{\Lambda_0}\sum_{n=0}^5\theta_n^{[00]}(-x)^{n/4}}{(-x)^{5/4}\,\left[(4+\frac{1}{x})\,(\frac{\gamma_{\,22}}{|\bm{\gamma}\,|}\,-\,\mathtt{c}_{\,0}^{[0]})\,-\,2\,\mathtt{c}_{\,0}^{[1]}\right]}\right]}{\Bigg[\Big(1-\mathtt{c}_{\slashed{\pi}}^{[0]}\,C_0(\Lambda)\Big)\,+\,\Lambda_0\,C_0(\Lambda)\Big(\frac{2\,\sqrt{\Lambda_0}\,\mathtt{d}^{[0]}}{(-x)^{1/4}}\,-\,\frac{\Lambda_0^2\,(\mathtt{c}_0^{[0]}-\frac{\gamma_{\,22}}{|\bm{\gamma}\,|})}{\sqrt{-x}}\Big)\Bigg]^2}\,, \label{r00M}\\
r_{\,02} &=& -\frac{4\pi\,\Lambda_0^3}{\sqrt{-x}}\,\frac{\left[\Big(6+\frac{1}{x}\Big)\,\frac{\gamma_{\,02}}{|\bm{\gamma}\,|}\right]\left[\Big(1-\mathtt{c}_{\slashed{\pi}}^{[0]}\,C_0(\Lambda)\Big)^2 \,+\, \frac{\frac{C_0(\Lambda)}{\Lambda_0}\sum_{n=0}^6\theta_n^{[02]}(-x)^{n/4}}{(-x)^{7/4}\,\left[(6+\frac{1}{x})\,\frac{\gamma_{\,02}}{|\bm{\gamma}\,|}\right]}\right]}{\Bigg[\Big(1-\mathtt{c}_{\slashed{\pi}}^{[0]}\,C_0(\Lambda)\Big)\,+\,\Lambda_0\,C_0(\Lambda)\Big(\frac{2\,\sqrt{\Lambda_0}\,\mathtt{d}^{[0]}}{(-x)^{1/4}}\,-\,\frac{\Lambda_0^2\,(\mathtt{c}_0^{[0]}-\frac{\gamma_{\,22}}{|\bm{\gamma}\,|})}{\sqrt{-x}}\Big)\Bigg]^2}\quad\!\!\!=\!\!\!\quad r_{\,20}\,, \label{r02M}\\
r_{\,22} &=& \frac{4\pi\,\Lambda_0^5}{\sqrt{-x}}\frac{\left[\Big(8+\frac{1}{x}\Big)\,\Big(\frac{\gamma_{\,00}}{|\bm{\gamma}\,|}\,-\,\mathtt{c}_{\,2}^{[0]}\Big)\,-\,2\,\mathtt{c}_{\,2}^{[1]}\right]\left[\Big(1-\mathtt{c}_{\slashed{\pi}}^{[0]}\,C_0(\Lambda)\Big)^2 \,+\, \frac{\frac{C_0(\Lambda)}{\Lambda_0}\sum_{n=0}^7\theta_n^{[22]}(-x)^{n/4}}{x^2\,\left[(8+\frac{1}{x})\,(\frac{\gamma_{\,00}}{|\bm{\gamma}\,|}\,-\,\mathtt{c}_{\,2}^{[0]})\,-\,2\,\mathtt{c}_{\,2}^{[1]}\right]}\right]}{\Bigg[\Big(1-\mathtt{c}_{\slashed{\pi}}^{[0]}\,C_0(\Lambda)\Big)\,+\,\Lambda_0\,C_0(\Lambda)\Big(\frac{2\,\sqrt{\Lambda_0}\,\mathtt{d}^{[0]}}{(-x)^{1/4}}\,-\,\frac{\Lambda_0^2\,(\mathtt{c}_0^{[0]}-\frac{\gamma_{\,22}}{|\bm{\gamma}\,|})}{\sqrt{-x}}\Big)\Bigg]^2}\,. \label{r22M}
\end{eqnarray} 
The dimensionless coefficients $\theta_n^{[\ell'\ell]}$ ($n=0,1,\ldots,5+\overline{\ell}$) upon which $r_{\ell'\ell}$ depends are given in App. \ref{app.231227.1}.
Since such coefficients \eqref{thetafirst}--\eqref{thetalast} are all finite for $C_0(\Lambda)\equiv0$, it is apparent that, if one successively turns off the contact term in the potential \eqref{withczero} and removes the cutoff,  
the form of Eqs. \eqref{a00}--\eqref{r22} is  consistently recovered from Eqs. \eqref{a00M}--\eqref{r22M}. 

\bigskip
The running equation for $C_0(\Lambda)$ can be explicitly found by imposing the usual condition that the $S$-wave scattering length is exactly cutoff independent. Solving Eq. \eqref{a00M} for $C_0(\Lambda)$ and then expanding around $\Lambda\to\infty$ gives
\begin{eqnarray}
\frac{1}{C_0(\Lambda)} &=& \frac{\Lambda}{2\pi^2}\left[1\,-\,\frac{2\pi^2}{\Lambda}\,\frac{4\pi\,a_{\,00}{\mathfrak{d}^{[0]}}^2\,-\,\frac{2\,(-x)^{1/4}}{\Lambda_0^{3/2}}\mathfrak{d}^{[0]}\,+\,\mathfrak{c}_0^{[0]}\,+\,\frac{1-\mathfrak{c}_{\,2}^{[0]}\gamma_{\,22}}{|\bm{\gamma}\,|\mathfrak{c}_2^{[0]}-\gamma_{\,00}}}{4\pi\,a_{\,00}\,\Big(\mathfrak{c}_0^{[0]}\,+\,\frac{1-\mathfrak{c}_{\,2}^{[0]}\gamma_{\,22}}{|\bm{\gamma}\,|\mathfrak{c}_2^{[0]}-\gamma_{\,00}}\Big)\,-\,\frac{\sqrt{-x}}{\Lambda_0^3}}\,+\,\ldots\right]\,, \label{running}
\end{eqnarray}
where the ellipses stand for smaller finite-cutoff contributions, $\mathfrak{c}_\ell^{[0]}$ was given in Eq. \eqref{cl0}, and
\begin{eqnarray}
\mathfrak{d}^{[0]} &\equiv& \lim_{\Lambda\to\infty}\mathtt{d}^{[0]} \quad\!\!\!=\!\!\!\quad \frac{1}{2\pi^2}\int_0^{\,\infty} dq\,g_0(q^2) \quad\!\!\!=\!\!\!\quad \frac{1}{4\pi\sqrt{\Lambda_0}} \Big[\sqrt{2}\left(1+x\right)^{1/4}\,+\,\begin{pmatrix}-\tfrac{3}{4}\\-\tfrac{1}{2}\end{pmatrix}\frac{(-x)^{3/4}}{3}\,{_2F_1}(\tfrac{1}{2},1;\tfrac{7}{4};-{x})\Big]\,. \notag\\
\end{eqnarray}
The only unbound term in Eq. \eqref{running}, which is linear in $\Lambda$, cancels out with the divergence brought by the integral \eqref{Ipiless} so that the piece $[\,{C_0(\Lambda)}^{-1}-\mathcal{I}_{\slashed{\pi}}(A)\,]$ that enters the amplitude [see Eqs. \eqref{Tell'ell}--\eqref{t22}] is kept finite. 	That is to say, our solution is properly renormalized.

\bigskip

\subsection{Nonexistence of a matrix $N/D$ representation}
\def\theequation{\arabic{section}.\arabic{equation}}
\label{incompatibility}

\bigskip

Here we apply a similar rationale as in Sec. \ref{sec.240101.1} to discuss the impossibility to find an $N/D$ representation for the amplitude \eqref{Tell'ell}. Notice that this $N/D$ representation exists if and only if there is an $N/D$ form for the matrix 
\begin{eqnarray}
\bm{t}(A) &=& \begin{pmatrix}
t_{00}(A,A;A) && t_{02}(A,A;A) \\ t_{20}(A,A;A) && t_{22}(A,A;A) 
\end{pmatrix}
\end{eqnarray}
[c.f. Eqs. \eqref{t00}--\eqref{t22}], 
i.e. if and only if there are matrices $\bm{n}(A)$ and $\bm{d}(A)$ with only LHC and RHC, respectively, fulfilling
\begin{eqnarray}
  \label{ntd}
{\bm n}(A)\,{\bm t}(A)&=&{\bm d}(A)\,.
\end{eqnarray}
As $\bm{t}(A)=\bm{t}(A)^\top$ it follows that 
\begin{eqnarray}
\bm{t}(A)&=&\bm{n}(A)^{-1}\bm{d}(A)\quad\!\!\!=\!\!\!\quad\bm{d}(A)^\top\bm{n}(A)^{\top\,-1}\,, \label{illustration}
\end{eqnarray} 
which corresponds to an $N/D$ decomposition for $\bm{t}(A)$.

\bigskip
We rewrite the $\bm{t}(A)$ matrix elements  
in the more compact form
\begin{eqnarray}
  \label{240101.3}
\bm{t}(A)&=&\begin{pmatrix}
  \alpha_0\,+\,(A-L)^\frac{1}{4}\,\alpha_1\,\widehat{g}_0\,+\,(A-L)^\frac{1}{2}\,\alpha_2 \,\widehat{g}_0^{\,2} & (A-L)^\frac{1}{4}\,\beta_1\, \widehat{g}_2\,+\,(A-L)^\frac{1}{2}\,\beta_2\,\widehat{g}_0\,\widehat{g}_2\\(A-L)^\frac{1}{4} \,\beta_1\, \widehat{g}_2\,+\,(A-L)^\frac{1}{2}\,\beta_2\,\widehat{g}_0\,\widehat{g}_2 & (A-L)^\frac{1}{2}\, \xi_2\, \widehat{g}_2^{\,2}  \end{pmatrix}\,,\end{eqnarray}
where the $A$ dependence of the cut-free function $\widehat{g}_\ell(A)\equiv(A-L)^{-{1}/{4}}g_\ell(A)=A^{\,\ell/2}/{(A+\Lambda_0^2)^{\,1+\ell/2}}$, as well as the $A$ dependence of the six functions denoted by a Greek letter, has been made implicit for brevity. On the other hand, combining Eqs. \eqref{t00}--\eqref{t22} with Eq. \eqref{complexexpand}, one learns that: \textit{a)} $t_{02}(A,A;A)$ is actually RHC-free; \textit{b)} there are two RHC-free functions $\chi(A)$ and $\zeta(A)$ such that $t_{00}(A,A;A) = \chi(A) + \zeta(A)\,\mathcal{I}_2(A)$;
\textit{c)} conversely, the RHC structure of $t_{22}(A,A;A)$ is more involved, stemming from terms both linear and quadratic in the loop integrals. 

\bigskip
Our starting point is a parameterization of $\bm{n}(A)$ that is similar to that used in Eq.~\eqref{Omat2} for $\bm{O}(A)$, say
\begin{eqnarray}
\label{240101.2}
\bm{n}(A)&=&\begin{pmatrix}
\eta_{00}(A)\,(A-L)^{\phi_{\,00}} & \eta_{02}(A)\,(A-L)^{\phi_{\,02}} \\ \eta_{20}(A)\,(A-L)^{\phi_{\,20}} & \eta_{22}(A)\,(A-L)^{\phi_{\,22}} 
\end{pmatrix}\,,
\end{eqnarray}
with the constraint that all the $\eta_{\ell'\ell}(A)$ functions must be free of cuts; in addition, Eq. \eqref{illustration} illustrates that 
\begin{eqnarray}|\bm{n}(A)|&\neq&0\,.\label{invertible}\end{eqnarray}
Then, from Eq. \eqref{ntd} we get the matrix element
\begin{eqnarray}
d_{22}(A) &=& (A-L)^{\frac{1}{4}+\phi_{20}}\beta_1(A)\eta_{20}(A)\widehat{g}_{2}(A)\notag\\
&&+\,(A-L)^{\frac{1}{2}+\phi_{20}}\beta_2(A)\eta_{20}(A)\widehat{g}_0(A)\widehat{g}_2(A)\notag\\
&&+\,(A-L)^{\frac{1}{2}+\phi_{22}}\xi_2(A)\eta_{22}(A)\widehat{g}_{2}(A)^2\,. \label{d22noLHC}
\end{eqnarray}
The two first exponents above differ in a non-integer number, so it is clear that the three exponents above cannot be integers at the same time. We thus distinguish three mutually exclusive scenarios:
\begin{itemize}
\item \textbf{None of the exponents is an integer.} Only two mutually exclusive options are compatible with this case:

  \begin{itemize}
  \item \textit{Each of the exponents is different from the two others.} 
  \begin{itemize} \item[] Then, the fact that Eq. \eqref{d22noLHC} is LHC-free leads to $\eta_{20}(A)=\eta_{22}(A)=0$, which is in turn in contradiction with Eq. \eqref{invertible}.\end{itemize} 

  \item \textit{Two exponents are equal and different from the other.} 
    \begin{itemize} \item[] Suppose e.g. that $\frac{1}{2}+\phi_{22}=\frac{1}{2}+\phi_{20}$. Then, from Eq. \eqref{d22noLHC} being LHC-free it follows that $\eta_{22}(A)\,=\,-\beta_2(A)\,\eta_{20}(A)\,\widehat{g}_0(A)\,/\,[\xi_2(A)\,\widehat{g}_2(A)]$, where $\eta_{22}(A)$ is RHC-free by definition. However, the piece $\beta_2(A)/\xi_2(A)$ cannot be RHC-free, since the RHC structure is different for $\beta_2(A)$ and $\xi_2(A)$, as discussed above.  A similar contradiction  emerges if the equality between exponents involves $\frac{1}{4}+\phi_{20}$.\end{itemize}
    \end{itemize}

\item \textbf{Only one exponent is an integer.} Three possibilities are to be considered:
  \begin{itemize}
  \item \textit{The first exponent $\frac{1}{4}+\phi_{20}$ is an integer; the second and the third exponents are non-integers.}
    \begin{itemize}
    \item[$\diamond$] If $\,\frac{1}{2}+\phi_{20}\neq \frac{1}{2}+\phi_{22}$ then $\eta_{20}(A)=\eta_{22}(A)=0$.  This is in contradiction with Eq. \eqref{invertible}.
      \item[$\diamond$] If $\frac{1}{2}+\phi_{20}=\!\frac{1}{2}\!+\phi_{22}$ then $\eta_{22}(A)\,=\,-\beta_2(A)\,\eta_{20}(A)\,\widehat{g}_0(A)\,/\,[\xi_2(A)\,\widehat{g}_2(A)]$. As discussed above, this result is incompatible with $\eta_{\ell'\ell}(A)$ being cut-free.
    \end{itemize}
  \item \textit{The second exponent $\frac{1}{2}+\phi_{20}$ is an integer; the first and the third exponents are non-integers.}
  \begin{itemize} \item[{\color{white}$\diamond$}] This case is analogous to the one just discussed. \end{itemize}
  \item  \textit{The third exponent $\frac{1}{2}+\phi_{22}$ is an integer; the first and the second exponents are non-integers.}
  \begin{itemize} \item[{\color{white}$\diamond$}] Here, $\eta_{20}(A)=0$. 
    We can consider the expression for $d_{20}(A)$ coming out from Eq. \eqref{ntd}, which taking $\eta_{20}(A) = 0$ becomes $d_{20}(A) = \widehat{g}_2(A)\eta_{22}(A)[(A-L)^{\frac{1}{4}+\phi_{22}}\beta_1(A)+(A-L)^{\frac{1}{2}+\phi_{22}}\beta_2(A)\widehat{g}_0(A)]$. While the second exponent within this square bracket is an integer by hypothesis, the first exponent is non-integer, which in turns implies that $\eta_{22}(A)=0$ given that $d_{20}(A)$ is LHC-free.
    In this way we conclude that $\eta_{20}(A)=\eta_{22}(A)=0$, again in contradiction with Eq. \eqref{invertible}. 
\end{itemize}
  \end{itemize}

\item \textbf{Only two exponents are integers.}
\begin{itemize}
\item[{\color{white}}] In such a situation the term involving the unequal exponent implies that either $\eta_{20}(A)$ or $\eta_{22}(A)$ is zero. Once such a constraint is imposed on  $d_{20}(A)$, one can easily conclude the other $\eta_{2\ell}(A)$ function must vanish as well. [In the point above we made this reasoning from the $\eta_{20}(A)=0$ condition. Analogous steps would follow starting from $\eta_{22}(A)=0$.]\end{itemize}
  
  \end{itemize}
Thus, a function $\bm{n}(A)$ of the type of Eq.~\eqref{240101.2} cannot give rise to an $N/D$ representation of $\bm{t}(A)$.

\bigskip
Yet one could try parameterizations for $\bm{n}(A)$ that are more general than Eq.~\eqref{240101.2}, namely
\begin{eqnarray}
  \label{240102.4}
\eta_{\ell'\ell}(A)&=&\sum_{\mathfrak{m}=1}^{\mathfrak{n}_{\ell'\ell}}\eta_{\ell'\ell}^{(\mathfrak{m})}(A)\,\,(A-L)^{\phi_{\ell'\ell}^{(\mathfrak{m})}}\,.
\end{eqnarray}
Indeed, given that $t_{00}(A,A;A)$, $t_{02}(A,A;A)$, and $t_{22}(A,A;A)$ contain respectively 3, 2, and 1 different exponents of $(A-L)$ [c.f. Eq. \eqref{240101.3}], the most general structure to be considered would require to take $\mathfrak{n}_{00}=\mathfrak{n}_{20}=3$ and $\mathfrak{n}_{02}=\mathfrak{n}_{22}=2$ in Eq. \eqref{240102.4}.
But would we really win anything by including more LHC structures in the $\eta_{\ell'\ell}(A)$ functions? The most interesting strategy would be to consider cancellations between terms with different exponents involved in the product of only one $\eta_{\ell'\ell}(A)$ times only one  $t_{\ell'\ell}(A)$. [Otherwise we would need cancellations between terms involving Greek-letter coefficients of different $t_{\ell'\ell}(A)$ functions that would require to reabsorb RHC structures in the $\eta_{\ell'\ell}^{(\mathfrak{m})}(A)$ functions, which cannot be.] But the drawback of such a strategy is that at the same time we generate new different LHC structures involving different exponents so that the problem gets even worse than before.

\bigskip
Let us illustrate this discussion with an example. For definiteness let us take 
\begin{eqnarray}
\eta_{20}(A)&=&\eta_{20}^{(1)}(A)\,\,(A-L)^{\phi^{(1)}_{20}}+\eta_{20}^{(2)}(A)\,\,(A-L)^{\phi^{(2)}_{20}} \,; 
\\  
\eta_{22}(A)&=&\eta_{22}^{(1)}(A)\,\,(A-L)^{\phi^{(1)}_{22}}+\eta_{22}^{(2)}(A)\,\,(A-L)^{\phi^{(2)}_{22}}\,,
\end{eqnarray} 
thus
\begin{eqnarray}
  \label{240102.5}
  d_{22}(A) &=& \beta_1(A)\,\widehat{g}_2(A)\,\,\Big[\eta_{20}^{(1)}(A)\,\,(A-L)^{\frac{1}{4}+\phi_{20}^{(1)}}\,+\,\eta_{20}^{(2)}(A)\,\,(A-L)^{\frac{1}{4}+\phi_{20}^{(2)}}\Big]\notag\\
  &&+\,\beta_2(A)\,\widehat{g}_0(A)\,\widehat{g}_2(A)\,\Big[\eta_{20}^{(1)}(A)\,\,(A-L)^{\frac{1}{2}+\phi_{20}^{(1)}}+\eta_{20}^{(2)}(A)\,\,(A-L)^{\frac{1}{2}+\phi_{20}^{(2)}}\Big]\notag\\
  &&+\,\xi_2(A)\,\widehat{g}_2(A)^{\,2}\,\Big[\eta_{22}^{(1)}(A)\,\,(A-L)^{\frac{1}{2}+\phi_{22}^{(1)}}+ \eta_{22}^{(2)}(A)\,\,(A-L)^{\frac{1}{2}+\phi_{22}^{(2)}} \Big]\,.
\end{eqnarray}
Next, we assume that $\frac{1}{4}+\phi_{20}^{(2)}=\frac{1}{2}+\phi_{20}^{(1)}$, i.e. $\phi_{20}^{(2)}=\phi_{20}^{(1)}+\frac{1}{4}$. Then, it follows that $\frac{1}{4}+\phi_{20}^{(1)}=\phi_{20}^{(2)}\neq \frac{1}{2}+\phi_{20}^{(2)}$, and Eq.~\eqref{240102.5} becomes
\begin{eqnarray}
  \label{240102.6}
d_{22}(A) &=&  (A-L)^{\frac{1}{4}+\phi_{20}^{(2)}}\,\widehat{g}_2(A)\,\Big[\beta_2(A)\,\eta_{20}^{(1)}(A)\,\widehat{g}_0(A)\,+\,\beta_1(A)\,\eta_{20}^{(2)}(A)\Big] \notag
  \\
&& +\,\,(A-L)^{\frac{1}{4}+\phi_{20}^{(1)}}\beta_1(A)\,\eta_{20}^{(1)}(A)\,\widehat{g}_2(A)\, +\,(A-L)^{\frac{1}{2}+\phi_{20}^{(2)}}\beta_2(A)\,\eta_{20}^{(2)}(A)\widehat{g}_0(A)\,\widehat{g}_2(A) \notag
\\
&& +\,\,\xi_2(A)\,\widehat{g}_2(A)^{2}\,\Big[\eta_{22}^{(1)}(A)\,\,(A-L)^{\frac{1}{2}+\phi_{22}^{(1)}}+\eta_{22}^{(2)}(A)\,\,(A-L)^{\frac{1}{2}+\phi_{22}^{(2)}}\Big]\,.
\end{eqnarray}
Therefore, now there are \textit{three} different LHC structures not involving $\xi_2(A)$ in Eq.~\eqref{240102.5}, while originally there were only two and one in $t_{20}(A,A;A)$ and $t_{22}(A,A;A)$, respectively. If we do not want to arrive to contradictions by giving relations between $\eta_{20}^{(\mathfrak{m})}(A)$ and $\eta_{22}^{(\mathfrak{m}')}(A)$ functions ($\mathfrak{m},\,\mathfrak{m}'=\lbrace1,\,2\rbrace$), having to reabsorb in them terms with RHC, we then have to conclude that $\eta_{20}^{(\mathfrak{\mathfrak{m}})}(A)=0$. But attending to $d_{20}(A)$ with $\eta_{20}^{(\mathfrak{m})}(A)=0$ we can also see by analyzing the different cases there that $\eta_{22}^{(\mathfrak{m})}(A)=0$, which is not allowed.

\bigskip
In summary, we conclude from these arguments that there is no matrix $N/D$ representation for the amplitude due to the momentum-space potential consisting of a constant plus the separable potential \eqref{orpot}.

\bigskip
\bigskip
\bigskip

\section{Dispersive relations for each separate \textit{T}-matrix element 
}
\setcounter{equation}{0}   
\def\theequation{\arabic{section}.\arabic{equation}}
\label{sec.240103.6}

\bigskip
In this section we recover the regular interaction \eqref{orpot} with the vertex functions $g_0(Q)$ and $g_2(Q)$ given by Eqs.~\eqref{newg0alt} and \eqref{newg2alt}, respectively. That is to say, we come back to the potential proposed in the last part of Sec. \ref{sec.240101.1}, where it was shown that the associated amplitude matrix cannot be accommodated to a matrix $N/D$ representation. This leads us to discard the employ of the matrix $N/D$ method for this particular interaction and motivates us to make use of an alternative strategy, yet in the framework of the $N/D$ method, to solve the scattering problem, as we detail below.  

\bigskip
Explicitly, 
\begin{eqnarray}
V_{\,\ell'\ell}(Q',Q) &=& \frac{\gamma_{\,\ell'\ell}}{M_N}\,g_{\,\ell'}(Q')\,g_{\,\ell}(Q)\,; \label{231227.7} \\
g_\ell(Q) &=& \frac{(Q-L)^{\theta_{\,\ell}}\,Q^{\,\ell/2}}{(Q+\Lambda_0^2)^{\,1+\ell/2}}\,, \label{genericvertex}
 \end{eqnarray}
 with  
 $0<\theta_{\,\ell}<1/2$.  
Then, the Lippmann-Schwinger equation has an algebraic solution that reads
\begin{eqnarray}
  \label{231227.3}
\bm{T}(A) &=& \begin{pmatrix}
\frac{(A-L)^{2\theta_{\,0}}}{M_N\,(A+\Lambda_0^2)^2\,\mathfrak{D}(A)}\left[\gamma_{\,00}-|\bm{\gamma}\,|I_2(A)\right] & \frac{A\,(A-L)^{\theta_{\,0}+\theta_{\,2}}}{M_N\,(A+\Lambda_0^2)^3\,\mathfrak{D}(A)}\,\gamma_{\,02} \\ \frac{A\,(A-L)^{\theta_{\,0}+\theta_{\,2}}}{M_N\,(A+\Lambda_0^2)^3\,\mathfrak{D}(A)}\,\gamma_{\,02} & \frac{A^2\,(A-L)^{2\theta_{\,2}}}{{M_N\,(A+\Lambda_0^2)^4\,\mathfrak{D}(A)}}\left[\gamma_{\,22}-|\bm{\gamma}\,|I_0(A)\right]
\end{pmatrix}\,,
\end{eqnarray}
where $I_{\,\ell}(A)$ keeps being given by the first line of Eq. \eqref{loop}, and $\mathfrak{D}(A)$ was introduced in Eq. \eqref{DA}. From here it is straightforward to find the discontinuity of each separate amplitude $t_{\,\ell'\ell}(A)$ along the LHC, ${\Delta}_{\,\ell'\ell}(A)\equiv \mathfrak{I}\left[t_{\,\ell'\ell}(A)\right]$, $A<L$,
\begin{eqnarray}
\bm{\Delta}(A) &=& \begin{pmatrix}
\frac{\sin[2\pi\theta_0]\,(L-A)^{2\theta_{\,0}}}{M_N\,(A+\Lambda_0^2)^2\,\mathfrak{D}(A)}\left[\gamma_{\,00}-|\bm{\gamma}\,|I_2(A)\right] & \frac{\sin[\pi(\theta_0+\theta_2)]\,A\,(L-A)^{\theta_{\,0}+\theta_{\,2}}}{M_N\,(A+\Lambda_0^2)^3\,\mathfrak{D}(A)}\,\gamma_{\,02} \\ \frac{\sin[\pi(\theta_0+\theta_2)]\,A\,(L-A)^{\theta_{\,0}+\theta_{\,2}}}{M_N\,(A+\Lambda_0^2)^3\,\mathfrak{D}(A)}\,\gamma_{\,02} & \frac{\sin[2\pi\theta_2]\,A^2\,(L-A)^{2\theta_{\,2}}}{{M_N\,(A+\Lambda_0^2)^4\,\mathfrak{D}(A)}}\left[\gamma_{\,22}-|\bm{\gamma}\,|I_0(A)\right]
\end{pmatrix}\,.
\end{eqnarray}
Furthermore, consider the LHC-free matrix element $\tau_{\,\ell'\ell}(A) \equiv {t_{\,\ell'\ell}(A)}\,{A^{-\overline{\ell}}\,(A-L)^{-\theta_{\,\ell'}-\theta_{\,\ell}}}$, whence
\begin{eqnarray}
\bm{\tau}(A) &=& \begin{pmatrix}
\frac{\gamma_{\,00}-|\bm{\gamma}\,|I_2(A)}{M_N\,(A+\Lambda_0^2)^2\,\mathfrak{D}(A)} & \frac{\gamma_{\,02}}{M_N\,(A+\Lambda_0^2)^3\,\mathfrak{D}(A)} \\ \frac{\gamma_{\,02}}{M_N\,(A+\Lambda_0^2)^3\,\mathfrak{D}(A)} & \frac{\gamma_{\,22}-|\bm{\gamma}\,|I_0(A)}{{M_N\,(A+\Lambda_0^2)^4\,\mathfrak{D}(A)}}\end{pmatrix}\,.
\end{eqnarray}

\bigskip
Now we can introduce the 
functions $n_{\,\ell'\ell}(A)$ and $d_{\,\ell'\ell}(A)$, respectively RHC- and LHC-free,
such that each matrix element $t_{\ell'\ell}(A)$ is separately written as 
\begin{eqnarray}
t_{\,\ell'\ell}(A) &=& \frac{n_{\,\ell'\ell}(A)}{d_{\,\ell'\ell}(A)}\,, \label{tnd}
\end{eqnarray}
with the requirement of the normalization condition
\begin{eqnarray}
d_{\,\ell'\ell}(0) &=& 1\,. \label{dnorm}
\end{eqnarray}
Explicitly,
\begin{eqnarray}
n_{\,\ell'\ell}(A) &=& A^{\overline{\ell}}\left(A-L\right)^{\theta_{\,\ell'}+\theta_{\,\ell}}\,\tau_{\,\ell'\ell}(0)\,; \\
d_{\,\ell'\ell}(A) &=& \frac{\tau_{\,\ell'\ell}(0)}{\tau_{\,\ell'\ell}(A)}\,, 
\end{eqnarray}
that we gather in the matrices  
\begin{eqnarray}
\bm{n}(A) &=& \begin{pmatrix}
\frac{(A-L)^{2\theta_{\,0}}}{M_N\,\Lambda_0^4\,\mathfrak{D}(0)}\left[\gamma_{\,00}-|\bm{\gamma}\,|I_2(A)\right] & \frac{A\,(A-L)^{\theta_{\,0}+\theta_{\,2}}}{M_N\,\Lambda_0^6\,\mathfrak{D}(0)}\,\gamma_{\,02} \\ \frac{A\,(A-L)^{\theta_{\,0}+\theta_{\,2}}}{M_N\,\Lambda_0^6\,\mathfrak{D}(0)}\,\gamma_{\,02} & \frac{A^2\,(A-L)^{2\theta_{\,2}}}{{M_N\,\Lambda_0^8\,\mathfrak{D}(0)}}\left[\gamma_{\,22}-|\bm{\gamma}\,|I_0(A)\right]
\end{pmatrix}\,; \label{nmatrix}\\
\bm{d}(A) &=& \begin{pmatrix}
\left(1+\frac{A}{\Lambda_0^2}\right)^2\frac{\mathfrak{D}(A)}{\mathfrak{D}(0)}\,\frac{\gamma_{\,00}-|\bm{\gamma}\,|I_2(0)}{\gamma_{\,00}-|\bm{\gamma}\,|I_2(A)} & \left(1+\frac{A}{\Lambda_0^2}\right)^3\frac{\mathfrak{D}(A)}{\mathfrak{D}(0)} \\ \left(1+\frac{A}{\Lambda_0^2}\right)^3\frac{\mathfrak{D}(A)}{\mathfrak{D}(0)} & \left(1+\frac{A}{\Lambda_0^2}\right)^4\frac{\mathfrak{D}(A)}{\mathfrak{D}(0)}\,\frac{\gamma_{\,22}-|\bm{\gamma}\,|I_0(0)}{\gamma_{\,22}-|\bm{\gamma}\,|I_0(A)}
\end{pmatrix}\,. \label{dmatrix} \end{eqnarray}
The following ultraviolet behaviors are apparent 
from the expressions above 
\begin{eqnarray}
n_{\,\ell'\ell}(A\to\infty) &\propto& A^{\overline{\ell}+\theta_{\,\ell'}+\theta_{\,\ell}}\,; \label{asyn} \\
d_{\,\ell'\ell}(A\to\infty) &\propto& A^{\overline{\ell}+2}\,. \label{asyd}
\end{eqnarray}
The LHC and RHC discontinuities are
\begin{eqnarray}
\mathfrak{I}\left[n_{\,\ell'\ell}(A)\right] \quad\!\!\!\,=& \Delta_{\,\ell'\ell}(A)\,d_{\,\ell'\ell}(A)\,, & A<L\,; \\  
\mathfrak{I}\left[d_{\,\ell'\ell}(A)\right] \quad\!\!\!\,=& -\nu_{\,\ell'\ell}(A)\,n_{\,\ell'\ell}(A)\,, & A>0\,, \end{eqnarray}
respectively, where
\begin{eqnarray}
  \nu_{\,\ell'\ell}(A)  \quad\!\!\!\,\equiv& -\,\mathfrak{I}\left[\dfrac{1}{t_{\,\ell'\ell}(A)}\right]\,, & A>0\,.
\label{231227.1}\end{eqnarray}
It can be checked that this behaves at low energies as
\begin{eqnarray}
\nu_{\,\ell'\ell}(A\to0) &\propto& A^{\frac{1}{2}-\overline{\ell}}\,; \label{nuth} 
\end{eqnarray}
see Ref.~\cite{Albaladejo:2012sa} for expressions of the $\nu_{\ell'\ell}(A)$ functions in terms of phase shifts and mixing angle, and its appendix for a derivation of such threshold behavior. 

\bigskip
Recalling that $(\theta_{\,\ell'}+\theta_{\,\ell})\in(0,1)$, from Eq. \eqref{asyn} one sees that the $n_{\,\ell'\ell}(A)$ DR demands at least $(\overline{\ell}+1)$ subtractions which we choose to locate at $C=0$. Taking the integration contour to be the positively oriented, LHC-engulfing, infinite-radius circumference centered at the origin of the complex plane, 
\begin{eqnarray}
\oint \frac{dz\,\,n_{\ell'\ell}(z)}{(z-A)\,(z-C)^{\,\overline{\ell}+1}} &=& 2i\,\int^{L}_{-\infty} \frac{d\omega_L\,\mathfrak{I}[n_{\ell'\ell}(\omega_L)]}{(\omega_L-A)\,\omega_L^{\overline{\ell}+1}} \quad\!\!\!=\!\!\!\quad \frac{2\pi\,i}{A^{\overline{\ell}+1}} \left[{n_{\,\ell'\ell}(A)} - \sum_{k=0}^{\overline{\ell}}\frac{A^k}{k!}n_{\,\ell'\ell}^{(k)}(0)\right]\,, \label{combined}
\end{eqnarray}
where the residue theorem was invoked in the last step. Consistently with Eq. \eqref{nmatrix}, we impose the constraints
\begin{eqnarray}
n_{02}(0) &=& 0\,; \\
n_{22}(0) &=& n_{22}'(0) \quad\!\!\!=\!\!\!\quad 0\,,
\end{eqnarray}
so that the only term that survives in the previous summation is the one with $k=\overline{\ell}$. It follows
\begin{eqnarray}
{n_{\,\ell'\ell}(A)} &=& \frac{A^{\overline{\ell}}}{\overline{\ell}!}\,n_{\,\ell'\ell}^{(\overline{\ell})}(0)\,+\,\frac{A^{\overline{\ell}+1}}{\pi}\int^{L}_{-\infty} d\omega_L\,\frac{\Delta_{\,\ell'\ell}(\omega_L)\,d_{\,\ell'\ell}(\omega_L)}{(\omega_L-A)\,\omega_L^{\overline{\ell}+1}}\,. \label{nDRsep}
\end{eqnarray}
In particular,
\begin{eqnarray}
n_{\,\ell'\ell}(A\to0) &\propto& A^{\overline{\ell}}\,. \label{nth} 
\end{eqnarray}

\bigskip
Given Eq. \eqref{asyd}, we impose $(\overline{\ell}+3)$ subtractions in the $d_{\,\ell'\ell}(A)$ DR, one of which we fix at $C=0$ 
to guarantee Eq.~\eqref{dnorm}, while the $(\overline{\ell}+2)$ remaining ones are taken at $\widetilde{C}=-\Lambda_0^2$. Let the integration path be the counterclockwise, RHC-engulfing, infinite circumference centered at the origin of the complex plane,
\begin{eqnarray}
\!\!\!\!\!\!\!\!\!\!\!\!\!\!\!\!\!\!\!\!\!\!\!\!\!\!\!\!\!\!\!\!\oint \frac{dz\,\,d_{\ell'\ell}(z)}{(z-A)\,(z-C)\,(z-\widetilde{C})^{{\,\overline{\ell}+2}}} &=& 2i\int_{0}^{\,\infty} \frac{d\omega_R\,{\mathfrak{I}[d_{\ell'\ell}(\omega_R)]}}{(\omega_R-A)\,\omega_R\,(\omega_R+\Lambda_0^2)^{\overline{\ell}+2}} \notag \\
&=& \frac{2\pi\,i}{A\,\Lambda_0^{2(\overline{\ell}+2)}} \left[-1\,+\,\frac{d_{\,\ell'\ell}(A)}{(1+\frac{A}{\Lambda_0^2})^{\overline{\ell}+2}}\,+\, \sum_{k=0}^{\overline{\ell}+1}\frac{\Lambda_0^{2k}}{k!}\left(1\,-\,{(1+\tfrac{A}{\Lambda_0^2})^{k-\overline{\ell}-2}}\right)d_{\,\ell'\ell}^{(k)}(-\Lambda_0^2)\right]\,. \label{combined2}
\end{eqnarray}
 Furthermore, each term in the previous summation happens to vanish provided that we implement the conditions
\begin{eqnarray}
d_{00}(-\Lambda_0^2) &=& d\,'_{\!00}(-\Lambda_0^2)\quad\!\!\!=\!\!\!\quad 0\,; \\
d_{02}(-\Lambda_0^2) &=& d\,'_{\!02}(-\Lambda_0^2)\quad\!\!\!=\!\!\!\quad d\,''_{\!02}(-\Lambda_0^2)\quad\!\!\!=\!\!\!\quad 0\,; \\
d_{22}(-\Lambda_0^2) &=& d\,'_{\!22}(-\Lambda_0^2)\quad\!\!\!=\!\!\!\quad d\,''_{\!22}(-\Lambda_0^2)\quad\!\!\!=\!\!\!\quad d\,'''_{\!22}(-\Lambda_0^2) \quad\!\!\!=\!\!\!\quad 0\,, \end{eqnarray}
which are consistent with Eq. \eqref{dmatrix}. Thus,
\begin{eqnarray}
d_{\,\ell'\ell}(A) &=& {\left(1+\frac{A}{\Lambda_0^2}\right)^{\overline{\ell}+2}}\,-\,\frac{A\,{({A}+{\Lambda_0^2})^{\overline{\ell}+2}}}{\pi}\int_{0}^{\,\infty}d\omega_R\, \frac{\nu_{\,\ell'\ell}(\omega_R)\,n_{\,\ell'\ell}(\omega_R)}{(\omega_R-A)\,\omega_R\,(\omega_R+\Lambda_0^2)^{\overline{\ell}+2}}\,. \label{dDRsep}
\end{eqnarray}
Note that, according to Eqs. \eqref{nuth} and \eqref{nth}, the previous integrand becomes $\propto1/\sqrt{\omega_R}$ when $\omega_R\to0$, which is indeed integrable in ${\omega_R}$.

\bigskip
Substituting Eqs. \eqref{nDRsep} and \eqref{dDRsep} in Eq. \eqref{tnd}, pushing $A\to0$, and recalling 	Eq. \eqref{threscond} then gives
\begin{eqnarray} n_{\,\ell'\ell}^{(\overline{\ell})}(0) &=& -\,\frac{4\pi\,{\overline{\ell}!}}{M_N}\,a_{\,\ell'\ell}\,. \end{eqnarray}
As a result, only the three scattering lengths are left as free parameters in the DRs for  $n_{\ell'\ell}(A)$ and $d_{\ell'\ell}(A)$. This clearly illustrates that the DRs derived in this method are simpler in comparison with the ones deduced for the matrix $N/D$ method of Secs.~\ref{sec.231228.1} and \ref{sec.231222.1}. 

\bigskip

 Plugging Eq. \eqref{dDRsep} into Eq. \eqref{nDRsep} yields the IE for ${n_{\,\ell'\ell}(A)}$ with $A\geqslant0$,
\begin{eqnarray}
  \label{231227.6}
{n_{\,\ell'\ell}(A)} &=& -\,\frac{4\pi}{M_N}\,a_{\,\ell'\ell}\,A^{\overline{\ell}}\,+\,\frac{A^{\overline{\ell}+1}}{\pi\,\Lambda_0^{2(\overline{\ell}+2)}}\int^{L}_{-\infty} d\omega_L\,\frac{{({\omega_L}+{\Lambda_0^2})^{\overline{\ell}+2}}\,\Delta_{\,\ell'\ell}(\omega_L)}{\omega_L^{\overline{\ell}+1}\,(\omega_L-A)}\notag\\
&&+\,\frac{A^{\overline{\ell}+1}}{\pi^2}\int_{0}^{\,\infty}d\omega_R\, \frac{\nu_{\,\ell'\ell}(\omega_R)\,n_{\,\ell'\ell}(\omega_R)}{\omega_R\,(\omega_R+\Lambda_0^2)^{\overline{\ell}+2}}\int^{L}_{-\infty} d\omega_L\,\frac{{{({\omega_L}+{\Lambda_0^2})^{\overline{\ell}+2}}}\,\Delta_{\,\ell'\ell}(\omega_L)}{\omega_L^{\overline{\ell}}\,(\omega_L-\omega_R)\,(\omega_L-A)}\,, \label{itsolved}
\end{eqnarray}
which is solved by means of the iterative algorithm that follows:

\begin{enumerate}
\item[\textbf{1.}] Given a starting input for the amplitude $t_{\ell'\ell}^{[0]}(A)$, the associated $\nu_{\ell'\ell}^{[0]}(A)$ is found by direct application of Eq.~\eqref{231227.1}, 
  \begin{eqnarray}
    \label{231020.13}
\nu_{\ell'\ell}^{[0]}(A)&=&-\mathfrak{I}\left[\frac{1}{t^{[0]}_{\ell'\ell}(A)}\right]\,,\quad A>0\,.
  \end{eqnarray}
In this way, the $\nu_{\ell'\ell}^{[0]}(A)$ matrix elements have the right threshold behavior, as long as the seed amplitude matrix made out of the $t_{\ell'\ell}^{[0]}(A)$ matrix elements satisfies unitarity with the right threshold behavior \eqref{threscond}. 

In practice, we have either used the exact solution provided by the Lippmann-Schwinger equation \eqref{231227.3} or an amplitude matrix obtained by unitarizing directly the potential $V_{\ell'\ell}(A)$.
In the matrix notation we have employed until now, the unitarization procedure reads \cite{Oller:2020guq,Oller:2019opk,Oller:2000ma}
  \begin{eqnarray}
\label{231020.19}
    \bm{T}^{[0]}(A) &=& [\bm{V}(A)^{-1}\,+\,h_1(A)\bm{1}]^{-1}\,,
  \end{eqnarray}
  with
  \begin{eqnarray}
    \label{231020.20}
    h_1(A)&=&-\frac{M_N\,(A-L)}{4\pi^2}\int_0^\infty \frac{d\omega_R\,\sqrt{\omega_R}}{(\omega_R-A)\,(\omega_R-L)}
    \quad\!\!\!=\!\!\!\quad
-\frac{M_N}{4\pi}\,(\sqrt{-L}\,+\,i\sqrt{A})\,,
  \end{eqnarray}
  such that $h_1(L)=0$, at the onset of the LHC. In this way the influence of the LHC is mitigated in the calculation of $\bm{T}^{[0]}(A)$, which is convenient for a unitarization procedure that only resums the RHC.

\item[\textbf{2.}] Solve Eq. \eqref{itsolved} for $n_{\ell'\ell}^{[1]}(A)$, $A\in\text{RHC}$, and then use Eq. \eqref{dDRsep} to calculate $d_{\ell'\ell}^{[1]}(A)$ also in the physical region, $A>0$. In terms of both functions we obtain $t_{\ell'\ell}^{[1]}(A)=n_{\ell'\ell}^{[1]}(A)/d_{\ell'\ell}^{[1]}(A)$. However, in general the amplitude matrix resulting from such partial-wave amplitudes, namely   \begin{eqnarray}
    \label{231020.15}
    \bm{T}^{[1]}(A)&=&\left(\begin{array}{ll} t_{00}^{[1]}(A) & t_{02}^{[1]}(A)\\
      t_{02}^{[1]}(A) & t_{22}^{[1]}(A)
    \end{array}\right)\,, \label{T1ofA}
    \end{eqnarray}
does not fulfill unitarity anymore. Therefore, as an intermediate step it is convenient to obtain unitarized partial-wave amplitudes, so as to guarantee the right general properties of the novel $\nu_{\ell'\ell}(A)$ functions.
  
\item[\textbf{3.}] The unitarization procedure for the partial-wave amplitudes is accomplished by means of the general formula $\bm{T}(A)=[\bm{M}(A)-i\frac{M_N\sqrt{A}}{4\pi}\bm{1}]^{-1}$.
 For applying it we construct $\bm{M}^{[1]}(A)$ from $\bm{T}^{[1]}(A)$ \eqref{T1ofA} as
  \begin{eqnarray}
    \label{231020.16}
\bm{M}^{[1]}(A)&=&{\mathfrak{R}}\,[\bm{T}^{[1]}(A)^{-1}]\,.
  \end{eqnarray}
 Then, the matrix
  \begin{eqnarray}
    \label{231020.17}
    \bm{T}^{[U1]}(A)&=&\Big(\bm{M}^{[1]}(A)\,-\,i\,\dfrac{M_N\sqrt{A}}{4\pi}\bm{1}\Big)^{-1}
  \end{eqnarray}
does fulfill unitarity by construction.
  
\item[\textbf{4.}] The new $\nu_{\ell'\ell}(A)$ functions are then directly calculated as
  \begin{eqnarray}
    \label{231020.18}
    \nu_{\ell'\ell}^{[1]}(A)&=&-\mathfrak{I}\left[\frac{1}{t_{\ell'\ell}^{[U1]}(A)}\right]\,,\quad~A>0\,.
  \end{eqnarray}
\item[\textbf{5.}] These steps are repeatedly implemented until convergence is reached.  That is to say, after $n$ iterations we obtain $t_{\ell'\ell}^{[n]}(A)$ and then using it from the step 1 onwards one can calculate $t_{\ell'\ell}^{[n+1]}(A)$, and so on.  
  
\end{enumerate}

\bigskip

\subsection{Check of the iterative procedure}
\label{sec.231020.3}

\bigskip

Now we put in practice the recipe just described for solving Eq. \eqref{231227.6} using the parameter values of Table~\ref{tab.231020.2}. We take as our starting input the $\nu^{[0]}_{\ell'\ell}(A)$ functions corresponding to the exact solution \eqref{231227.3} and proceed with the iterative algorithm, computing the phase shifts and mixing angle at the end of each iteration. For that, we parameterize the $S$ matrix so that its unitary character is manifest, i.e. we employ the Stapp-Ypsilantis-Metropolis form \cite{SYM}
\begin{eqnarray}
  \label{231227.8}
\bm{S}(A)&=&\bm{1}\,+\,i\,\frac{M_N\sqrt{A}}{2\pi}\bm{T}(A) \quad\!\!\!=\quad\!\!\! \begin{pmatrix}
  \cos\!\left(2\ep\right) e^{2i\delta_0} & i\sin\!\left(2\ep\right) e^{i(\delta_0+\delta_2)} \\
  i\sin\!\left(2\ep\right) e^{i(\delta_0+\delta_2)} &  \cos\!\left(2\ep\right) e^{2i\delta_2}
  \end{pmatrix}\,,
\end{eqnarray}
where $\delta_\ell$ and $\ep$ are the $\ell$-wave phase shift and the mixing angle, respectively. After $n$ iterations the $S$ matrix is calculated using ${\bm T}^{[Un]}(A)$, cf. Eqs.~\eqref{231020.15}--\eqref{231020.17} replacing $\bm{T}^{[1]}(A)$, $\bm{M}^{[1]}(A)$, and $\bm{T}^{[U1]}(A)$ by $\bm{T}^{[n]}(A)$, $\bm{M}^{[n]}(A)$, and $\bm{T}^{[Un]}(A)$, respectively. 

\begin{center}
  \begin{figure}
    \begin{tabular}{ll}
    \includegraphics[width=.45\textwidth]{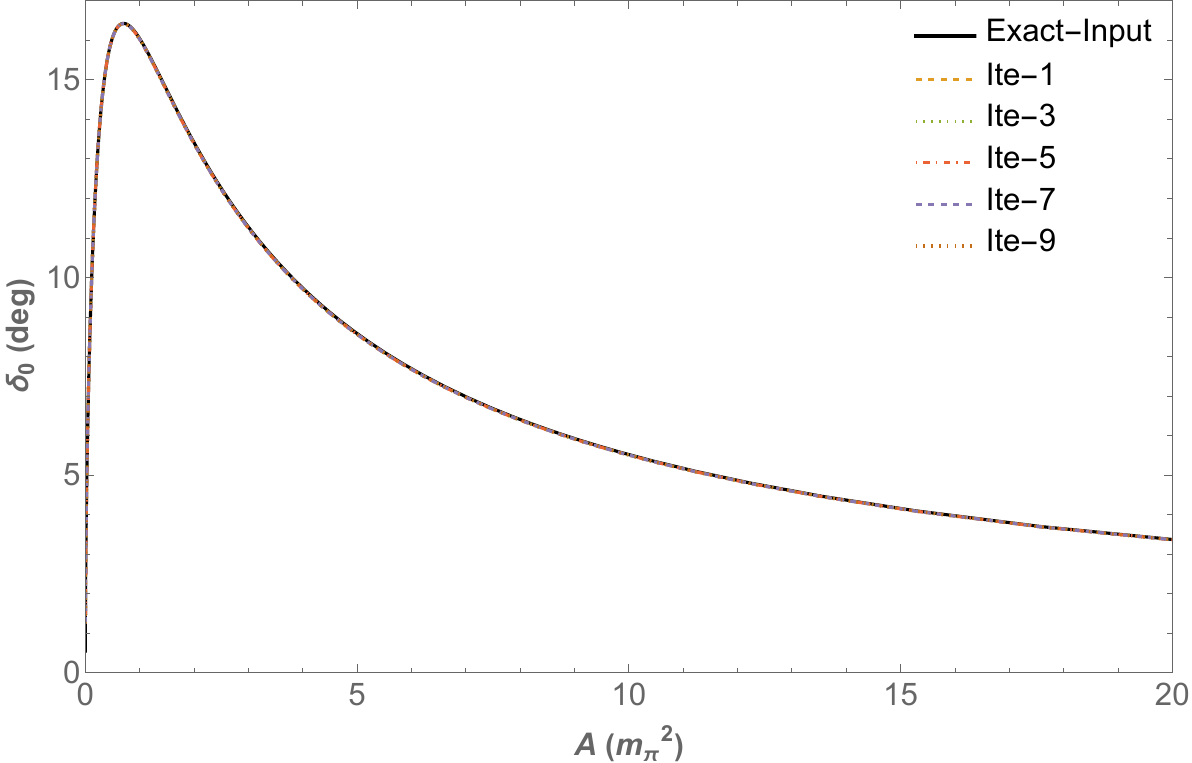} & 
    \includegraphics[width=.45\textwidth]{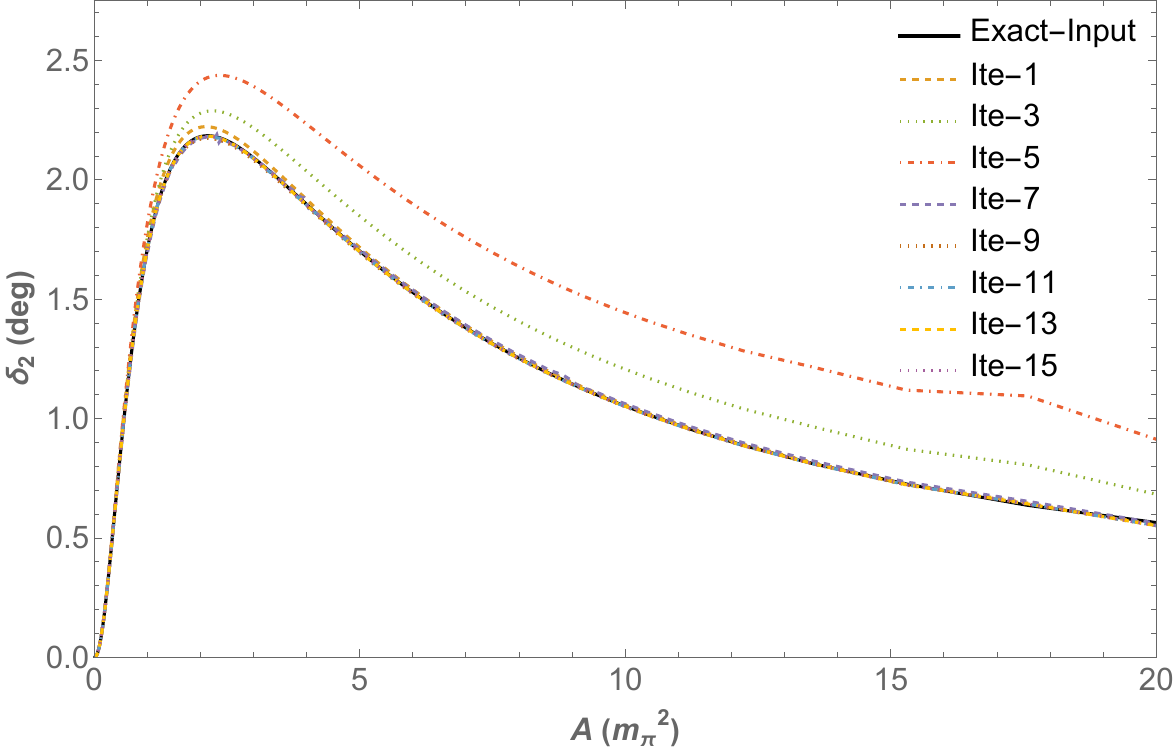} \\ 
    \includegraphics[width=.45\textwidth]{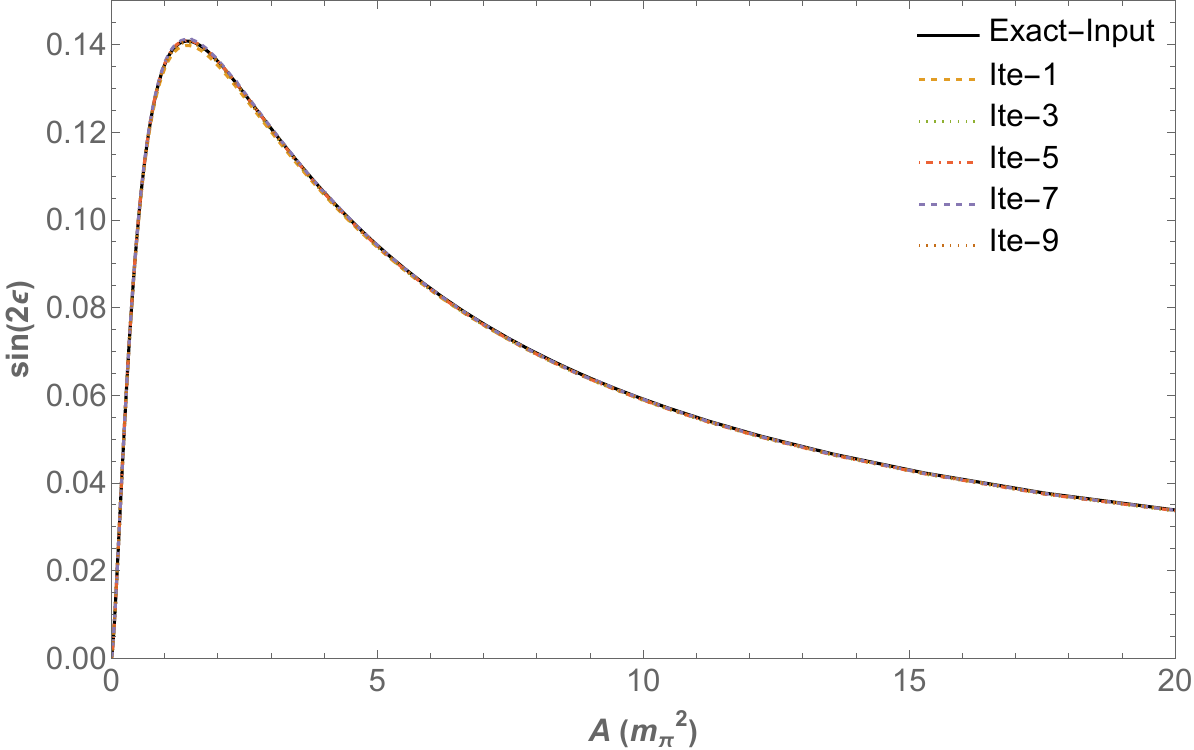} & 
    \includegraphics[width=.45\textwidth]{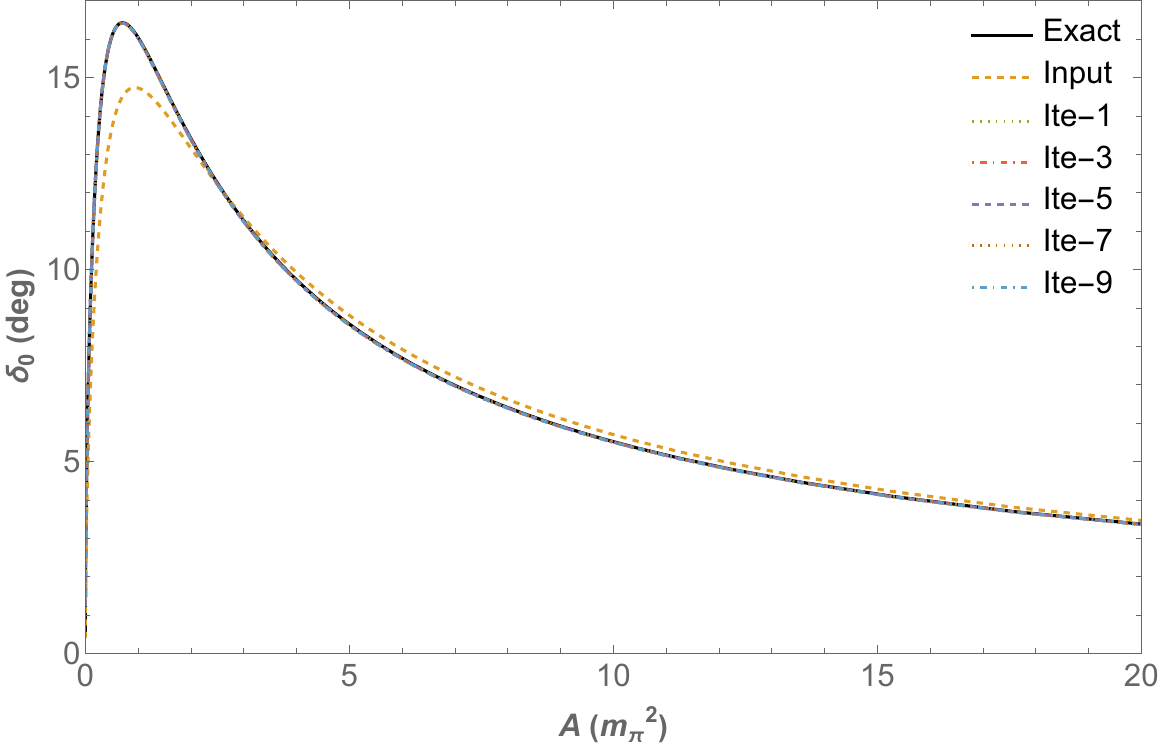} \\ 
    \includegraphics[width=.45\textwidth]{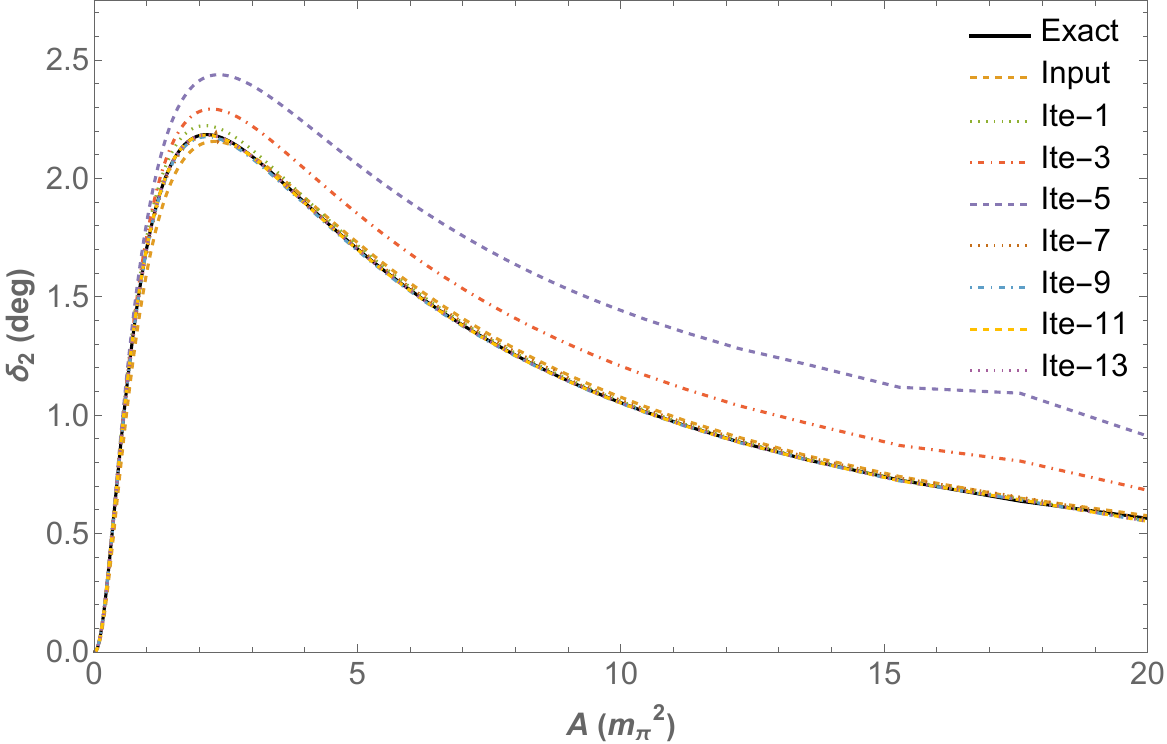} & 
    \includegraphics[width=.45\textwidth]{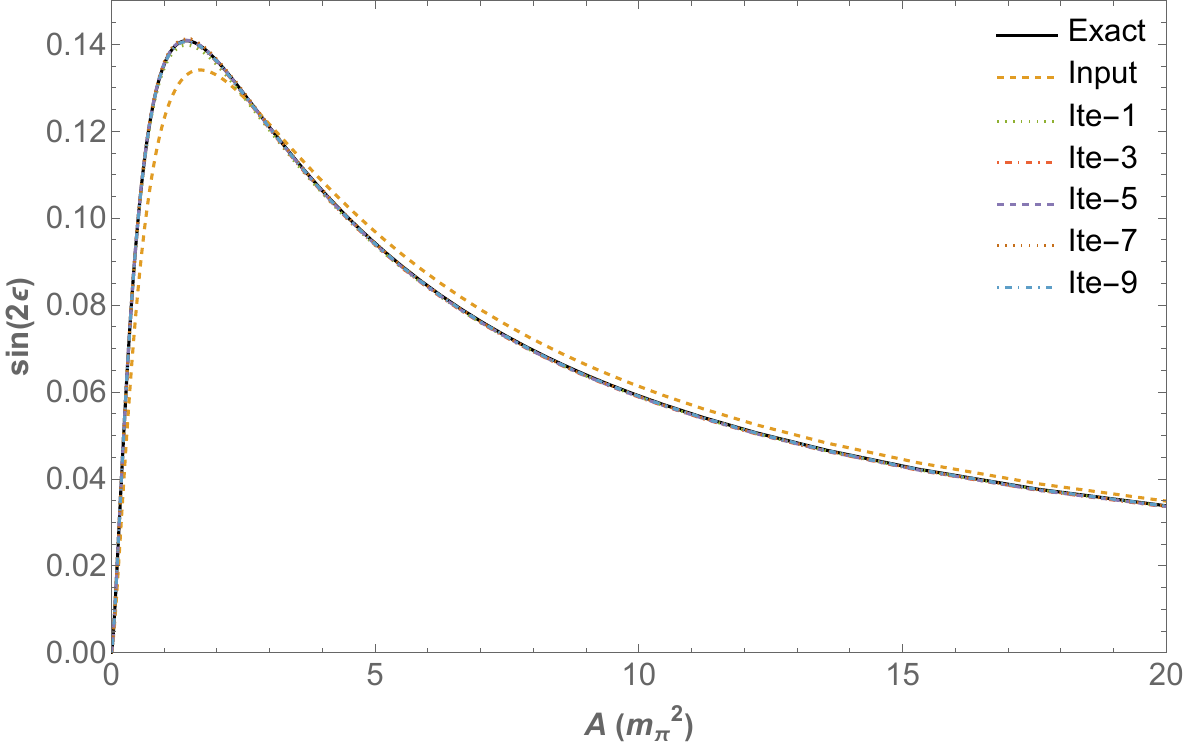} 
  \end{tabular}
  \caption{{\small The ordering of panels is from left to right and top to bottom; along the first (last) three panels, the seed amplitude is the one found through solving the Lippmann-Schwinger equation (unitarizing the potential). We display $\delta_{00}$ and $\delta_{22}$ (in degrees) together with $\sin(2\ep)$ as functions of the squared center-of-mass momentum $A$ (in units of $m_\pi^2=1$) along the iteration process. The convergence in the calculation of these quantities is shown with different lines that correspond to different iterative outputs according to the legend, and eventually agree with the exact solution.}\label{fig.231021.1}}
\end{figure}
\end{center}

\bigskip
  In the first three panels of  Fig.~\ref{fig.231021.1}, from left to right and top to bottom, we plot the functions $\delta_0(A)$, $\delta_2(A)$, and $\sin[2\ep(A)]$. The different lines corresponding to different iterative outputs of these quantities overlap with the exact results coming out from Eq. \eqref{231227.3}, either right from the beginning or after just a few iterations for the case of the small $\delta_2$. We thus see that the method is consistent, and that the exact solution for the amplitude matrix is itself solution (as it must) of the IE \eqref{231227.6}, calculated with the iterative process. 

\bigskip
We also plot in the last three panels of Fig.~\ref{fig.231021.1} the results obtained for $\delta_0(A)$, $\delta_2(A)$ and $\sin[2\ep(A)]$ when the input $\bm{T}^{[0]}(A)$ stems from unitarizing the potential according to Eq.~\eqref{231020.19}. Again, we see that the exact phase shifts and mixing angle are reproduced after convergence is reached.

\begin{table}
  \begin{center}
    \begin{tabular}{llllllll}
      \hline
Set 1: & $\theta_0=1/3$ & $\theta_2=1/6$ & $\ga_{00}=10.0$ & $\ga_{22}=5.0$ & $\ga_{02}=-5.0$  & $L=-1/4$ & $\Lambda_0=1$\\
        \end{tabular}
\caption{{\small Values of the parameters that enter the potential \eqref{231227.7}, and used to generate Fig.~\ref{fig.231021.1}. The units are taken such that $m_\pi^2=1$.}\label{tab.231020.2}}
    \end{center}
  \end{table}

\bigskip

\section{Conclusions and outlook}
\setcounter{equation}{0}   
\def\theequation{\arabic{section}.\arabic{equation}}
\label{sec.240103.7}

\bigskip
Along this manuscript we have mainly studied the application of the $N/D$ method to coupled-wave two-body systems. We have considered two variants of the method. On the one hand, we have the straightforward generalization of the $N/D$ method to coupled waves through a matrix formalism \cite{Bjorken:1960zz}, the so-called matrix $N/D$ method. On the other hand, we have the $N/D$ method applied to each partial wave separately, in which case the coupled nature of the problem implies that the $N/D$ method is implemented iteratively. After every iteration coupled-channel unitarity is imposed, and this allows to calculate the right-hand cut input  for the next round. The scattering problem is solved once results from the iterations converge \cite{Noyes:1960abc,Albaladejo:2012sa}. 

\bigskip
We have demonstrated that there is an infinite family of separable potentials rendering amplitude matrices that do not admit a matrix $N/D$ representation. Yet this is not an issue when one makes use of the $N/D$ strategy for each partial wave separately --- which, indeed, is an analyticity and unitarity requirement for any amplitude matrix. It also turns out that the integral equations for the partial-wave separate $N/D$ representation are typically much simpler than those one can derive from the matrix $N/D$ method. We then conclude that, generally speaking, the type of $N/D$ method to be  applied to a coupled-wave or coupled-channel scattering problem is the separate $N/D$ form, since the  matrix $N/D$  method might not even be applicable. This fact could potentially handicap the application of the matrix $N/D$ method to realistic situations in which case one has to face the risk of implementing heavy integral equations of increasing complexity that actually would not make sense. 

\bigskip
 Concerning our plans for the near future, we intend to use the partial-wave separate $N/D$ method to solve realistic coupled-wave scattering problems such as two-nucleon scattering. With this purpose, we will employ the integral equation derived in Ref.~\cite{Oller:2018zts} to calculate the exact discontinuity of the amplitude matrix along the LHC for a given finite-range potential, and then solving for the partial-wave amplitudes with increasing degree of sophistication and accuracy, as it was similarly done for the uncoupled $^1S_0$ partial-wave amplitude in Ref.~\cite{Entem:2016ipb}.

\bigskip
\bigskip
\bigskip

\section*{Acknowledgements}

 MSS and JAO would like to acknowledge partial financial support to the Grant PID2022-136510NB-C32 funded by MCIN/AEI/10.13039/501100011033/ and FEDER, UE. DRE acknowledges partial financial support from Grant PID2022-141910NB-I00 funded by MCIN/AEI/10.13039/501100011033/. This work has been also  partially funded by EU Horizon
2020 research and innovation program, STRONG-2020 project, under grant agreement no. 824093.  

\bigskip
\bigskip
\bigskip

\appendix

\section{{More on the absence of a matrix $\pmb{N/D}$ representation}}
\def\theequation{\Alph{section}.\arabic{equation}}
\setcounter{equation}{0}  
\label{sec.appa}

\bigskip

In this appendix we stick to a separable form of the potential such as Eq. \eqref{orpot}, but the definition of the $g_\ell(A)$ functions given in Eq. \eqref{gi} is replaced by more general functions $g_0(A)$ and $g_2(A)$. We already know that the on-shell amplitude matrix may be put as
\begin{eqnarray}
\bm{T}(A) &=& \bm{n}(A)\,\bm{m}(A)^{-1}\,\bm{n}(A)\,, \label{nmsplit2}
\end{eqnarray}
with the matrices
\begin{eqnarray}
\bm{n}(A) &=& 
\begin{pmatrix}
g_0(A) & 0 \\ 0 & g_2(A)
\end{pmatrix}\,; \label{genn2} \\
\bm{m}(A) &=& \frac{M_N}{|\bm{\gamma}\,|}\,\begin{pmatrix}
{\gamma_{\,22}}\,-\,{|\bm{\gamma}\,|}\,I_{\,0}(A) & -\,{\gamma_{\,02}} \\ -\,{\gamma_{\,02}} & {\gamma_{\,00}}\,-\,{|\bm{\gamma}\,|}\,I_{\,2}(A) 
\end{pmatrix}\,, \label{genm2}
\end{eqnarray}
where $I_{\,\ell}(A)$ keeps being given by the first line of Eq. \eqref{loop}. Provided that both matrices \eqref{genn2} and \eqref{genm2} commute, one gets 
\begin{eqnarray}
\!\!\!\!\!\!\qquad\!\!\!\!\!\!
\bm{T}(A) \,\,\,=\,\,\, \bm{n}(A)^2\,\bm{m}(A)^{-1}\,\,\,=\,\,\, \bm{m}(A)^{-1}\,\bm{n}(A)^2\,\,\,=\,\,\,\bm{T}(A)^\top\,.
\end{eqnarray}
 Next, given two  arbitrary (non-vanishing) functions of $A$ $\mathfrak{f}_0(A)$ and $\mathfrak{f}_2(A)$, we consider the diagonal matrix
\begin{eqnarray}
\bm{\phi}(A) &=& 
\begin{pmatrix}
\mathfrak{f}_0(A) & 0 \\ 0 & \mathfrak{f}_2(A)
\end{pmatrix}\,, 
\end{eqnarray}
which of course commutes with the diagonal matrix $\bm{n}(A)$. Therefore, from Eq.~\eqref{nmsplit2} we also have the equivalent new composition 
\begin{eqnarray}
\bm{T}(A) &=& \bm{\widetilde{n}}(A)\,\bm{\widetilde{m}}(A)^{-1}\,\bm{\widetilde{n}}(A)\,, 
\label{withwidetilde} \end{eqnarray}
where
\begin{eqnarray}
\bm{\widetilde{n}}(A) &=& \bm{n}(A)\,\bm{\phi}(A)\,; \\
\bm{\widetilde{m}}(A) &=& \bm{\phi}(A)\,\bm{m}(A)\,\bm{\phi}(A)\,.
\end{eqnarray}
Assume that these two matrices commute,
\begin{eqnarray}
\left[\,\bm{\widetilde{n}}(A), \bm{\widetilde{m}}(A)\,\right] \quad\!\!\! = \!\!\!\quad \frac{M_N\,\gamma_{\,02}}{|\bm{\gamma}\,|}\,\mathfrak{f}_0(A)\,\mathfrak{f}_2(A)\left[ \mathfrak{f}_2(A)\,g_2(A)\,-\, \mathfrak{f}_0(A)\,g_0(A) \right] \begin{pmatrix}
0 & 1 \\ -1 & 0
\end{pmatrix} \quad\!\!\! = \!\!\!\quad \bm{0}\,,
\end{eqnarray}
namely
\begin{eqnarray}
\mathfrak{f}_2(A) &=& \mathfrak{f}_0(A)\,\frac{g_0(A)}{g_2(A)}\,,
\end{eqnarray}
in which case Eq. \eqref{withwidetilde} becomes
\begin{eqnarray}
\bm{T}(A) &=& \bm{N}(A)\,\bm{D}(A)^{-1} \quad\!\!\!=\!\!\!\quad \bm{D}(A)^{-1}\,\bm{N}(A)\,,
\end{eqnarray}
where
\begin{eqnarray}
\bm{N}(A) \quad\!\!\!= & \bm{\widetilde{n}}(A)^2 & =\quad\!\!\! \mathfrak{f}_0(A)^2\,g_0(A)^2\,\bm{1}\,; \label{nn} \\
\bm{D}(A) \quad\!\!\!= & \bm{\widetilde{m}}(A) & =\quad\!\!\! \frac{M_N\,\mathfrak{f}_0(A)^2}{|\bm{\gamma}\,|} \,\begin{pmatrix}
{\gamma_{\,22}}\,-\,{|\bm{\gamma}\,|}\,I_{\,0}(A) & -\,\frac{g_0(A)}{g_2(A)}\,\gamma_{\,02} \\ -\,\frac{g_0(A)}{g_2(A)}\,\gamma_{\,02} & \left[\frac{g_0(A)}{g_2(A)}\right]^2\left[{\gamma_{\,00}}\,-\,{|\bm{\gamma}\,|}\,I_{\,2}(A)\right] \end{pmatrix}\,. \label{dd}
\end{eqnarray}
For example, coming back to the case of the vertex functions \eqref{gi}, and taking $\mathfrak{f}_0(A) = (A+\Lambda_0^2)/(-L)^{1/4}$ so that $\bm{N}(0)=\bm{1}$, we recover the split given by Eqs. \eqref{nmat2} and \eqref{dmat2}.

\bigskip
But recall that, by definition, $\bm{D}(A)$ has to be LHC-free so as to a matrix $N/D$ representation exists. Using the original definitions of the $g_\ell$'s \eqref{gi}, it is clear that
${g_0(A)}/{g_2(A)} = {1+\Lambda_0^2/A}$ does not give rise to any discontinuity along the $A\in(-\infty,L)$ interval, so it is checked that $\bm{N}(A)$ and $\bm{D}(A)$ carry all the LHC- and RHC-discontinuity, respectively. Nevertheless, consider 
$g_0(A)$ and $g_2(A)$ given in Eqs.~\eqref{newg0alt} and \eqref{newg2alt}. 
Hence, taking $\mathfrak{f}_0(A)=(A+\Lambda_0^2)/(-L)^{\phi_0}$ so that $\bm{N}(0)=\bm{1}$, Eqs. \eqref{nn} and \eqref{dd} become
\begin{eqnarray}
\bm{N}(A) &=& \frac{(A-L)^{2\phi_0}}{(-L)^{2\phi_0}}\,\bm{1}\,; \\
\bm{D}(A) &=& \frac{M_N\,(A+\Lambda_0^2)^2}{|\bm{\gamma}\,|\,(-L)^{2\phi_0}} \, \begin{pmatrix}
{\gamma_{\,22}}\,-\,{|\bm{\gamma}\,|}\,I_{\,0}(A) & -\,\frac{A+\Lambda_0^2}{A}\,(A-L)^{\phi_0-\phi_2}\,\gamma_{\,02} \\ -\,\frac{A+\Lambda_0^2}{A}\,(A-L)^{\phi_0-\phi_2}\,\gamma_{\,02} & \frac{(A+\Lambda_0^2)^2}{A^2}\,(A-L)^{2(\phi_0-\phi_2)}\,\left[{\gamma_{\,00}}\,-\,{|\bm{\gamma}\,|}\,I_{\,2}(A)\right] \end{pmatrix}\,,
\end{eqnarray}
from where it is manifest that $\bm{D}(A)$ exhibits a discontinuity along the LHC. 

\bigskip
Nevertheless, we stress that 
actually the condition that $\bm{\widetilde{n}}(A)$ and $\bm{\widetilde{m}}(A)$ commute is sufficient, but it has not been shown that it is necessary for the existence of a matrix $N/D$ representation. We refer the reader back to Sec. \ref{sec.240101.1} for the complete proof.

\bigskip
\bigskip

\section{Large-$\pmb{{\Lambda}_0}$ expansion of the separable interaction}
\def\theequation{\Alph{section}.\arabic{equation}}
\setcounter{equation}{0}  
\label{sec.240103.4}

\bigskip

In this appendix we discuss another potential amenable to an $N/D$ representation whose finite-range part is non-regular. 
For the sake of simplicity, here the $D$-wave component of the ${^3S_1}\!-{^3D_1}$ two-nucleon channel will be neglected. We thus start from our definition of the $S$-wave separable potential
\begin{eqnarray}
V(Q',Q) &=& \frac{\gamma}{M_N}\,g(Q')\,g(Q)\,; \label{potS}\\
g(Q) &=& \frac{(Q-L)^{1/4}}{Q+\Lambda_0^2}\,, \label{vertexS}
\end{eqnarray}
c.f. Eq. \eqref{orpot}.
When $\Lambda_0$ is large, the vertex function is amenable to an expansion as a geometric series
\begin{eqnarray}
g(Q) 
&=& \frac{(Q-L)^{1/4}}{\Lambda_0^2}\,\Big(1 - {Q}/{\Lambda_0^2} + \mathcal{O}(Q^2/\Lambda_0^4)\Big)\,. 
\end{eqnarray}
Omitting the small corrections, the potential now reads
\begin{eqnarray}
V(Q',Q) &=& \frac{1}{M_N}\, (Q'-L)^{1/4}\,(Q-L)^{1/4} \left[ c_0 + c_2\left(Q'+Q\right)\right]\,, \label{potential} 
\end{eqnarray}
where one has to take two independent parameters $c_0$ and $c_2$ for renormalization purposes (see below). This potential is plugged into the fully non-perturbative Lippmann-Schwinger equation
\begin{eqnarray}
T(Q',Q;A) &=& V(Q',Q) \,+\, \frac{M_N}{2\pi^2} \int_0^\infty dq\,V(Q',q^2) \frac{q^2}{q^2-A} T(q^2,Q;A)\,, \label{LS}
\end{eqnarray}
whose solution is
\begin{eqnarray}
T(Q',Q;A) &=& \frac{1}{M_N}\, (Q'-L)^{1/4}\,(Q-L)^{1/4} \, \frac{\Big[I_6(A)+\frac{c_0}{c_2^2}\Big]-\Big[I_4(A)-\frac{1}{c_2}\Big]\left(Q'+Q\right)+I_{\,2}(A)\,Q'Q}{\Big[I_4(A)-\frac{1}{c_2}\Big]^2-I_{\,2}(A)\Big[I_6(A)+\frac{c_0}{c_2^2}\Big]}\,. \label{T}
\end{eqnarray}
Since the potential \eqref{potential} is badly divergent in the ultraviolet (i.e. singular), the use of some regularization technique, say a momentum cutoff $\Lambda$, is mandatory for the solution \eqref{T} to make sense. We have thus introduced the sharp-cutoff regularized integrals ($n=1,2,3$)
\begin{eqnarray}
I_{2n}\,(A) &=& \frac{1}{2\pi^2}\int_0^\Lambda dq\,\frac{q^{2n}\,{\sqrt{q^2-L}}}{q^2-A} \notag \\
&=& \frac{1}{2\pi^2} \Bigg[-\,A^{n-1}\sqrt{\left(L-A\right)A}\,\,\arctan\Bigg(\frac{\sqrt{A-L}}{\sqrt{-\,A\left(1-{L}/{\Lambda^2}\right)}}\Bigg)\,+\,\Lambda^{2n} \sum_{j=0}^n \beta_{\,2n,j}(\Lambda)\,A^j  \,\Bigg]\,. \label{I2n}
\end{eqnarray}
The $A$-independent coefficient $\beta_{\,2n,j}$ has mass dimension $-2j$ and runs with the cutoff so that it converges to 
$\delta_{j0}/(2n)$ 
when $\Lambda\to\infty$. Explicitly,
\begin{eqnarray}
\beta_{\,2,0}(\Lambda) &=& -\frac{L}{2\Lambda^2}\lambda(\Lambda)\,+\,\frac{1}{2}\mu(\Lambda)\,, \label{beta20} \\ 
\beta_{\,2,1}(\Lambda) &=& +\frac{1}{\Lambda^2}\lambda(\Lambda)\,; \\ \notag \\
\beta_{\,4,0}(\Lambda) &=& -\frac{L^2}{8\Lambda^4}\lambda(\Lambda)\,+\,\left(\frac{1}{4}-\frac{L}{8\Lambda^2}\right)\mu(\Lambda)\,, \\
\beta_{\,4,1}(\Lambda) &=&-\frac{L}{2\Lambda^4}\lambda(\Lambda)\,+\,\frac{1}{2\Lambda^2}\mu(\Lambda)\,, \\
\beta_{\,4,2}(\Lambda)&=&+\frac{1}{\Lambda^4}\lambda(\Lambda)\,;  \\ \notag \\
\beta_{\,6,0}(\Lambda) &=& - \frac{L^3}{16\Lambda^6}\lambda(\Lambda) \,+\, \left(\frac{1}{6}-\frac{L}{24\Lambda^2}-\frac{L^2}{16\Lambda^4}\right)\mu(\Lambda)\,, \\
\beta_{\,6,1}(\Lambda) &=& - \frac{L^2}{8\Lambda^6}\lambda(\Lambda)\,+\,\left(\frac{1}{4\Lambda^2}-\frac{L}{8\Lambda^4}\right)\mu(\Lambda)\,, \\
\beta_{\,6,2}(\Lambda) &=& - \frac{L}{2\Lambda^6}\lambda(\Lambda)\,+\,\frac{1}{2\Lambda^4}\mu(\Lambda)\,, \\
\beta_{\,6,3}(\Lambda) &=& +\frac{1}{\Lambda^6}\lambda(\Lambda)\,, \label{beta63}
\end{eqnarray}
with the abbreviations
\begin{eqnarray}
\lambda(\Lambda) &=& \log\left(\frac{\Lambda}{\sqrt{-L}}\,\Big(1+\mu(\Lambda)\,\Big)\right)\,; \\
\mu(\Lambda) &=& \sqrt{1-\frac{L}{\Lambda^2}}\,.
\end{eqnarray}
Note that, had one taken other momentum-space regulator --- for instance, an either local or non-local Gau{\ss}ian prescription --- the coefficients \eqref{beta20}--\eqref{beta63} and the very functional dependence of $I_{2n}\,(A)$ \eqref{I2n} would have differed in general. Nonetheless, the large-$\Lambda$ behavior $I_{2n}\,(A) \propto \Lambda^{2n}$ must be preserved. 

\bigskip
Having $\Lambda_0$ in the full vertex function \eqref{vertexS} might remind us of some kind of `form factor'. Yet it should be emphasized that $\Lambda_0$ and $\Lambda$ stand for different meanings.  While $\Lambda_0$ is a physical parameter of the original potential, $\Lambda$ is a mathematical tool introduced to ensure the convergence of the loop integrals. Renormalization amounts to make sure that the resulting amplitude is free of any cutoff dependence that does not vanish in the cutoff-removal limit, see Refs. \cite{VanKolck:2020eyy,VanKolck:2020hiu} and references therein for discussions in the contexts of contact and chiral EFTs respectively. However, we mention that other authors have advocated alternative interpretations of what renormalization means and how it should be carried out, see e.g. Refs. \cite{Epelbaum:2009jkm,Epelbaum:2013nou}.

\bigskip
Here, renormalization is achieved by imposing certain conditions on the low-energy behavior of the observable on-shell amplitude 
\begin{eqnarray}
T(A) &\equiv& T(A,A;A) \notag \\
&=& \frac{N(A)}{D(A)}\,, \label{nd}
\end{eqnarray}
$N(A)$ and $D(A)$ having respectively the LHC and the RHC only. By choosing the normalization $D(0)=1$, we take 
\begin{eqnarray}
N(A) &=& \frac{\sqrt{A-L}}{M_N\,\mathcal{K}\!(\Lambda)}\,;  \label{n}\\ 
D(A) &=& \frac{1}{\mathcal{K}\!(\Lambda)} \frac{\Big[I_4(A)-\frac{1}{c_2}\Big]^2-I_{\,2}(A)\Big[I_6(A)+\frac{c_0}{c_2^2}\Big]}{\Big[I_6(A)+\frac{c_0}{c_2^2}\Big]-2A\Big[I_4(A)-\frac{1}{c_2}\Big]+A^2\,I_{\,2}(A)} \,, \label{d} \end{eqnarray}
with
\begin{eqnarray}
\mathcal{K}\!(\Lambda)&=& \frac{\Big[I_4(0)-\frac{1}{c_2}\Big]^2}{\Big[I_6(0)+\frac{c_0}{c_2^2}\Big]} - I_{\,2}(0) \notag \\ 
 &=& \frac{1}{2\pi^2} \Big[\frac{(c_2\,\beta_{\,4,0}(\Lambda) \,-\, 2\pi^2)^2}{c_2^2\,\beta_{\,6,0}(\Lambda) + 2\pi^2\,c_0} - \beta_{\,2,0}(\Lambda)\Big]\,. \label{K}
\end{eqnarray}
In particular, $T(A)$ has to fit exactly (i.e. at any cutoff) the $S$-wave ERE at low energies $0<A\ll -L$,
\begin{eqnarray}
\Big[\frac{M_N}{4\pi}\,T(A)\Big]^{-1}\,+\,i\,\sqrt{A}&=& \sqrt{A}\,\cot\delta(A)  \notag\\
&=& -\frac{1}{a}\,+\,\frac{r}{2}\,A\,+\,\ldots \label{kcd}
\end{eqnarray}
--- the knowledge of these two free inputs, the scattering length $a$ and the effective range $r$, allows one to find the running equations for the two couplings $c_0$ and $c_2$. The full result is pretty cumbersome and will not be given here. However, for large cutoffs it turns out
\begin{eqnarray}
c_0^{(\pm)}(\Lambda) &=& -\frac{4\pi^2}{3\Lambda^2}\, \Bigg[\,4\,\pm\,\frac{8}{\sqrt{\log\,(\frac{\Lambda_{\text{crit}}^{(\text{trunc})}}{\Lambda})}}\,+\,\frac{1}{\log\,(\frac{\Lambda_{\text{crit}}^{(\text{trunc})}}{\Lambda})}\,+\,\dots\,\Bigg]\,; \label{c0}\\
c_2^{(\pm)}(\Lambda) &=& +\frac{8\pi^2}{\Lambda^4}\, \Bigg[\,1\,\pm\,\frac{1}{\sqrt{\log\,(\frac{\Lambda_{\text{crit}}^{(\text{trunc})}}{\Lambda})}}\,+\,\dots\,\Bigg]\,, \label{c2}
\end{eqnarray}
where the ellipses stand for smaller cutoff corrections, and
\begin{eqnarray}
\Lambda_{\text{crit}}^{(\text{trunc})} \quad=\quad \frac{\sqrt{-L}}{2}\exp\left(2\,+\,\frac{\pi}{4}\,\Big(\frac{1}{a\,\sqrt{-L}}\,-\,r\,\sqrt{-L}\Big)\,\right) \quad=\quad 236\,\text{MeV}\,, \label{star}
\end{eqnarray}
having used the physical values $\sqrt{-L}\equiv m_\pi/2=70\,\text{MeV}$, $a = 5.4\,\text{fm}$, $r = 1.8\,\text{fm}$. Such a result matches the condition $\Lambda_{\text{crit}}^{(\text{trunc})}>m_\pi/2$. It is a reasonable, typical cutoff value in a contact theory (so-called `pionless EFT' in the nuclear sector) where nucleons are the only explicit degrees of freedom \cite{VanKolck:1999zys}, as one indeed has $\Lambda_{\text{crit}}^{(\text{trunc})}\gtrsim M_{\text{hi}} \sim m_\pi$, with $M_{\text{hi}}$ the expected high-momentum (or breakdown) scale of such an EFT. 

\bigskip
However,  Eqs. \eqref{c0}--\eqref{c2} anticipate that taking $\Lambda\gtrsim\Lambda_{\text{crit}}^{(\text{trunc})}$ will induce a non-vanishing imaginary part of the potential. That is to say, using too hard cutoffs would render a non-Hermitian bare Hamiltonian. As it is well-known, such a feature is potentially problematic. [The situation is even worse in the $^1S_0$ channel, as taking $a = -23.7\,\text{fm}$, $r = 2.7\,\text{fm}$ in Eq. \eqref{star} gives $\Lambda_{\text{crit}}^{(\text{trunc})}=111\,\text{MeV}<m_\pi$.] At the same time, this issue reminisces the `Wigner bound'  \cite{Wigner:1955zz} derived from general principles for potentials that vanish identically beyond some radius. A similar situation would persist after suppressing the LHC of the interaction invoked here, which nominally implies $L\to-\infty$, --- as done in `standard' pionless EFT \cite{VanKolck:1999zys} --- unless one assumed the condition $r<0$ \cite{Phillips:1996ae,Phillips:1997xu}.

\begin{figure*}[t]
  \begin{center}
\includegraphics[scale=0.35]{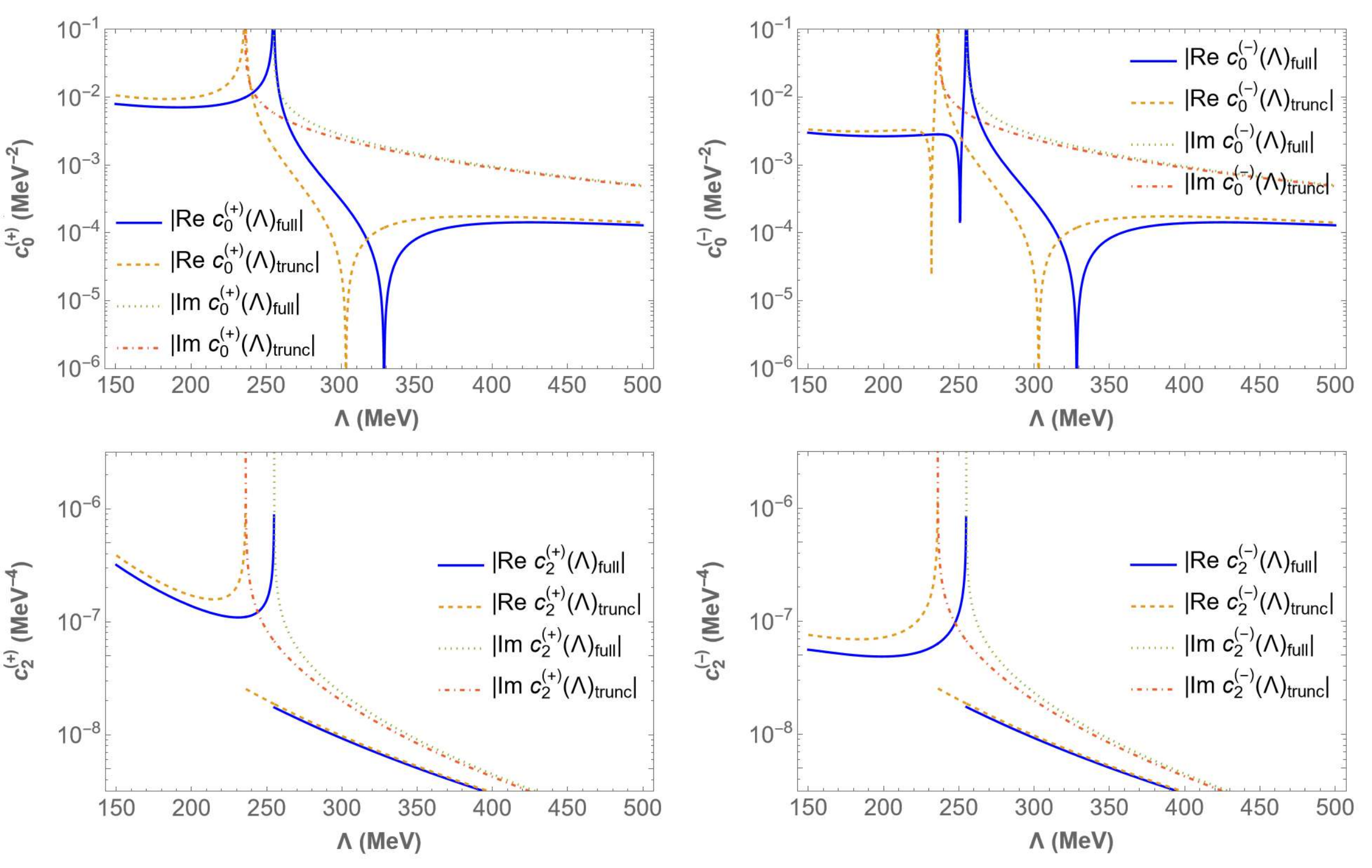}
\caption{{\small From left to right, top to bottom, counterterms $c_0^{(+)}$ and $c_0^{(-)}$ (in MeV$^{-2}$) and $c_2^{(+)}$ and $c_2^{(-)}$ (in MeV$^{-4}$) as a function of the cutoff $\Lambda$ (in MeV). The solid and dashed (dotted and dotdashed) lines on the upper [lower] panels refer to the running of the absolute values of the real (imaginary) parts of the full and truncated results for $c_0$ [$c_2$].}}
\label{c0c2}
  \end{center}
\end{figure*}

\bigskip
In Fig. \ref{c0c2} we give a plot of the functions $c_0^{(\pm)}(\Lambda)$ and $c_2^{(\pm)}(\Lambda)$, both full and truncated, where the latter are given by omitting the ellipses in Eqs. \eqref{c0}--\eqref{c2}. We highlight the following elements:
\begin{itemize}
\item The full case converges to the truncated one when $\Lambda$ is increased. 
\item The imaginary parts of the full $c_0^{(\pm)}(\Lambda)$ and $c_2^{(\pm)}(\Lambda)$ become non-zero at $\Lambda=\Lambda_{{\text{crit}}}^{(\text{full})}=255\,\text{MeV}$, which is slightly ($8\,\%$) shifted to the right as compared to $\Lambda_{\text{crit}}^{(\text{trunc})}$ due to the small corrections neglected in the truncated case.
\item The real parts of the full (truncated) $c_0^{(\pm)}(\Lambda)$ and $c_2^{(\pm)}(\Lambda)$ exhibit discontinuities at $\Lambda=\Lambda_{{\text{crit}}}^{(\text{full})}$ ($\Lambda=\Lambda_{\text{crit}}^{(\text{trunc})}$).
\end{itemize}

\bigskip

\begin{figure*}[htt!]
  \begin{center}
\includegraphics[scale=0.43]{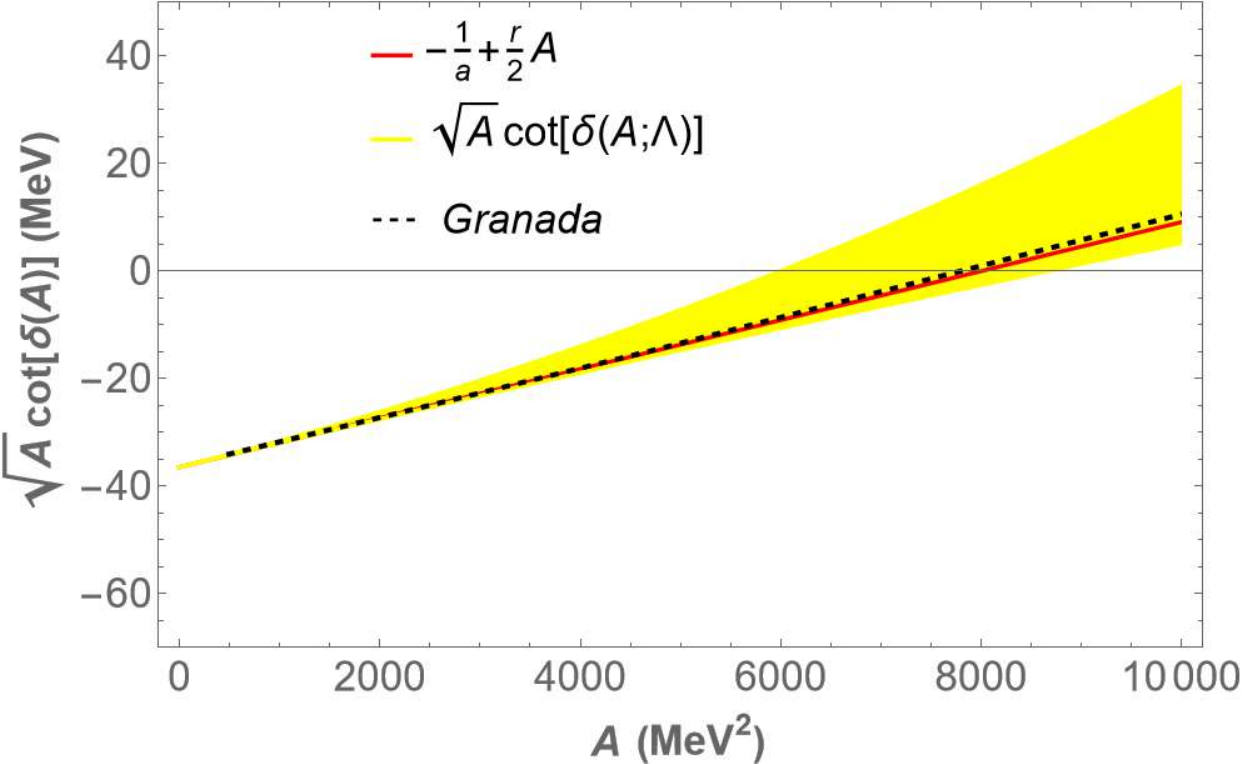}
\caption{\begin{small}
Effective-range function $\sqrt{A}\,\cot\delta$ (in MeV) as a function of the squared
center-of-mass momentum $A$ (in MeV$^2$). The solid straight line is the truncated ERE. The dashed line is the phenomenological curve given by the \textit{Granada} partial-wave analysis \cite{Navarro:2013zvu,Navarro:2013ujy}. The band stands for the full effective-range function coming out from our potential; the width of this band corresponds to a cutoff interval $\Lambda\in[250\,\text{MeV},\infty)$.
\end{small} } \label{erf}
\end{center}
  \end{figure*}

Now that the counterterms at a given cutoff are fixed, the effective-range function $\sqrt{A}\,\cot\delta(A)$ is readily found from the first equality of Eq. \eqref{kcd}. In Fig. \ref{erf} we provide a plot of such a function in the low-momentum regime ($\sqrt{A}\leqslant100\,\text{MeV}$) corresponding to three cases: \textit{a)} the straight line coming out from the second line of Eq. \eqref{kcd}; \textit{b)} the full result due to the potential \eqref{potS}, with a cutoff variation between $\Lambda=250\,\text{MeV}$ and infinity; \textit{c)} the phenomenology given by the \textit{Granada} database \cite{Navarro:2013zvu,Navarro:2013ujy} for $^3S_1$ neutron-proton scattering.
A few remarks are in order:
\begin{itemize}
\item The intercept and the slope of the three curves match when $A\to0$, confirming that the counterterms are indeed well fitted to reproduce the experimental values of $a$ and $r$.
\item In the low-momentum region we are looking at, the \textit{Granada} curve is nicely reproduced by the truncated ERE, indicating that shape-parameter corrections will only get significant at somewhat larger momenta, {coherently with the well-known fact that the $^3S_1$ phase shifts are quite well reproduced already at low orders of pionless EFT (especially in comparison with the case for the $^1S_0$ ones \cite{SanchezSanchez:2017tws})}.
\item In contrast, our result seems to overestimate such shape-parameter contributions, mainly when soft cutoffs are used, while pushing $\Lambda$ brings the curve closer to the phenomenological data.
\end{itemize}
Besides, it is worth noticing that the solution does exist (i.e. the phase shifts keep being well-defined) even when the cutoff is so large to render complex $c_0$ and $c_2$ \cite{Entem:2007jg}.

\bigskip
\bigskip

\section{The coefficients $\pmb{\theta_n^{[\ell'\ell]}}$ ($\pmb{n=0,1,\ldots,5+\overline{\ell}}$)}
\label{app.231227.1}
\def\theequation{\Alph{section}.\arabic{equation}}
\setcounter{equation}{0}   

\bigskip
We list below the expressions for the $\theta_n^{[\ell'\ell]}$ coefficients used in Eqs.~\eqref{r00M}--\eqref{r22M}:
\begin{eqnarray}
\theta_0^{[00]}&=&\Lambda_0^{5/2} \mathtt{d}^{[0]} \left({C_0(\Lambda)} (\mathtt{d}^{[0]})^2+(1-{C_0(\Lambda)} \mathtt{c}_{\slashed{\pi}}^{[0]}) (\mathtt{c}_0^{[0]}-{\widehat{\gamma}_{\,22}})\right)\,, \label{thetafirst} \\
\theta_1^{[00]}&=&(1-{C_0(\Lambda)} \mathtt{c}_{\slashed{\pi}}^{[0]}) {\Lambda_0} (\mathtt{d}^{[0]})^2\,, \\
\theta_2^{[00]}&=&0\,, \\
\theta_3^{[00]}&=&-2 {C_0(\Lambda)} \Lambda_0^4 \left(\mathtt{c}_0^{[1]} (\mathtt{d}^{[0]})^2-2 \mathtt{d}^{[1]} (\mathtt{c}_0^{[0]}-{\widehat{\gamma}_{\,22}}) \mathtt{d}^{[0]}+\mathtt{c}_{\slashed{\pi}}^{[1]} (\mathtt{c}_0^{[0]}-{\widehat{\gamma}_{\,22}})^2\right)\,, \\
\theta_4^{[00]}&=&4 \Lambda_0^{5/2} ((1-{C_0(\Lambda)} \mathtt{c}_{\slashed{\pi}}^{[0]}) (\mathtt{c}_0^{[0]} (\mathtt{d}^{[1]}-\mathtt{d}^{[0]})-\mathtt{c}_0^{[1]} \mathtt{d}^{[0]})+{C_0(\Lambda)} \mathtt{d}^{[0]} (\mathtt{c}_0^{[0]} \mathtt{c}_{\slashed{\pi}}^{[1]}-\mathtt{d}^{[0]} (\mathtt{d}^{[0]}+\mathtt{d}^{[1]}))\notag\\
&&+\,((1-{C_0(\Lambda)} \mathtt{c}_{\slashed{\pi}}^{[0]}) (\mathtt{d}^{[0]}-\mathtt{d}^{[1]})-{C_0(\Lambda)} \mathtt{c}_{\slashed{\pi}}^{[1]} \mathtt{d}^{[0]}) {\widehat{\gamma}_{\,22}})\,, \\
\theta_5^{[00]}&=&-2 {\Lambda_0} \mathtt{d}^{[0]} ({C_0(\Lambda)} \mathtt{c}_{\slashed{\pi}}^{[1]} \mathtt{d}^{[0]}+2 (1-{C_0(\Lambda)} \mathtt{c}_{\slashed{\pi}}^{[0]}) (\mathtt{d}^{[0]}+\mathtt{d}^{[1]}))\,; \\
\theta_0^{[02]}&=&\frac{1}{2} {C_0(\Lambda)} \Lambda_0^{11/2} \mathtt{d}^{[0]} {\widehat{\gamma}_{\,02}} (\mathtt{c}_0^{[0]}-{\widehat{\gamma}_{\,22}})\,, \\
\theta_1^{[02]}&=&-2 {C_0(\Lambda)} \Lambda_0^4 (\mathtt{d}^{[0]})^2 {\widehat{\gamma}_{\,02}}\,, \\
\theta_2^{[02]}&=&-\frac{5}{2} \Lambda_0^{5/2} (1-{C_0(\Lambda)} \mathtt{c}_{\slashed{\pi}}^{[0]}) \mathtt{d}^{[0]} {\widehat{\gamma}_{\,02}}\,, \\ 
\theta_3^{[02]}&=& 0\,, \\ 
\theta_4^{[02]}&=& 2 {C_0(\Lambda)} \Lambda_0^{11/2} {\widehat{\gamma}_{\,02}} (\mathtt{c}_0^{[1]} \mathtt{d}^{[0]}+({\widehat{\gamma}_{\,22}}-\mathtt{c}_0^{[0]}) (2 \mathtt{d}^{[0]}+\mathtt{d}^{[1]}))\,, \\ 
\theta_5^{[02]}&=& 2 \Lambda_0^4 {\widehat{\gamma}_{\,02}} \Bigg[6 {C_0(\Lambda)} (\mathtt{d}^{[0]})^2+\mathtt{c}_0^{[1]} (1-{C_0(\Lambda)} \mathtt{c}_{\slashed{\pi}}^{[0]})-\mathtt{c}_0^{[0]} (1-{C_0(\Lambda)} (\mathtt{c}_{\slashed{\pi}}^{[0]}+\mathtt{c}_{\slashed{\pi}}^{[1]}))\notag\\
&&-\,{C_0(\Lambda)} (\mathtt{c}_{\slashed{\pi}}^{[0]}+\mathtt{c}_{\slashed{\pi}}^{[1]}) {\widehat{\gamma}_{\,22}}+{\widehat{\gamma}_{\,22}}\Bigg]\,, \\ 
\theta_6^{[02]}&=&2 \Lambda_0^{5/2} ((-8 {C_0(\Lambda)} \mathtt{c}_{\slashed{\pi}}^{[0]}-{C_0(\Lambda)} \mathtt{c}_{\slashed{\pi}}^{[1]}+8) \mathtt{d}^{[0]}-(1-{C_0(\Lambda)} \mathtt{c}_{\slashed{\pi}}^{[0]}) \mathtt{d}^{[1]}) {\widehat{\gamma}_{\,02}}\,; \\
\theta_0^{[22]} &=&{C_0(\Lambda)} \Lambda_0^7 (\mathtt{c}_0^{[0]}-{\widehat{\gamma}_{\,22}}) \left(\mathtt{c}_0^{[0]} (\mathtt{c}_2^{[0]}-{\widehat{\gamma}_{\,00}})-\mathtt{c}_2^{[0]} {\widehat{\gamma}_{\,22}}+\frac{1}{{|\bm{\gamma}\,|}}\right)\,, \\
\theta_1^{[22]} &=&-{C_0(\Lambda)} \Lambda_0^{11/2} \mathtt{d}^{[0]} \left(4 \mathtt{c}_0^{[0]} (\mathtt{c}_2^{[0]}-{\widehat{\gamma}_{\,00}})-(4 \mathtt{c}_2^{[0]}-{\widehat{\gamma}_{\,00}}) {\widehat{\gamma}_{\,22}}+\frac{3}{{|\bm{\gamma}\,|}}\right)\,, \\
\theta_2^{[22]} &=&2 \Lambda_0^4 \left(2 {C_0(\Lambda)} (\mathtt{c}_2^{[0]}-{\widehat{\gamma}_{\,00}}) (\mathtt{d}^{[0]})^2+(1-{C_0(\Lambda)} \mathtt{c}_{\slashed{\pi}}^{[0]}) \left(-\mathtt{c}_0^{[0]} \mathtt{c}_2^{[0]}+{\widehat{\gamma}_{\,22}} \mathtt{c}_2^{[0]}+\mathtt{c}_0^{[0]} {\widehat{\gamma}_{\,00}}-\frac{1}{{|\bm{\gamma}\,|}}\right)\right)\,, \\
\theta_3^{[22]}&=&4 (1-{C_0(\Lambda)} \mathtt{c}_{\slashed{\pi}}^{[0]}) \Lambda_0^{5/2} \mathtt{d}^{[0]} (\mathtt{c}_2^{[0]}-{\widehat{\gamma}_{\,00}})\,, \\
\theta_4^{[22]}&=&2 {C_0(\Lambda)} \Lambda_0^7 \Bigg[-\left((4 \mathtt{c}_2^{[0]}+\mathtt{c}_2^{[1]}) ({\widehat{\gamma}_{\,22}}-\mathtt{c}_0^{[0]})^2\right)+4 \left(\mathtt{c}_0^{[0]} {\widehat{\gamma}_{\,00}}-\frac{1}{{|\bm{\gamma}\,|}}\right) \mathtt{c}_0^{[0]}+\frac{\mathtt{c}_0^{[1]}}{{|\bm{\gamma}\,|}}\notag\\
&&+\,\left(\frac{4}{{|\bm{\gamma}\,|}}-(4 \mathtt{c}_0^{[0]}+\mathtt{c}_0^{[1]}) {\widehat{\gamma}_{\,00}}\right) {\widehat{\gamma}_{\,22}}\Bigg]\,, \\
\theta_5^{[22]}&=&4 {C_0(\Lambda)} \Lambda_0^{11/2} \Bigg[-\frac{\mathtt{d}^{[1]}}{{|\bm{\gamma}\,|}}+\mathtt{d}^{[0]} \left(8 \mathtt{c}_0^{[0]} \left(\mathtt{c}_2^{[0]}+\frac{1}{4} \mathtt{c}_2^{[1]}-{\widehat{\gamma}_{\,00}}\right)+\frac{5}{{|\bm{\gamma}\,|}}\right)\notag\\
&&+\,((-8 \mathtt{c}_2^{[0]}-2 \mathtt{c}_2^{[1]}+3 {\widehat{\gamma}_{\,00}}) \mathtt{d}^{[0]}+\mathtt{d}^{[1]} {\widehat{\gamma}_{\,00}}) {\widehat{\gamma}_{\,22}}\Bigg]\,, \\
\theta_6^{[22]}&=& 2 {C_0(\Lambda)} \Lambda_0^4 \left[16 \left(-\mathtt{c}_2^{[0]}+{\widehat{\gamma}_{\,00}}-\frac{1}{4} \mathtt{c}_2^{[1]}\right) (\mathtt{d}^{[0]})^2+\mathtt{c}_{\slashed{\pi}}^{[1]} \left(\frac{1}{{|\bm{\gamma}\,|}}-{\widehat{\gamma}_{\,00}} {\widehat{\gamma}_{\,22}}\right)\right] \notag\\
&&+\,12\Lambda_0^4 (1-{C_0(\Lambda)} \mathtt{c}_{\slashed{\pi}}^{[0]}) \left[\frac{4}{3} \left(\mathtt{c}_2^{[0]}+\frac{1}{4} \mathtt{c}_2^{[1]}-{\widehat{\gamma}_{\,00}}\right) (\mathtt{c}_0^{[0]}-{\widehat{\gamma}_{\,22}})-{\widehat{\gamma}_{\,00}} {\widehat{\gamma}_{\,22}}+\frac{1}{{|\bm{\gamma}\,|}}\right]\,, \\
\theta_7^{[22]}&=& 32 (1-{C_0(\Lambda)} \mathtt{c}_{\slashed{\pi}}^{[0]}) \Lambda_0^{5/2} \mathtt{d}^{[0]} \left(-\mathtt{c}_2^{[0]}+{\widehat{\gamma}_{\,00}}-\frac{1}{4} \mathtt{c}_2^{[1]}\right)\,, \label{thetalast}
\end{eqnarray}
with the shorthand
\begin{eqnarray}
\widehat{\gamma}_{\,\ell'\ell} &\equiv& \frac{{\gamma}_{\,\ell'\ell}}{|\bm{\gamma}\,|}\,.
\end{eqnarray}

\bigskip
\bigskip
\bigskip

\bibliographystyle{unsrt}
\bibliography{references}

\end{document}